\documentclass[aoas]{imsart}

\RequirePackage{amsthm,amsmath,amsfonts,amssymb}
\RequirePackage[authoryear]{natbib}
\RequirePackage[colorlinks,citecolor=blue,urlcolor=blue]{hyperref}
\RequirePackage{graphicx} 
\usepackage{amscd}
\usepackage{epsfig,color, epstopdf, bm}
\usepackage{amsbsy}
\usepackage{diagbox}
\usepackage{rotating}
\usepackage[ruled,vlined]{algorithm2e}
\usepackage{algpseudocode}
\usepackage[english]{babel}
\usepackage{booktabs} 
\usepackage{adjustbox}
\usepackage{comment}
\usepackage{xcolor}
\usepackage{url}
\usepackage{afterpage}
\usepackage{lipsum}
\usepackage{multirow}
\usepackage[flushleft]{threeparttable}
\usepackage{pdflscape}
\usepackage{bbm}

\startlocaldefs
\theoremstyle{plain}

\def\sym#1{\ifmmode^{#1}\else\(^{#1}\)\fi}

\theoremstyle{definition}
\newtheorem{assumption}{Assumption}

\def\H{\bm H}
\def\X{\bm X}

\def\Z{\bm Z}

\newcommand{\PP}{\mathbb P}

\DeclareMathOperator*{\argmin}{arg\,min}
\DeclareMathOperator*{\argmax}{arg\,max}

\endlocaldefs

\begin{document}

\begin{frontmatter}
\title{Risk-Aware Quantile Learning for Personalized Dynamic Treatment Regimes}

\begin{aug}
\author[A]{\fnms{Chunyin}~\snm{Lei}\ead[label=e1]{chunyin\_lei@ucsb.edu}}
\author[A]{\fnms{Annie}~\snm{Qu}\ead[label=e2]{aqu2@ucsb.edu}}

\address[A]{Department of Statistics and Applied Probability, University of California, Santa Barbara\printead[presep={,\ }]{e1,e2}}
\end{aug}

\begin{abstract}
Sequential clinical decision-making often involves more than maximizing average efficacy. Clinicians may need to simultaneously optimize clinically relevant tails of the outcome distribution, control treatment-related risk, and choose among multiple treatment options. Existing quantile dynamic treatment regime (DTR) methods capture distributional features of treatment outcomes but remain largely restricted to efficacy-only objectives and binary treatments. To address these limitations, we propose Risk-Aware Quantile Dynamic Treatment Regimes (RQDTR), a unified framework that optimizes a prespecified quantile of the cumulative potential outcome while explicitly incorporating treatment-related risk. We also develop an angle-based formulation for jointly learning decision rules across multiple treatment categories. Our framework includes three interpretable subclasses: efficacy-only quantile learning, constraint-based learning with population-level risk control, and utility-based learning through a composite benefit-risk utility. Theoretically, we establish identification and oracle equivalence, surrogate and estimator consistency under increasing quadratic penalties, asymptotic risk feasibility, and finite-sample performance error bounds. Extensive simulation studies and applications to All of Us major depressive disorder and MIMIC-III sepsis data demonstrate that RQDTR improves tail-oriented efficacy and achieves more favorable benefit-risk trade-offs than existing quantile DTR methods.
\end{abstract}


\end{frontmatter}

\section{Introduction}

\subsection{Motivation}
Personalized treatment often involves a sequence of decisions. Clinicians repeatedly update treatment choices in response to evolving patient histories, intermediate outcomes, and adverse-event information. Dynamic treatment regimes (DTRs) provide a rigorous statistical framework for such sequential individualized decision-making and have become an important tool for personalized chronic-disease management. At each decision stage, a DTR maps a patient's observed history to a recommended treatment, with the goal of optimizing long-term clinical outcomes \citep{murphy2003optimal, chakraborty2014dynamic}. 

A central question in DTR learning is how the clinical value of a treatment regime should be defined. In many applications, efficacy alone provides an incomplete measure of treatment value because improvements in the primary outcome may be accompanied by adverse events or other treatment-related burdens. Antidepressant management for major depressive disorder (MDD) illustrates this benefit-risk tension. Treatment response and tolerability are both highly heterogeneous, and antidepressant classes have distinct adverse-event profiles, including metabolic, cardiovascular, gastrointestinal, neurologic, sleep-related, anxiety-related, hepatic, and sexual-function outcomes \citep{wang2018addressing}. In the iSPOT-D study of 1,008 adults with MDD, 46.9\% of participants experienced an adverse event during the 8-week treatment period, and 79.1\% of the reported adverse events were considered related to antidepressant treatment \citep{braund2021antidepressant}. Sexual-function burden also differs substantially across antidepressant choices. For example, \citet{modell1997comparative} reported adverse sexual effects in 73\% of patients treated with selective serotonin reuptake inhibitors, compared with 14\% of those treated with bupropion. These findings suggest that treatment-related risk should enter the learning objective rather than be treated only as a secondary evaluation metric. Sepsis resuscitation provides a complementary acute-care example. Fluids and vasopressors are central therapies, but treatment intensity must be calibrated against physiological risk signals. The Surviving Sepsis Campaign recommends restoring adequate organ perfusion, with mean arterial pressure and lactate serving as clinically meaningful indicators of hemodynamic support and tissue perfusion \citep{evans2021surviving}. Sepsis-3 also uses elevated lactate together with vasopressor requirement to define septic shock \citep{singer2016third}. A regime for sepsis resuscitation should therefore not only promote short-term organ-function recovery but also account for clinically anchored risk burdens, such as low mean arterial pressure and persistent hyperlactatemia.

Most existing DTR methods optimize the mean potential outcome, but the mean can hide clinically important distributional heterogeneity. Evidence from the Sequenced Treatment Alternatives to Relieve Depression (STAR*D) trial illustrates this limitation. Even under structured care, remission rates across the four acute treatment steps were 36.8\%, 30.6\%, 13.7\%, and 13.0\%, respectively \citep{rush2006acute}. Thus, a regime with a favorable average symptom reduction may still provide inadequate benefit for patients in the lower tail of the response distribution. Conversely, in recovery-oriented settings such as sepsis resuscitation, upper-tail quantiles may be scientifically relevant when the goal is to identify regimes associated with strong short-term recovery of organ function. Quantile-based criteria allow the learning target to focus directly on a prespecified, clinically relevant part of the potential outcome distribution rather than only on its mean. This flexibility is particularly important in sequential settings, where repeated treatment assignments can amplify heterogeneity in both efficacy and risk over time.

Another practical challenge is that clinical decisions often involve choosing among several treatment options rather than making a simple binary choice. In our motivating applications, clinicians choose among four antidepressant classes or four levels of resuscitation intensity at each decision point. Reducing these decisions to a series of binary comparisons can discard clinically meaningful distinctions among treatment options, whereas fitting separate pairwise rules does not directly produce a coherent sequential policy over all available treatments. 

The considerations discussed above motivate a single treatment-regime learning framework that simultaneously addresses four practical requirements: targeting a clinically relevant tail of the outcome distribution, controlling treatment-related risk, choosing among multiple treatment options, and optimizing decisions over multiple stages. To address this joint problem, we propose a novel quantile DTR method. It optimizes a prespecified quantile of cumulative efficacy while incorporating treatment-related risk through population-level feasibility constraints and/or a stagewise benefit-risk utility function. Within our framework, an angle-based formulation jointly learns all treatment options at each decision stage.

\subsection{Literature Review}
A broad literature has developed model-based and value-search methods for estimating optimal treatment regimes \citep[e.g.,][]{murphy2003optimal, murphy2005generalization, zhang2012robust, zhao2012estimating, zhao2015new}. Despite their different formulations, most existing methods optimize the mean cumulative outcome. Such a target can be misaligned with clinical objectives that prioritize protection against poor response in the lower tail or promotion of strong recovery in the upper tail.

Quantile DTR methods address this mismatch by optimizing a prespecified quantile of the potential-outcome distribution. \citet{linn2017interactive} extended interactive Q-learning to optimize probabilities and quantiles of the outcome distribution, but its performance depends on the correct specification of the outcome regression model. \citet{wang2018quantile} developed a model-free approach that formulates quantile-optimal regime estimation as an optimization problem with a nuisance parameter and proposed a doubly robust estimator. Later, \citet{fang2023fairness} introduced a fairness-oriented framework that maximizes the expected outcome subject to a lower-tail performance constraint. More recently, \citet{xia2026successive} reformulated quantile optimization as a sequence of weighted classification tasks through successive classification learning.

However, these advances do not resolve the joint decision problem arising in our motivating applications. Two gaps are particularly relevant. First, existing  quantile DTRs evaluate treatment policies based on efficacy alone, even though more aggressive treatment may improve efficacy at the cost of increased adverse events or cumulative treatment burden. Risk-constrained methods \citep{wang2018learning, zhu2024risk, liu2024controlling, liu2024learning} and utility-based benefit-risk trade-offs \citep{lee2015bayesian, butler2018incorporating, luckett2021estimation} have been studied for mean-based DTRs, whereas comparable tools for quantile-oriented DTR learning remain limited. Second, existing quantile DTR methods are restricted to binary treatments. Although angle-based learning has been successfully applied for multicategory treatment-rule learning under mean-based objectives \citep{fu2019robust, qi2020multi, zhang2020multicategory, xue2022multicategory}, these methods do not jointly address quantile optimization and treatment-related risk control.

Overall, an important methodological gap remains: existing methods do not jointly address multicategory sequential regime learning, tail-oriented efficacy optimization, and treatment-related risk control. Supplementary Material Section~S.1 provides a detailed comparison with existing treatment-regime learning methods.

\subsection{Contribution}
In this article we propose Risk-Aware Quantile Dynamic Treatment Regimes (RQDTR), a unified framework that optimizes a prespecified quantile of the potential outcome while explicitly incorporating treatment-related risk in a multicategory treatment setting. The main contributions are threefold. Methodologically, we develop a unified risk-aware quantile DTR framework whose specifications yield three interpretable subclasses: efficacy-only quantile learning, hard risk-constrained learning, and utility-based benefit-risk learning. A common angle-based procedure estimates all treatment categories jointly at each stage, avoiding the need to combine independently fitted pairwise rules. 
Theoretically, we establish identification and oracle equivalence, surrogate and estimator consistency under increasing quadratic penalties, and asymptotic risk feasibility. We further derive finite-sample performance bounds and convergence rates relative to the restricted-class oracle, and characterize conditions for recovering the unrestricted optimum. Empirically, simulation studies and applications to All of Us major depressive disorder and MIMIC-III sepsis demonstrate improved tail-oriented efficacy and clinically interpretable benefit-risk trade-offs compared with existing quantile DTR methods. 

The remainder of this article is organized as follows. In Section~\ref{sc:Methodology}, we introduce the main methodology, including notation, problem setup, the proposed framework, and the implementation algorithm. Section~\ref{sc:Theory} establishes the theoretical properties of the proposed method. Extensive simulation studies and two real-data applications are presented in Sections~\ref{sc:Simulation} and \ref{sc:Application}. Section~\ref{sc:Discussion} concludes with a discussion and directions for future research. Technical proofs and additional results are provided in the Supplementary Material.

\section{Methodology}
\label{sc:Methodology}


\subsection{Notation and Setup}
\label{sc:Notation and Setup}
Consider a two-stage longitudinal study with observed data
$\{(\bm X_{1i},A_{1i},Y_{1i},\bm Z_{1i},\bm X_{2i},A_{2i},Y_{2i},\bm Z_{2i})\}_{i=1}^n$. For stage $t=1,2$, $\bm X_t\in\mathcal X_t$ denotes the covariates summarizing the patient's health status, and $A_t\in\mathcal A=\{1,\ldots,K\}$ denotes the treatment assigned based on the observed history $\bm{H}_t \in \mathcal{H}_t$. Let $\bm{H}_1 = (\bm{X}_1)$ and $\bm{H}_2 = (\bm{X}_1, A_1, Y_1, \bm{Z}_1, \bm{X}_2)$. At each stage, we observe the efficacy outcome $Y_t$ and the vector of risk attributes $\bm Z_t=(Z_t^{(1)},\ldots,Z_t^{(M)})^\top$. For $m=1,\ldots,M$, $Z_t^{(m)}$ represents the $m$th risk attribute. Note that the stage-1 risk vector $\bm Z_1$ is observed after the stage-1 treatment and before the stage-2 treatment assignment, allowing interim risk information to inform the subsequent treatment choice. Define the stagewise propensity score as $\pi_t(a | h) = \mathbb{P}(A_t = a | \bm{H}_t = h)$ and the clipped plug-in estimator used in implementation as $\widetilde\pi_t(a | h)=\max\{\widehat\pi_t(a | h),\pi_{\min}\}$. Let $\mathcal D$ be the class of measurable two-stage regimes $d=(d_1,d_2)$ with $d_t: \mathcal{H}_t \to \mathcal{A}$.

In many clinical applications, treatment selection is driven not only by efficacy but also by multiple side effects or treatment burdens. To accommodate such multi-attribute risk profiles, we quantify the stagewise side-effect burden using a risk score 
$$\mathcal{R}_t=\frac{\sum_{m=1}^M \omega_m Z_t^{(m)}}{\sum_{m=1}^M \omega_m}, \ t=1,2,$$ 
where $\omega_m \geq 0$ reflects the clinical severity of the $m$th risk attribute and $\sum_{m=1}^{M} \omega_m > 0$. Each $Z_t^{(m)}$ is normalized to $[0,1]$ before aggregation, which ensures cross-attribute comparability and provides a stable scale for the utility function introduced in Section~\ref{sc:Methodology-Multicategory Risk-Aware Quantile DTRs}.


For a prespecified quantile level $\tau \in (0, 1)$, our primary goal is to identify the optimal two-stage regime $d^{*}$ that maximizes the $\tau$th quantile of the cumulative outcome while explicitly incorporating treatment-related risk. To define counterfactual targets, write $\bar{A}_t = (A_1, \ldots, A_t)$ for the observed treatment history up to stage $t$ and $\bar{a}_t = (a_1, \ldots, a_t) \in \mathcal{A}^{t}$ for any fixed treatment sequence. We assume that the following causal assumptions hold.

\begin{assumption}[Stable Unit Treatment Value Assumption \citep{rubin1980randomization}]
\label{ass:sutva}
There is no interference between subjects, and the potential variables are well-defined under each treatment history. In addition, for each stage $t=1,2$, if $\bar A_t=\bar a_t$, then $(Y_t,\bm Z_t)=(Y_t(\bar a_t),\bm Z_t(\bar a_t))$ almost surely.
\end{assumption}

\begin{assumption}[Sequential Ignorability \citep{robins1986new}]
\label{ass:nuc}
For all treatment histories $\bar a_2=(a_1,a_2)$, treatment assignment is sequentially ignorable conditional on the observed history: $A_1 \perp\!\!\!\perp \{\bm H_2(a_1),Y_2(\bar a_2),\bm Z_2(\bar a_2)\} | \bm H_1$ and $A_2 \perp\!\!\!\perp \{Y_2(\bar a_2),\bm Z_2(\bar a_2)\} | \bm H_2$.
\end{assumption}

\begin{assumption}[Positivity]
\label{ass:positivity}
There exists a constant $\pi_{\min}>0$ such that
$\pi_t(a | \H_t)\ge \pi_{\min}$ for all $a\in\mathcal A$ and $t=1,2$ almost surely.
\end{assumption}

\subsection{Angle-based Learning}
\label{sec:method_Angle-based}
A direct way to handle multicategory treatments is to decompose the problem into multiple binary comparisons, such as one-versus-one or one-versus-rest schemes. However, such decompositions optimize each pairwise rule in isolation and discard the joint structure across treatments, which can be inefficient when the treatment effects share common predictive directions across categories. To learn all $K$ treatments jointly within a single optimization, we adopt angle-based learning \citep{zhang2014multicategory}. The key idea is to encode the $K$ treatments as symmetric vertices of a simplex in $\mathbb R^{K-1}$. A $(K-1)$-dimensional score function is then learned, and treatment assignment is determined by the vertex most aligned with the estimated score function.

\begin{figure}[tb]
\centering
\includegraphics[width=0.75\linewidth]{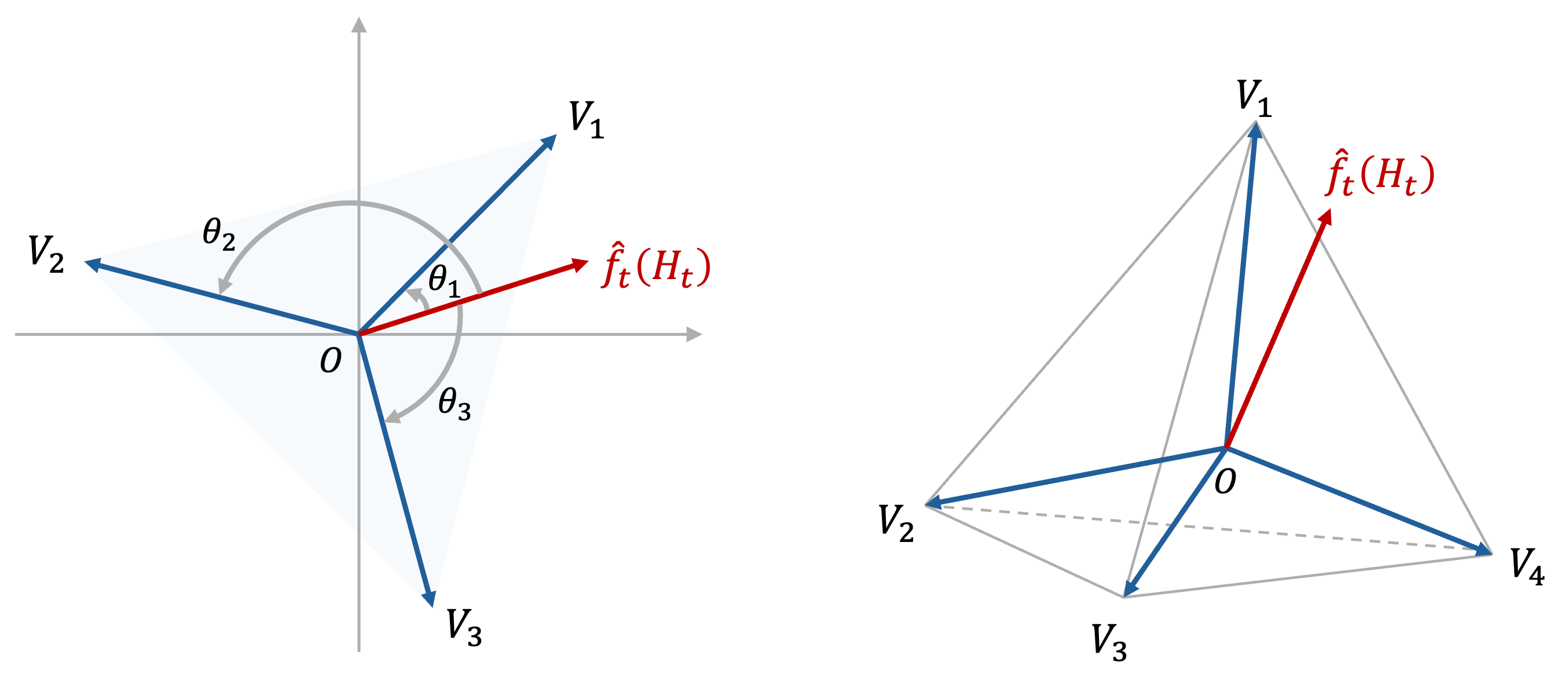}
\caption{Illustration of the angle-based approach with $K=3$ (left) and $K=4$ (right). In both examples, the recommended treatment is treatment 1.}
\label{fig:angle-based}
\end{figure}

Specifically, define a simplex $V$ with $K$ vertices $V_1, \ldots,V_j, \ldots, V_K \in \mathbb{R}^{K-1}$ by
$$V_j =
\begin{cases}
\tfrac{1}{\sqrt{K-1}}\, \bm{1}_{K-1}, & j = 1, \\
-\tfrac{1 + \sqrt{K}}{(K-1)^{3/2}}\, \bm{1}_{K-1} + \sqrt{\tfrac{K}{K-1}}\, \bm{e}_{j-1}, & 2 \leq j \leq K,
\end{cases}$$
where $\bm{1}_{K-1}\in\mathbb{R}^{K-1}$ is the vector of ones and $\bm{e}_j \in \mathbb{R}^{K-1}$ is the standard basis vector with a $1$ in position $j$ and $0$ elsewhere. The vertex $V_j$ represents the $j$th treatment option. By construction, the vertices satisfy $\sum_{j=1}^{K}V_j=\bm 0$ and $\|V_j\|=1$ for each $j$, yielding a balanced and exchangeable representation of the treatment labels. Figure~\ref{fig:angle-based} illustrates the angle-based treatment representation for $K=3$ and $K=4$.

For each stage $t=1,2$, we learn a score function $f_t(\H_t)=\{f_{t,1}(\H_t),\ldots,f_{t,K-1}(\H_t)\}^\top\in\mathbb R^{K-1}$. Each coordinate is represented as $f_{t,r}(h)=b_{t,r}+g_{t,r}(h)$, where $b_{t,r}$ is an intercept and $g_{t,r}$ belongs to a reproducing kernel Hilbert space (RKHS) $\mathcal H_{k_t}$ generated by a positive-definite kernel $k_t$ on the history space $\mathcal H_t$. For a fixed constant $0<C_{\mathcal F}<\infty$, define the RKHS norm-bounded ambient class
\begin{align*}
\mathcal F_t(C_{\mathcal F})=\bigl\{f_t:\;&f_{t,r}(h)=b_{t,r}+g_{t,r}(h),\ |b_{t,r}|\le C_{\mathcal F},\\
&g_{t,r}\in\mathcal H_{k_t},\ \|g_{t,r}\|_{\mathcal H_{k_t}}\le C_{\mathcal F},\ r=1,\ldots,K-1\bigr\}.
\end{align*}
The complexity of $f_t$ is measured by $\|f_t\|^2=\sum_{r=1}^{K-1}\|g_{t,r}\|_{\mathcal H_{k_t}}^2$. This formulation accommodates both linear and nonlinear score functions through the choice of kernel. All simulation studies and real-data applications use the linear kernel $k_t(u,v)=\langle u,v\rangle$, for which $g_{t,r}(h)=w_{t,r}^{\top}h$ and $\|f_t\|^2=\sum_{r=1}^{K-1}\|w_{t,r}\|_2^2$. This specification facilitates coefficient-based interpretation of the fitted stage-specific scores. Nonlinear kernels, such as the Gaussian kernel, can be incorporated within the same formulation.

For the theoretical analysis in Section~\ref{sc:Theory}, we work with fixed, nonempty, deterministic restricted classes $\mathcal F_t^\dagger\subseteq\mathcal F_t(C_{\mathcal F})$, $t=1,2$, and write $\mathcal F=\mathcal F_1^\dagger\times\mathcal F_2^\dagger$. The feasibility, boundedness, compactness, and uniform treatment-score margin conditions for this restricted class are specified in Supplementary Material Section~S.4.

Given any $f = (f_1, f_2)\in \mathcal{F}$, the induced two-stage regime $d_f = (d_{f_1}, d_{f_2})$ is defined by
$$d_{f_t}(\H_t) = \argmax_{a \in \mathcal{A}} \langle V_a, f_t(\H_t)\rangle, \ t = 1, 2,$$
with ties resolved by a fixed measurable rule. For nonzero scores, the selected treatment is the one whose simplex vertex has the smallest angle with the estimated score. 



\subsection{Risk-Aware Quantile DTR Framework}
\label{sc:Methodology-Multicategory Risk-Aware Quantile DTRs}
To incorporate treatment-related risk into policy learning, we introduce two complementary strategies: a constraint-based formulation that imposes explicit population-level controls on adverse outcomes, and a utility-based formulation that balances efficacy and risk through a composite outcome. Combining them in a single penalized objective yields a unified framework that contains efficacy-only quantile learning, hard risk-constrained learning, and utility-based benefit-risk learning as interpretable subclasses.

\textit{\textbf{Constraint-based formulation.}} 
For any regime $d = (d_1, d_2) \in \mathcal{D}$, define the counterfactual attribute-level risk burden by
\begin{equation*}
\label{eq:population risk burden, regime d}
\mathcal{B}_m(d) = \mathbb{E} \left[\psi_{\mathrm{agg}}\left(Z_1^{(m)}(d),Z_2^{(m)}(d)\right)\right], \ m=1,\dots,M,
\end{equation*}
where $\psi_{\mathrm{agg}}$ aggregates the two stagewise outcomes for the $m$th attribute. In this article we take $\psi_{\mathrm{agg}}(x,y)=x+y$, so that $\mathcal B_m(d)$ represents the expected cumulative burden of the $m$th risk attribute over the two stages. Hard constraints are imposed by restricting the policy search space to the feasible regime class $\mathcal D_b=\{d\in\mathcal D: \mathcal{B}_m(d)\le b_m,\ m=1,\dots,M\}$, where the tolerance levels $b_m$ can be prespecified from clinical safety requirements. In the absence of established clinical thresholds, as in our simulations and applications, we employ the data-adaptive choice $b_m=c\,\mathcal{B}_m^{\mathrm{obs}}$, where $\mathcal{B}_m^{\mathrm{obs}}=n^{-1}\sum_{i=1}^{n}\psi_{\mathrm{agg}}(Z_{1i}^{(m)},Z_{2i}^{(m)})$ is the empirical average of the cumulative risk burden under the observed treatment assignments and $c\in(0,1)$ is a prespecified contraction factor. Unless otherwise stated, we set $c=0.85$, targeting a 15\% reduction in expected risk burden relative to the observed treatment assignments. 


\textit{\textbf{Utility-based formulation.}} 
As an alternative to hard constraints, we introduce a stagewise composite outcome, called the utility, 
$$U_t = Y_t - \lambda_\mathcal{R} \phi_{\mathrm{SE}}(\mathcal{R}_t), \ t=1,2,$$
where $\lambda_{\mathcal{R}} \geq 0$ controls the efficacy-risk trade-off and $\phi_{\mathrm{SE}}: [0, 1] \to [0, \infty)$ is an increasing penalty function for risk. To reflect the clinical principle that heavier side-effect burden should be penalized progressively more aggressively, we adopt the normalized exponential form
$$\phi_{\mathrm{SE}}(x;\rho,\xi) = \rho\,\frac{\exp(\xi x)-1}{\exp(\xi)-1}, \ x\in[0,1],\ \rho>0,\ \xi>0,$$
where $\rho$ controls the total magnitude of the penalty and $\xi$ controls its curvature. By construction, $\phi_{\mathrm{SE}}(0)=0$ and $\phi_{\mathrm{SE}}(1)=\rho$. The utility parameters $(\lambda_{\mathcal R},\rho,\xi)$ encode benefit-risk preferences and are specified before policy learning. Motivated by prior work on incorporating clinical preferences into treatment decision-making \citep{lee2015bayesian,butler2018incorporating}, we treat the utility parameters as prespecified inputs rather than tuning parameters and examine alternative specifications through sensitivity analyses. Define the cumulative utility as $U=U_1+U_2$. When $\lambda_{\mathcal R}=0$, $U$ reduces to the cumulative efficacy outcome $Y=Y_1+Y_2$, and the framework recovers efficacy-only quantile DTR learning as a special case.

\textit{\textbf{Unified optimization target.}}
Under any regime $d \in \mathcal{D}$ and any $q \in \mathbb{R}$, the counterfactual survival function of the cumulative utility is defined as $S(q,d)=\PP\{U(d)>q\}$. For a quantile level $\tau \in (0, 1)$, the optimal risk-aware quantile DTR is given by
\begin{equation}
\label{eq:target problem}
    \begin{aligned}
    d^* &= \argmax_{d\in \mathcal{D}} Q_\tau\{U(d)\} \ \text{s.t.} \ \mathcal{B}_m(d)\le b_m,\ m=1,\dots,M, \\
    &=\argmax_{d\in \mathcal{D}} \sup\Bigl\{ q:\; S(q,d)\ge 1-\tau,\ \mathcal{B}_m(d)\le b_m,\ m=1,\ldots,M \Bigr\},
\end{aligned}
\end{equation}
where $Q_\tau\{U(d)\} = \sup\{q : S(q, d) \geq 1 - \tau\}$ denotes the $\tau$th quantile of $U(d)$.

Under the quantile-attainment and oracle-existence conditions in Theorem~S.1 of Supplementary Material Section~S.4, Eq.~\eqref{eq:target problem} is equivalent to the following constrained optimization problem:
\begin{equation}
\label{eq:constrained optimization problem}
\max_{q \in \mathbb{R},\, d \in \mathcal{D}} q \quad \text{subject to} \quad S(q, d) \geq 1 - \tau,\ \mathcal{B}_m(d) \leq b_m,\ m = 1, \ldots, M.
\end{equation}
This reformulation replaces the implicit quantile target by an explicit scalar variable and is the form most suitable for subsequent penalization. Then, we recast Eq.~\eqref{eq:constrained optimization problem} as a penalized optimization problem $\max_{q\in[q_L,q_U],\,f}\{\mathcal M(q,d_f)-\lambda(\|f_1\|^2+\|f_2\|^2)\}$, where $[q_L,q_U]$ is a sufficiently wide interval containing the population-optimal quantile $q^*$, and
\begin{equation*}
\label{eq:limiting population penalized objective}
\mathcal{M}(q, d) = q - \mu_U\bigl[(1-\tau) - S(q, d)\bigr]_+^2 - \sum_{m=1}^M \mu_m\bigl[\mathcal{B}_m(d) - b_m\bigr]_+^2.
\end{equation*}
Here $\mu_U > 0$ controls the quantile constraint, $\mu_m \geq 0$ controls the $m$th risk constraint, $\lambda > 0$ is a regularization parameter to prevent overfitting, and $[x]_{+} = \max\{x, 0\}$. 

Different choices of $(\lambda_{\mathcal R},\mu_m)$ lead to three practically interpretable subclasses. Table~\ref{tab:estimands} summarizes the target estimand for each subclass, as well as a hybrid formulation that maximizes a risk-adjusted utility quantile subject to explicit population-level risk constraints. The hybrid formulation requires simultaneous specification of benefit-risk preferences and risk tolerances, which introduces additional interpretational complexity of the results. We therefore focus throughout the remainder of this article on the three interpretable subclasses: RQDTR-E, RQDTR-C, and RQDTR-U.

\begin{table}[tb]
\centering
\caption{Target estimands and parameter configurations of the proposed RQDTR framework: the three interpretable subclasses and the general hybrid formulation.}
\label{tab:estimands}
\footnotesize
\begin{tabular}{@{}lll@{}}
\toprule
Method & Parameters & Target estimand \\
\midrule
Efficacy-only (RQDTR-E)
& $\lambda_{\mathcal{R}} = 0$,\ $\mu_m = 0$, $\forall m$
& $\displaystyle \max_{d \in \mathcal{D}} Q_\tau\{Y(d)\}$ \\[2pt]
Constraint-based (RQDTR-C)
& $\lambda_{\mathcal{R}} = 0$,\ $\exists m:\mu_m > 0$
& $\displaystyle \max_{d \in \mathcal{D}} Q_\tau\{Y(d)\}\ \text{s.t.}\ \mathcal{B}_m(d) \leq b_m$ \\[2pt]
Utility-based (RQDTR-U)
& $\lambda_{\mathcal{R}} > 0$,\ $\mu_m = 0$, $\forall m$
& $\displaystyle \max_{d \in \mathcal{D}} Q_\tau\{U(d)\},\ U_t = Y_t - \lambda_{\mathcal{R}} \phi_{\mathrm{SE}}(\mathcal{R}_t)$ \\
\midrule
Hybrid
& $\lambda_{\mathcal{R}} > 0$,\ $\exists m:\mu_m > 0$
& $\displaystyle \max_{d \in \mathcal{D}} Q_\tau\{U(d)\}\ \text{s.t.}\ \mathcal{B}_m(d) \leq b_m$ \\
\bottomrule
\end{tabular}
\end{table}


\subsection{Estimation and Algorithm}
\label{sc:Methodology-Estimation and Algorithm}
Directly optimizing Eq.~\eqref{eq:constrained optimization problem} is computationally challenging because the IPW representations of $S(q,d)$ and $\mathcal B_m(d)$ involve the discontinuous treatment-assignment indicators $I\{d_t(\H_t)=A_t\}$ and the tail indicator $I(U>q)$. To facilitate optimization, we replace these indicators with smooth surrogates.

For the treatment-assignment indicator, we introduce the angle-based softmax smoothing
$$p_{t, A_t}(\bm{H}_t; f_t, h) = \frac{\exp\bigl(\langle f_t(\bm{H}_t), V_{A_t}\rangle / h\bigr)}{\sum_{k=1}^{K} \exp\bigl(\langle f_t(\bm{H}_t), V_k\rangle / h\bigr)},$$
with bandwidth $h > 0$. For the tail indicator, we employ the logistic surrogate $g_\delta(U - q) = \{1 + \exp(-(U - q)/\delta)\}^{-1}$ with bandwidth $\delta > 0$. As $h\to0$ and $\delta\to0$, both surrogates converge pointwise to their corresponding hard indicators when the maximizing treatment score is unique and $U\ne q$, respectively. Formal statements are given in Supplementary Material Sections~S.2--S.3. For asymptotic analysis, we let $h_n \to 0$ and $\delta_n \to 0$ at rates specified in Supplementary Material Section~S.4. In finite-sample implementation we adopt the default choices $h = 0.2 / \log n$ and $\delta = 0.1\, \mathrm{SD}(U)$, where $\mathrm{SD}(\cdot)$ denotes the empirical standard deviation. The sensitivity analysis in Supplementary Material Section~S.9 confirms that performance is stable across practical ranges of $(h, \delta)$.

After replacing both indicators in $S(q, d)$ and $\mathcal{B}_m(d)$ by their smoothed counterparts and weighting by the inverse propensities, the smoothed population survival function takes the form
\begin{equation*}
\label{eq:Population smoothed survival_score f}
S_{h,\delta}(q,f) = \mathbb{E} \left[ g_\delta(U-q) \frac{p_{1,A_1}(\H_1;f_1,h)\,p_{2,A_2}(\H_2;f_2,h)}{\pi_1(A_1 | \H_1)\pi_2(A_2 | \H_2)}
\right],
\end{equation*}
and the smoothed population risk-burden function is written as
\begin{equation*}
\label{eq:Population smoothed risk burden_score f}
\mathcal B_{m,h}(f) = \mathbb{E} \left[
\psi_{\mathrm{agg}}(Z_1^{(m)},Z_2^{(m)})
\frac{p_{1,A_1}(\H_1;f_1,h)\,p_{2,A_2}(\H_2;f_2,h)}{\pi_1(A_1 | \H_1)\pi_2(A_2 | \H_2)} \right], \ m=1,\dots,M.
\end{equation*}
The corresponding smoothed penalized population objective is
\begin{equation}
\label{eq:Population smoothed objective_score f}
\mathcal M_{h,\delta}(q,f) = q-\mu_U\bigl[(1-\tau)-S_{h,\delta}(q,f)\bigr]_+^2 -\sum_{m=1}^M \mu_m\bigl[\mathcal B_{m,h}(f)-b_m\bigr]_+^2.
\end{equation}
The sample analogues are obtained by replacing the expectations by empirical averages and the true propensities by their clipped estimators $\widetilde{\pi}_t$. We write the smoothed empirical estimator of the attribute-level risk burden as
\begin{equation*}
\label{eq:emp_smoothed empirical risk burden estimator}
\widehat{\mathcal{B}}_{m,h}(f) = \frac{1}{n} \sum_{i=1}^n \psi_{\mathrm{agg}} (Z_{1i}^{(m)},Z_{2i}^{(m)}) \, \frac{p_{1,A_{1i}}(\bm H_{1i};f_1,h)\, p_{2,A_{2i}}(\bm H_{2i};f_2,h)}{\widetilde\pi_1(A_{1i} | \bm H_{1i})\, \widetilde\pi_2(A_{2i} | \bm H_{2i})},
\end{equation*}
and the smoothed empirical survival estimator as
\begin{equation*}
\label{eq:emp_smoothed empirical survival estimator}
    \widehat{S}_{h,\delta}(q,f) = \frac{1}{n} \sum_{i=1}^n g_\delta(U_i-q)\, \frac{p_{1,A_{1i}}(\bm H_{1i};f_1,h)\, p_{2,A_{2i}}(\bm H_{2i};f_2,h)}{\widetilde\pi_1(A_{1i} | \bm H_{1i})\, \widetilde\pi_2(A_{2i} | \bm H_{2i})}.
\end{equation*}
By substituting these into Eq.~\eqref{eq:Population smoothed objective_score f} and adding an RKHS regularization penalty, the empirical optimization problem becomes
\begin{equation}
\label{eq:penalized_objective}
\max_{q, f} \ \widehat{\mathcal{M}}_{h,\delta}(q, f) - \lambda(\|f_1\|^2 + \|f_2\|^2),
\end{equation}
where $\widehat{\mathcal{M}}_{h,\delta}(q, f) = q - \mu_U\bigl[(1-\tau) - \widehat{S}_{h,\delta}(q, f)\bigr]_+^2 - \sum_{m=1}^M \mu_m\bigl[\widehat{\mathcal{B}}_{m,h}(f) - b_m\bigr]_+^2$. The regularization parameter $\lambda$ is selected by cross-validation. The positive penalty weights $\mu_U, \mu_1, \ldots, \mu_M$
are initialized at prespecified values and increased componentwise
when the corresponding empirical constraint tolerances are not met,
subject to prespecified upper caps.

In practical implementation, we solve Eq.~\eqref{eq:penalized_objective} by block coordinate updates over $q$ and $f$. Given a current $q$, $f$ is updated by solving
$$\max_f\;-\mu_U\Big[(1-\tau)-\widehat{S}_{h,\delta}(q, f)\Big]_+^2 -\sum_{m=1}^M \mu_m\Big[\widehat{\mathcal{B}}_{m,h}(f)-b_m\Big]_+^2 -\lambda\Big(\|f_1\|^2+\|f_2\|^2\Big).$$
Given a current $f$, $q$ is updated by solving $\max_{q \in [q_L,q_U]} q-\mu_U \Big[(1-\tau)- \widehat{S}_{h,\delta}(q,f)\Big]_+^2$. Since the risk-burden terms do not depend on $q$, the $q$-update has the same form regardless of the number of risk attributes. The alternating updates are iterated to convergence. Full pseudocode is provided in Algorithm~\ref{alg:RQDTR-main}.

\begin{algorithm}[tb]
\caption{Risk-Aware Quantile Dynamic Treatment Regimes (RQDTR)}
\label{alg:RQDTR-main}
\KwIn{Two-stage data $\{(\bm X_{1i},A_{1i},Y_{1i},\bm Z_{1i},\bm X_{2i},A_{2i},Y_{2i},\bm Z_{2i})\}_{i=1}^n$; treatment options $\{1,\ldots,K\}$; quantile level $\tau$; propensity scores $\pi_1,\pi_2$ (true or estimated); function classes $\mathcal F_1,\mathcal F_2$; risk tolerances $\{b_m\}_{m=1}^M$; prespecified utility parameters $(\lambda_{\mathcal R},\rho,\xi)$; tuning grid.}
\KwOut{Fitted surrogate threshold $\widehat q$ and estimated regime $\widehat d=(\widehat d_1,\widehat d_2)$.}

\textbf{Step 1:} Construct simplex vertices $\{V_a\}_{a=1}^K\subset\mathbb R^{K-1}$ for angle-based multicategory treatment coding.

\textbf{Step 2:} Truncate the propensities from below by $\pi_{\min}$.

\textbf{Step 3:} Using the prespecified utility parameters, construct the stagewise risk scores $\mathcal R_{ti}$, the stagewise utilities $U_{ti}=Y_{ti}-\lambda_{\mathcal R}\phi_{\mathrm{SE}}(\mathcal R_{ti};\rho,\xi)$, $t=1,2$, and the cumulative utility $U_i=U_{1i}+U_{2i}$.

\textbf{Step 4:} Using the angle-based softmax-smoothed probabilities $p_{t,a}(\H_t;f_t,h)$, compute the smoothed survival estimator $\widehat S_{h,\delta}(q,f)$ and the smoothed risk-burden estimators $\widehat{\mathcal B}_{m,h}(f)$, $m=1,\ldots,M$.

\textbf{Step 5:} Select the tuning parameter $\lambda$ by cross-validation, holding the utility specification fixed across all candidate values of $\lambda$.

\textbf{Step 6:} With the selected $\lambda$, optimize  Eq.~\eqref{eq:penalized_objective} over $q\in[q_L,q_U]$ and $f=(f_1,f_2)\in\mathcal F_1\times\mathcal F_2$ by alternating updates of $f$ and $q$. Increase each positive penalty weight by a fixed multiplicative factor when its corresponding empirical constraint tolerance is not met, subject to a prespecified upper cap.

\textbf{Step 7:} Set $\widehat d_t(\H_t)=\argmax_{a\in\{1,\dots,K\}}\langle V_a,\widehat f_t(\H_t)\rangle$, $t=1,2$.
\end{algorithm}

\section{Theory}
\label{sc:Theory}
In this section, we summarize the theoretical properties of RQDTR. Formal assumptions, propositions, theorem statements, and proofs are provided in Supplementary Material Section~S.4. 

\textit{\textbf{Identification and Oracle Equivalence.}}
Under standard causal assumptions, the counterfactual survival function $S(q,d)$ and attribute-specific cumulative risk burdens $\mathcal B_m(d)$ admit inverse-probability-weighted (IPW) representations, allowing candidate regimes to be evaluated using observed longitudinal data. Theorem~S.1 further establishes that maximizing the utility quantile over feasible regimes is equivalent to maximizing a scalar threshold subject to survival and risk constraints. This equivalence links the clinical target of risk-aware quantile optimization to the penalized optimization formulation used in our estimation procedure. 

\textit{\textbf{Surrogate Consistency with Increasing Penalties.}}
The surrogate analysis is conducted over the fixed restricted score class $\mathcal F=\mathcal F_1^\dagger\times\mathcal F_2^\dagger$ introduced in Section~\ref{sec:method_Angle-based}. Let $q_\dagger$ denote the largest utility quantile among feasible regimes represented by this class. Theorem~S.2 establishes uniform approximation of the hard-regime survival function and risk burdens by their smoothed counterparts. As the smoothing parameters vanish and the quadratic penalty weights increase at compatible rates, the population surrogate maximizers converge to the restricted oracle set. This result justifies the smoothed penalized formulation as an asymptotically valid approximation to constrained quantile optimization.

\textit{\textbf{Consistency of the Estimated Regime.}}
Let $(\widehat q_n,\widehat f_n)$ be an approximate global maximizer of the empirical regularized criterion. Theorem~S.3 establishes convergence to the restricted oracle set and asymptotic risk feasibility, accounting for estimated propensity scores and risk tolerances, regularization, and optimization error. Under an additional oracle-crossing condition, the induced regime quantile also converges to $q_\dagger$. In particular,
$Q_\tau\{U(d_{\widehat f_n})\}\xrightarrow{p}q_\dagger$, $[\mathcal B_m(d_{\widehat f_n})-b_m]_+\xrightarrow{p}0$
for each imposed risk constraint $m$. Thus, the estimated regime asymptotically attains the best feasible quantile represented by the chosen score class.

\textit{\textbf{Finite-Sample Performance Error Bounds.}}
Theorem~S.4 establishes finite-sample bounds on threshold shortfall and population risk-constraint violations. Under additional local regularity and oracle-crossing conditions, we further derive convergence rates for the distance to the restricted oracle set and the regime quantile error: $|Q_\tau\{U(d_{\widehat f_n})\}-q_\dagger|=O_p(\Delta_n^{\alpha/\gamma})$. Here $\Delta_n$ captures errors from empirical fluctuations, propensity-score and risk-tolerance estimation, smoothing, regularization, optimization, and finite penalties. $\alpha$ and $\gamma$ characterize the treatment-score margin and local oracle separation, respectively. We show that there is a trade-off in penalty selection: increasing the penalty weights reduces finite-penalty feasibility errors but amplifies constraint approximation errors. When the score class represents an unrestricted oracle, the performance guarantees extend to the unrestricted optimum.

\section{Simulation}
\label{sc:Simulation}
\subsection{Simulation Design}
We conduct extensive simulation studies to evaluate three aspects of the proposed method: finite-sample performance for quantile treatment-rule learning, its extension to multicategory treatments, and its ability to balance efficacy and risk. Four two-stage scenarios are considered (see Table~\ref{tab:simulation_design}). Scenarios A and B are risk-free settings designed to compare quantile rule learning under binary and multicategory treatments, respectively. Scenarios C and D are risk-aware settings in which treatments affect both efficacy and side-effect burden. The complete data generating processes, including all propensity-score models, covariate-transition models, treatment-effect functions, and risk models, are given in Supplementary Material Sections~S.5--S.6.


Scenarios A and B are designed to assess efficacy-driven quantile learning. Scenario A considers binary treatments, $A_t\in\{0,1\}$, with linear treatment effects and treatment-dependent covariate transitions. In this setting, the quantile-optimal regime differs from the mean-optimal regime, as shown in Supplementary Material Section~S.5.1. Scenario B extends the setting to four treatment options, $A_t\in\{1,2,3,4\}$, with multinomial treatment assignment and heterogeneous treatment effects that vary across quartiles of $X_{1,1}$. Unlike Scenario A, the quantile-optimal regime coincides with the mean-optimal regime at all quantile levels. This setting evaluates whether the angle-based formulation can effectively learn multicategory dynamic treatment rules without reducing the problem to multiple binary comparisons.

Scenarios C and D introduce treatment-related risk. In each stage, two binary side-effect indicators, $Z_t^{(1)}$ and $Z_t^{(2)}$, are generated from logistic models depending on the current history and treatment, and the stagewise risk score is defined as $\mathcal R_t=(Z_t^{(1)}+Z_t^{(2)})/2$. Scenario C uses binary treatments and modifies the stage-2 treatment and outcome models so that risk affects subsequent treatment assignment and efficacy. Scenario D combines the multicategory structure of Scenario B with treatment-dependent risk generation, producing a more challenging setting with four treatment options and risk-dependent histories.

\begin{table}
\caption{Summary of simulation scenarios.}
\label{tab:simulation_design}
\centering
\footnotesize
\begin{tabular}{@{}llll@{}}
\toprule
Scenario & Treatment space & Risk-aware & Key challenge \\
\midrule
A & Binary, $K=2$ & No & Treatment-dependent covariate transitions \\
B & Multicategory, $K=4$ & No & Heterogeneous treatment effects \\
C & Binary, $K=2$ & Yes & Efficacy-risk tradeoff \\
D & Multicategory, $K=4$ & Yes & Risk-dependent histories \\
\bottomrule
\end{tabular}
\end{table}

\subsection{Competing Methods and Evaluation}
To evaluate the performance of the proposed framework, we compare the three RQDTR variants (RQDTR-E, RQDTR-C, and RQDTR-U) with three existing quantile DTR methods: Successive Classification Learning (SCL; \citealp{xia2026successive}), QIQ-learning \citep{linn2017interactive}, and the method of \citet{wang2018quantile}.

For multicategory scenarios, existing binary-treatment competitors are not directly applicable. We therefore construct one-versus-one (OVO; \citealp{allwein2000reducing}) extensions of SCL, QIQ-learning, and Wang's method. Specifically, for each of the $\binom{4}{2}=6$ treatment pairs $(k,k')$, we restrict the training data to subjects who received either treatment $k$ or treatment $k'$, and fit the corresponding binary-treatment method on this subset. For a new patient, the six fitted pairwise rules produce pairwise treatment preferences, and the final treatment is selected by majority vote across all pairwise comparisons. This OVO construction provides a practical multicategory benchmark, although it does not learn all treatment categories jointly in a unified optimization problem. 

Each simulation setting is evaluated over 500 replications, with a training sample size of $n_{\mathrm{train}}=500$ and a test sample size of $n_{\mathrm{test}}=5000$ per replication. Replications in which a method fails to converge are excluded from the performance summaries for that method. The numbers of valid replications are reported in Supplementary Material Tables~S.8 and S.9. For an estimated regime $\hat d$, we report three evaluation metrics: the efficacy-based quantile value $Q_\tau^Y=Q_\tau\{Y(\hat d)\}$, the utility-based quantile value $Q_\tau^U=Q_\tau\{U^{\mathrm{ref}}(\hat d)\}$, and the average risk burden $V_{\mathrm{risk}}(\hat d)=\mathbb E\{\mathcal R_1(\hat d)+\mathcal R_2(\hat d)\}$. Here the reference cumulative utility $U^{\mathrm{ref}}=\sum_{t=1}^2\{Y_t-\phi_{\mathrm{SE}}(\mathcal R_t;1,2)\}$ corresponds to $\lambda_{\mathcal R}=1$ and $(\rho,\xi)=(1,2)$. This reference utility defines the RQDTR-U learning objective in the main analyses and provides a common evaluation criterion for all methods within each setting. In the real-data applications, it represents an assumed benefit-risk preference. For the risk-aware settings, we also report the constraint satisfaction rate $\mathrm{CSR}_m=\frac{1}{|\mathcal I_{\mathrm{valid}}|}\sum_{r\in\mathcal I_{\mathrm{valid}}}I\{\widehat{\mathcal B}_{m,\mathrm{test}}^{(r)}(\hat d^{(r)})\le b_m^{(r)}\}$, $m=1,\ldots,M$, where $\mathcal I_{\mathrm{valid}}$ denotes the set of valid replications. The risk burden $\widehat{\mathcal B}_{m,\mathrm{test}}^{(r)}(\hat d^{(r)})$ is estimated by direct Monte Carlo averaging of the outcomes generated under the learned regime on an independent test set, while the risk tolerance $b_m^{(r)}$ is calculated from the training sample. Further implementation details are provided in Supplementary Material Section~S.15.

Additional simulation results and sensitivity analyses are provided in the Supplementary Material. Section~S.8 compares RQDTR with four mean DTR methods in a heterogeneous-subgroup setting deliberately constructed so that the mean-optimal and lower-quantile-optimal regimes differ. Within this setting, the results demonstrate an objective-specific trade-off: the mean-optimal regime attains the highest mean value, whereas the lower-quantile-optimal regimes attain substantially better lower-tail values. Section~S.9 evaluates the effect of the smoothing parameters $h$ and $\delta$ used in the treatment-assignment and tail-indicator approximations. Section~S.10 studies the impact of the risk-tolerance level in RQDTR-C by varying $c\in\{0.75,0.85,0.95\}$ in the risk-aware scenarios. 
Section~S.11 examines the sensitivity of RQDTR-U to nine prespecified utility specifications $(\rho,\xi)\in\{0.5,1,1.5\}\times\{0.5,1,2\}$, with the resulting regimes evaluated under the common reference utility. Section~S.12 examines sensitivity to the training sample size, with $n_{\mathrm{train}} \in\{200,500,1000\}$. These analyses indicate that the proposed methods exhibit stable performance across these sensitivity settings.

\subsection{Results for Risk-Free Scenarios}
Scenarios A and B were designed as efficacy-only settings, in which no risk burden was incorporated into the data-generating mechanism. Thus, these scenarios evaluate whether the proposed efficacy-only method, RQDTR-E, can effectively recover treatment heterogeneity when the target is lower-tail protection.

Table~\ref{tab:Scenarios_AB} shows that, in the binary setting of Scenario A, RQDTR-E achieves the highest efficacy quantile value among the estimated methods and is consistently closest to the oracle benchmark at all three quantile levels. Specifically, relative to the best-performing competing method at each quantile level, RQDTR-E improves $Q_\tau^Y$ by $2.84\%$, $27.04\%$, and $8.51\%$ at $\tau=0.10$, $0.25$, and $0.50$, respectively. Moreover, the proposed method achieves the smallest Monte Carlo standard deviation among the estimated methods at all three quantile levels. The performance advantage is more pronounced in the multicategory setting of Scenario B. Compared with OVO-QIQ, the best-performing one-versus-one competitor at all three quantile levels, RQDTR-E improves $Q_\tau^Y$ by $43.26\%$, $27.93\%$, and $16.06\%$ at $\tau=0.10$, $0.25$, and $0.50$, respectively. This pattern is consistent with the design of the angle-based formulation: RQDTR-E learns all treatment categories jointly within a single optimization problem, whereas the OVO methods estimate separate rules using pairwise subsets of the data and subsequently aggregate the resulting treatment preferences. Thus, within this simulation setting, the results demonstrate a practical advantage of joint multicategory quantile optimization over the constructed pairwise reductions.

\begin{table}[tb]
\centering
\footnotesize
\setlength{\tabcolsep}{3pt}
\caption{Results for Scenarios A and B. Means of the value functions, with standard deviations in parentheses, are reported at quantile levels $\tau \in \{0.10, 0.25, 0.50\}$. ``Imp-rate'' denotes the improvement rate of RQDTR-E over each competing method, defined as $\mathrm{Imp} = (Q^{Y}_{\tau,\mathrm{RQDTR\text{-}E}}-Q^{Y}_{\tau,\mathrm{Baseline}})/|Q^{Y}_{\tau,\mathrm{Baseline}}|\times 100\%$. Population quantile oracles are included as benchmarks. The best result in each scenario is highlighted in bold.}
\label{tab:Scenarios_AB}

\begin{tabular}{@{}lcccccc@{}}
\toprule
& \multicolumn{2}{c}{$\tau=0.10$}
& \multicolumn{2}{c}{$\tau=0.25$}
& \multicolumn{2}{c}{$\tau=0.50$} \\
\cmidrule(lr){2-3}
\cmidrule(lr){4-5}
\cmidrule(lr){6-7}
Method
& $Q_\tau^Y$ (SD) $\uparrow$ & Imp-rate $\uparrow$
& $Q_\tau^Y$ (SD) $\uparrow$ & Imp-rate $\uparrow$
& $Q_\tau^Y$ (SD) $\uparrow$ & Imp-rate $\uparrow$ \\
\midrule

\multicolumn{7}{@{}l}{\textit{Scenario A: Binary treatment}} \\
\addlinespace[2pt]
Oracle
& -1.745 (0.070) & -
&  0.008 (0.056) & -
&  1.971 (0.049) & - \\
SCL
& -2.315 (0.286) & 5.44\%
& -0.355 (0.265) & 27.04\%
&  1.668 (0.234) & 8.51\% \\
QIQ
& -2.253 (0.261) & 2.84\%
& -0.449 (0.223) & 42.32\%
&  1.524 (0.205) & 18.77\% \\
Wang
& -2.414 (0.305) & 9.32\%
& -0.485 (0.228) & 46.60\%
&  1.551 (0.243) & 16.70\% \\
RQDTR-E
& \textbf{-2.189 (0.215)} & -
& \textbf{-0.259 (0.164)} & -
& \textbf{ 1.810 (0.105)} & - \\

\midrule
\multicolumn{7}{@{}l}{\textit{Scenario B: Multicategory treatment}} \\
\addlinespace[2pt]
Oracle
& 2.698 (0.036) & -
& 3.665 (0.029) & -
& 4.740 (0.028) & - \\
OVO-SCL
& 0.667 (0.340) & 99.10\%
& 1.877 (0.289) & 40.54\%
& 3.207 (0.250) & 20.55\% \\
OVO-QIQ
& 0.927 (0.267) & 43.26\%
& 2.062 (0.238) & 27.93\%
& 3.331 (0.214) & 16.06\% \\
OVO-Wang
& 0.785 (0.335) & 69.17\%
& 2.020 (0.295) & 30.59\%
& 3.273 (0.292) & 18.12\% \\
RQDTR-E
& \textbf{1.328 (0.267)} & -
& \textbf{2.638 (0.191)} & -
& \textbf{3.866 (0.156)} & - \\

\bottomrule
\end{tabular}
\end{table}

\subsection{Results for Risk-Aware Scenarios}
In Scenarios C and D, treatment decisions affect both efficacy and risk-related outcomes, inducing a nontrivial benefit-risk trade-off. These settings are designed to evaluate whether explicitly incorporating risk improves the clinically relevant utility of the estimated regimes, beyond optimizing efficacy alone.

Table~\ref{tab:Scenario_C} summarizes the results for the binary-treatment setting. The three RQDTR variants exhibit complementary and interpretable behavior. RQDTR-E, which optimizes efficacy alone, attains the largest or nearly largest efficacy-based quantile value $Q_\tau^Y$ at all three quantile levels, but its observed risk burden is relatively high. RQDTR-U outperforms others in terms of the utility-based quantile value $Q_\tau^U$ at all three quantile levels. Its efficacy remains close to that of RQDTR-E, and its risk burden is lower than the efficacy-only alternative. RQDTR-C provides the strongest empirical risk control, achieving the lowest average risk burden $V_{\mathrm{risk}}$ and the highest CSRs for both risk attributes across all quantile levels. For instance, at $\tau=0.25$, RQDTR-C reduces $V_{\mathrm{risk}}$ by 10.86\% relative to QIQ-learning, the lowest-risk competing method. It also increases $\mathrm{CSR}_1$ and $\mathrm{CSR}_2$ by 13.77 and 10.72 percentage points relative to SCL, the strongest competing method in constraint satisfaction. Although the absolute CSRs remain modest, the improvements in risk reduction and constraint satisfaction highlight the practical value of RQDTR-C in settings where explicit population-level risk control is important.

Table~\ref{tab:Scenario_D} reports the results for the multicategory setting. The proposed methods again show strong efficacy and utility performance relative to the competing methods. RQDTR-E attains the largest $Q_\tau^Y$ among the estimated methods at all three quantile levels, whereas RQDTR-U consistently achieves the largest $Q_\tau^U$. Compared with RQDTR-E, RQDTR-U maintains competitive efficacy while reducing average risk burden. Although OVO-SCL and OVO-QIQ attain lower risk burdens or higher constraint-satisfaction rates in some settings, these results are accompanied by substantially lower efficacy and utility quantile values. This pattern suggests that the risk reductions are achieved largely by selecting overly conservative regimes rather than by improving the overall benefit-risk trade-off. In contrast, RQDTR-C reduces risk and improves constraint satisfaction relative to RQDTR-E, with only a small loss in efficacy. In Scenario D, RQDTR-U provides the most favorable benefit-risk trade-off: it consistently achieves the highest utility quantile, retains competitive efficacy, and avoids excessive risk burden. Thus, the utility-based formulation achieves a more favorable utility-risk balance than both the hard-constraint formulation and the existing efficacy-only quantile competitors.

\begin{table}[tb]
\centering
\caption{Results for Scenario C. Means of the efficacy-based value $Q_\tau^Y$, utility-based value $Q_\tau^U$, average observed risk burden $V_{\mathrm{risk}}$, and constraint satisfaction rates $\mathrm{CSR}_1$ and $\mathrm{CSR}_2$ are reported at $\tau\in\{0.10,0.25,0.50\}$. Standard deviations for $Q_\tau^Y$, $Q_\tau^U$, and $V_{\mathrm{risk}}$ are given in parentheses. The best result in each scenario is highlighted in bold.}
\label{tab:Scenario_C}
\footnotesize
\setlength{\tabcolsep}{3pt}
\begin{tabular}{@{}lccccc@{}}
\toprule
Method & $Q_\tau^Y$ (SD) $\uparrow$ & $Q_\tau^U$ (SD) $\uparrow$ & $V_{\text{risk}}$ (SD) $\downarrow$ & $\mathrm{CSR}_1$ $\uparrow$ & $\mathrm{CSR}_2$ $\uparrow$ \\
\midrule

& \multicolumn{5}{c}{$\tau = 0.10$} \\ \cmidrule(lr){2-6}
SCL      & -2.326 (0.229) & -2.906 (0.232) & 0.790 (0.076) & 1.85\% & 0.00\% \\
QIQ      & -2.402 (0.249) & -2.977 (0.220) & 0.740 (0.074) & 4.60\% & 3.20\% \\
Wang     & -2.533 (0.290) & -3.139 (0.287) & 0.786 (0.041) & 0.00\% & 0.00\% \\
RQDTR-E  & -2.298 (0.204) & -2.923 (0.209) & 0.812 (0.037) & 0.00\% & 0.00\% \\
RQDTR-U & \textbf{-2.282 (0.218)} & \textbf{-2.875 (0.206)} & 0.791 (0.036) & 0.00\% & 0.00\% \\
RQDTR-C  & -2.434 (0.294) & -2.935 (0.260) & \textbf{0.681 (0.039)} & \textbf{21.34\%} & \textbf{17.07\%} \\
\midrule

& \multicolumn{5}{c}{$\tau = 0.25$} \\ \cmidrule(lr){2-6}
SCL      & -0.484 (0.273) & -1.091 (0.266) & 0.797 (0.091) & 5.02\% & 5.64\% \\
QIQ      & -0.594 (0.227) & -1.182 (0.209) & 0.764 (0.072) & 2.00\% & 1.60\% \\
Wang     & -0.599 (0.196) & -1.207 (0.209) & 0.786 (0.042) & 0.00\% & 0.00\% \\
RQDTR-E  & \textbf{-0.368 (0.147)} & -0.990 (0.159) & 0.804 (0.036) & 0.00\% & 0.00\% \\
RQDTR-U & -0.378 (0.159) & \textbf{-0.955 (0.145)} & 0.774 (0.036) & 0.00\% & 0.00\% \\
RQDTR-C  & -0.576 (0.215) & -1.071 (0.186) & \textbf{0.681 (0.037)} & \textbf{18.79\%} & \textbf{16.36\%} \\
\midrule

& \multicolumn{5}{c}{$\tau = 0.50$} \\ \cmidrule(lr){2-6}
SCL      & 1.545 (0.249) & 0.940 (0.268) & 0.801 (0.082) & 3.73\% & 2.70\% \\
QIQ      & 1.392 (0.213) & 0.790 (0.222) & 0.795 (0.069) & 1.40\% & 0.40\% \\
Wang     & 1.409 (0.190) & 0.817 (0.193) & 0.787 (0.042) & 0.00\% & 0.00\% \\
RQDTR-E  & \textbf{1.671 (0.110)} & 1.072 (0.116) & 0.787 (0.044) & 0.00\% & 0.20\% \\
RQDTR-U & 1.655 (0.123) & \textbf{1.096 (0.108)} & 0.748 (0.040) & 1.00\% & 1.20\% \\
RQDTR-C  & 1.421 (0.203) & 0.912 (0.180) & \textbf{0.702 (0.046)} & \textbf{10.80\%} & \textbf{10.60\%} \\
\bottomrule
\end{tabular}
\end{table}

\begin{table}[tb]
\centering
\caption{Results for Scenario D. Means of the efficacy-based value $Q_\tau^Y$, utility-based value $Q_\tau^U$, average observed risk burden $V_{\mathrm{risk}}$, and constraint satisfaction rates $\mathrm{CSR}_1$ and $\mathrm{CSR}_2$ are reported at $\tau\in\{0.10,0.25,0.50\}$. Standard deviations for $Q_\tau^Y$, $Q_\tau^U$, and $V_{\mathrm{risk}}$ are given in parentheses. The best result in each scenario is highlighted in bold.}
\label{tab:Scenario_D}
\footnotesize
\setlength{\tabcolsep}{3pt}
\begin{tabular}{@{}lccccc@{}}
\toprule
Method & $Q_\tau^Y$ (SD) $\uparrow$ & $Q_\tau^U$ (SD) $\uparrow$ & $V_{\text{risk}}$ (SD) $\downarrow$ & $\mathrm{CSR}_1$ $\uparrow$ & $\mathrm{CSR}_2$ $\uparrow$ \\
\midrule

& \multicolumn{5}{c}{$\tau = 0.10$} \\ \cmidrule(lr){2-6}
OVO-SCL  & 0.654 (0.345) & -0.026 (0.375) & 0.793 (0.092) & \textbf{29.80\%} & \textbf{25.80\%} \\
OVO-QIQ  & 0.868 (0.285) & 0.186 (0.326) & \textbf{0.782 (0.066)} & 27.80\% & 22.20\% \\
OVO-Wang & 0.667 (0.309) & -0.083 (0.324) & 0.845 (0.037) & 1.40\% & 0.20\% \\
RQDTR-E  & \textbf{1.322 (0.265)} & 0.628 (0.272) & 0.825 (0.042) & 3.40\% & 1.60\% \\
RQDTR-U & 1.314 (0.268) & \textbf{0.652 (0.264)} & 0.802 (0.038) & 7.39\% & 5.22\% \\
RQDTR-C  & 1.115 (0.314) & 0.443 (0.322) & 0.796 (0.049) & 14.00\% & 11.20\% \\
\midrule

& \multicolumn{5}{c}{$\tau = 0.25$} \\ \cmidrule(lr){2-6}
OVO-SCL  & 1.836 (0.288) & 1.197 (0.310) & 0.801 (0.087) & 23.20\% & 21.80\% \\
OVO-QIQ  & 2.009 (0.256) & 1.387 (0.287) & \textbf{0.782 (0.066)} & \textbf{28.00\%} & \textbf{23.00\%} \\
OVO-Wang & 1.892 (0.302) & 1.202 (0.314) & 0.844 (0.038) & 0.60\% & 0.60\% \\
RQDTR-E  & \textbf{2.608 (0.184)} & 1.952 (0.186) & 0.825 (0.046) & 4.60\% & 2.80\% \\
RQDTR-U & 2.592 (0.175) & \textbf{1.971 (0.175)} & 0.795 (0.039) & 11.48\% & 7.31\% \\
RQDTR-C  & 2.327 (0.252) & 1.703 (0.254) & 0.792 (0.051) & 15.60\% & 11.80\% \\
\midrule

& \multicolumn{5}{c}{$\tau = 0.50$} \\ \cmidrule(lr){2-6}
OVO-SCL  & 3.192 (0.271) & 2.589 (0.286) & 0.821 (0.083) & 14.20\% & 14.40\% \\
OVO-QIQ  & 3.287 (0.230) & 2.721 (0.253) & \textbf{0.783 (0.069)} & 
\textbf{28.40\%} & \textbf{23.40\%} \\
OVO-Wang & 3.137 (0.247) & 2.509 (0.256) & 0.844 (0.037) & 0.60\% & 0.20\% \\
RQDTR-E  & \textbf{3.778 (0.161)} & 3.159 (0.166) & 0.828 (0.053) & 6.20\% & 3.60\% \\
RQDTR-U & 3.774 (0.156) & \textbf{3.185 (0.156)} & 0.799 (0.048) & 13.05\% & 8.84\% \\
RQDTR-C  & 3.557 (0.217) & 2.966 (0.212) & 0.802 (0.054) & 13.40\% & 9.80\% \\
\bottomrule
\end{tabular}
\end{table}

\section{Real Data Analyses}
\label{sc:Application}


\subsection{Application to Major Depressive Disorder Data}
\label{sc:app_MDD}

We apply RQDTR to a major depressive disorder (MDD) dataset collected by the National Institutes of Health's (NIH) All of Us Research Program (\url{https://www.researchallofus.org/}). This analysis focuses on lower-tail quantile levels $\tau \in \{0.10, 0.25\}$. The clinical motivation is to identify personalized treatment regimens that improve outcomes among patients who would otherwise fall in the lower tail of symptom improvement or experience unfavorable benefit-risk profiles.

The analytic cohort contains 3,683 two-stage treatment trajectories with treatments classified into four antidepressant categories: selective serotonin reuptake inhibitors (SSRIs), serotonin-norepinephrine reuptake inhibitors (SNRIs), tricyclic antidepressants (TCAs), and other antidepressants. MDD severity is quantified using the Patient Health Questionnaire-9 (PHQ-9, \cite{kroenke2001phq}). It is a widely used instrument whose total score ranges from 0 to 27, with higher values indicating more severe depressive symptoms. Let $\mathrm{PHQ9}_{t,\mathrm{pre}}$ and $\mathrm{PHQ9}_{t,\mathrm{post}}$ be the PHQ-9 scores before and after the treatment decision at stage $t$, respectively. The stagewise efficacy outcome is defined as the reduction in PHQ-9 score, $Y_t=\mathrm{PHQ9}_{t,\mathrm{pre}} -\mathrm{PHQ9}_{t,\mathrm{post}}$. The cumulative efficacy outcome is given by $Y=Y_1+Y_2$. To capture treatment-related risk,  we construct 12 post-treatment side-effect indicators according to the clinical literature on antidepressant side effects \citep{wang2018addressing}. Let $Z_t=\{Z_t^{(1)},\ldots,Z_t^{(12)}\}$, where $Z_t^{(m)}=1$ if a new diagnosis in the $m$th side-effect subgroup occurred during the post-treatment stage interval, and $Z_t^{(m)}=0$ otherwise. Since these variables are diagnosis-based indicators, the average observed risk burden $V_{\mathrm{risk}}$ should be interpreted as an observed side-effect proxy rather than a complete measure of latent clinical risk. The stage-1 covariates include demographic characteristics, lifestyle factors, baseline clinical measurements, and pre-existing comorbidity indicators. The stage-2 history further includes updated BMI and PHQ-9 score, the stage-1 treatment assignment, the stage-1 efficacy response, and the observed side-effect indicators before the stage-2 treatment decision. Table~\ref{tab:mdd_variable_summary} summarizes the variables used in our analysis.

\begin{table}[tb]
\centering
\caption{Summary of variables used in the major depressive disorder application.}
\label{tab:mdd_variable_summary}
\footnotesize
\begin{tabular}{@{}lp{11cm}@{}}
\toprule
Component & Variables \\
\midrule
Baseline covariates $X_1$
& Age, sex at birth, race, smoking frequency, baseline BMI, baseline PHQ-9 score, and pre-existing psychiatric, metabolic, cardiovascular, respiratory, and neurologic comorbidity indicators. \\
Treatment $A_t$
& Four antidepressant categories: SSRI, SNRI, TCA, and other antidepressants. \\
Updated covariates $X_2$
& Updated BMI and PHQ-9 score before the second-stage treatment decision, together with accumulated stage-1 history. \\
Efficacy outcome $Y_t$
& Stagewise reduction in PHQ-9 score; larger values indicate greater symptom improvement. \\
Risk outcome $Z_t$
& Twelve binary indicators for newly observed post-treatment side-effect subgroups, including hypertension, arrhythmia or QT-related events, heart failure or myocardial infarction, gastrointestinal bleeding, vascular injury or bleeding, obesity or weight-related events, diabetes, insomnia, anxiety or panic disorder, hepatotoxicity, sexual dysfunction, and migraine or seizure. \\
\bottomrule
\end{tabular}
\end{table}

\begin{table}[tb]
\centering
\caption{Results for the major depressive disorder dataset. Means of the efficacy-based value $Q_\tau^Y$, utility-based value $Q_\tau^U$, and average observed risk burden $V_{\mathrm{risk}}$, with standard deviations in parentheses, are reported at quantile levels $\tau \in \{0.10, 0.25\}$. The best result in each scenario is highlighted in bold.}
\label{tab:depression application_K=4}
\footnotesize
\setlength{\tabcolsep}{3pt}
\begin{tabular}{@{}lcccccc@{}}
\toprule
& \multicolumn{3}{c}{$\tau = 0.10$}
& \multicolumn{3}{c}{$\tau = 0.25$} \\
\cmidrule(lr){2-4} \cmidrule(lr){5-7}
Method
& $Q_\tau^Y$ (SD) $\uparrow$
& $Q_\tau^U$ (SD) $\uparrow$
& $V_{\text{risk}}$ (SD) $\downarrow$
& $Q_\tau^Y$ (SD) $\uparrow$
& $Q_\tau^U$ (SD) $\uparrow$
& $V_{\text{risk}}$ (SD) $\downarrow$ \\
\midrule
OVO-SCL
& -5.190 (3.348) & -5.351 (3.316) & 0.413 (0.186)
& -1.920 (3.277) & -2.103 (3.257) & 0.444 (0.224) \\
OVO-QIQ
& -4.270 (2.613) & -4.447 (2.628) & 0.467 (0.196)
& -1.690 (1.529) & -1.940 (1.492) & 0.496 (0.193) \\
OVO-Wang
& -4.510 (3.743) & -4.582 (3.760) & 0.170 (0.138)
& -2.470 (3.799) & -2.539 (3.806) & 0.168 (0.121) \\
RQDTR-E
& \textbf{-3.590 (3.720)} & \textbf{-3.681 (3.718)} & 0.164 (0.110)
& \textbf{-0.500 (3.371)} & \textbf{-0.550 (3.378)} & 0.170 (0.112) \\
RQDTR-U
& -4.120 (3.295) & -4.221 (3.285) & \textbf{0.151 (0.096)}
& -1.500 (3.292) & -1.577 (3.275) & \textbf{0.150 (0.098)} \\
RQDTR-C
& -6.530 (2.322) & -6.627 (2.307) & 0.293 (0.084)
& -2.318 (2.026) & -2.410 (2.000) & 0.341 (0.042) \\
\bottomrule
\end{tabular}
\end{table}

We randomly split the cohort into training and testing sets at a 7:3 ratio, and report results averaged over 100 independent random splits. Implementation choices follow Section~\ref{sc:Simulation}. Since only factual treatment paths and outcomes are observed, policy values for candidate regimes are estimated on the test set using self-normalized IPW estimators based on trajectories consistent with the candidate regime. 
Empirical quantiles are computed using IPW-weighted quantile regression, with the treatment of ties and implementation details described in Supplementary Material Sections~S.13.2--S.13.3.

Table~\ref{tab:depression application_K=4} shows that the proposed efficacy-only and utility-based formulations consistently outperform the one-versus-one competing methods across the two lower-tail quantile levels. RQDTR-E attains the highest $Q^Y_\tau$ and $Q^U_\tau$ at both $\tau=0.10$ and $\tau=0.25$, indicating that the proposed framework can effectively identify regimes with improved PHQ-9 score reduction in the lower tail of the outcome distribution. Compared with the strongest one-versus-one competing method (OVO-QIQ), RQDTR-E improves $Q^Y_\tau$ and $Q^U_\tau$ by 15.93\% and 17.23\%, respectively, at $\tau=0.10$, and by 70.41\% and 71.66\%, respectively, at $\tau=0.25$. The risk patterns further illustrate the benefit of incorporating treatment-related adverse burden. Although OVO-Wang has the lowest $V_{\mathrm{risk}}$ among the one-versus-one competing methods, its low-risk profile comes at a substantial cost in efficacy and utility. In comparison, RQDTR-U further reduces $V_{\mathrm{risk}}$ relative to OVO-Wang and achieves higher $Q_\tau^Y$ and $Q_\tau^U$. Specifically, at $\tau=0.10$, RQDTR-U reduces $V_{\mathrm{risk}}$ by 11.07\% and improves $Q_\tau^Y$ and $Q_\tau^U$ by 8.65\% and 7.87\%, respectively. At $\tau=0.25$, the corresponding reduction in $V_{\mathrm{risk}}$ is 10.73\%, while $Q_\tau^Y$ and $Q_\tau^U$ increase by 39.27\% and 37.90\%, respectively. These results demonstrate that the utility-based formulation can improve the overall benefit-risk trade-off in this real-data application, rather than simply selecting conservative regimes with low observed risk.

In Supplementary Material Section~S.13.1, we identify the variables that contributed most strongly to the learned RQDTR-U regime. Across the two quantile levels and treatment stages, race, age at baseline, and baseline BMI are among the top-ranked covariates. Smoking frequency, sex at birth, and metabolic or cardiovascular comorbidities contribute mainly to the stage-1 rule, whereas prior treatment indicators, particularly SNRI and SSRI use, are important at stage 2. These results indicate that the learned regime incorporated baseline patient characteristics, comorbidity profiles, and treatment history when making sequential antidepressant recommendations.

\subsection{Application to MIMIC-III Sepsis Data}
The Medical Information Mart for Intensive Care III (MIMIC-III) database (\cite{PhysioNet-mimiciii-1.4}, \url{https://physionet.org/content/mimiciii/1.4/}) contains deidentified electronic health record data for over 40,000 intensive care unit (ICU) patients who stayed in critical care units of the Beth Israel Deaconess Medical Center between 2001 and 2012. In contrast to the first application, this analysis is recovery-oriented and focuses on upper-tail quantile levels $\tau \in \{0.75, 0.90\}$. Our goal is to identify personalized treatment regimes that promote stronger organ-function recovery for ICU patients. 

Following the construction and preprocessing procedure of \citet{miao2025reinforcement}, we extract a cohort of adult patients satisfying sepsis criteria from the MIMIC-III database. Each ICU trajectory is divided into consecutive 4-hour decision blocks, where each block summarizes the patient's clinical status, treatment exposures, and outcome variables. To formulate a two-stage DTR problem, we consider the first three decision blocks of each trajectory, resulting in an analytic cohort of 16,265 patients. At each stage, treatment exposure is characterized by intravenous fluid and vasopressor administration. For multicategory policy learning, we define a four-level treatment variable $A_t\in\{1,2,3,4\}$ representing overall resuscitation intensity, corresponding to no treatment, low intensity, medium intensity, and high intensity, respectively.

The Sequential Organ Failure Assessment (SOFA) score \citep{vincent1996sofa} is a standard measure of organ dysfunction for ICU patients. It assigns a score from 0 (normal) to 4 (severe failure) for each of six organ systems: neurological, cardiovascular, respiratory, hepatic, coagulation, and renal. Let $\mathrm{SOFA}_{t,\mathrm{pre}}$ and $\mathrm{SOFA}_{t,\mathrm{post}}$ denote the SOFA scores before and after the treatment decision at stage $t$, respectively. Define the stagewise efficacy outcome as the reduction in SOFA score, $Y_t=\mathrm{SOFA}_{t,\mathrm{pre}} -\mathrm{SOFA}_{t,\mathrm{post}}$. The corresponding cumulative efficacy outcome is $Y = Y_1 + Y_2$. According to the Surviving Sepsis Campaign 2021 guidelines \citep{evans2021surviving}, adequate organ perfusion is a central goal of sepsis resuscitation. Mean arterial pressure (MAP) and serum lactate provide complementary physiological indicators during sepsis resuscitation. Persistently low MAP indicates inadequate hemodynamic pressure, whereas elevated serum lactate reflects tissue hypoxia or metabolic dysfunction. 
Motivated by these clinical considerations, we construct two continuous post-treatment risk burden measures derived from the state at the end of decision stage $t$ (equivalently, the start of stage $t+1$):
\begin{equation*}
Z_t^{(1)} = \frac{\max\big(65 - \mathrm{MAP}_{t,\mathrm{end}},\, 0\big)}{65 - \mathrm{MAP}_{1\%}^{\mathrm{train}}}, \ Z_t^{(2)} = \frac{\max\big(\mathrm{Lactate}_{t,\mathrm{end}} - 2,\, 0\big)}{\mathrm{Lactate}_{99\%}^{\mathrm{train}} - 2},
\end{equation*}

where $\mathrm{MAP}_{t,\mathrm{end}}$ and $\mathrm{Lactate}_{t,\mathrm{end}}$ denote mean arterial pressure and lactate level at the end of stage $t$; $\mathrm{MAP}_{1\%}^{\mathrm{train}}$ and $\mathrm{Lactate}_{99\%}^{\mathrm{train}}$ are the first and 99th percentiles within the training data. The threshold of 65 mmHg is the MAP resuscitation target recommended by the Surviving Sepsis Campaign 2021 guidelines \citep{evans2021surviving}, and 2 mmol/L is the lactate threshold defining septic shock under the Sepsis-3 criteria \citep{singer2016third}. Both indicators are clipped to $[0, 1]$ after normalization, so that $Z_t^{(m)} \in [0, 1]$ with larger values indicating greater risk burden. The two risk measures capture complementary aspects of sepsis-related burden: hemodynamic inadequacy due to low MAP and persistent tissue hypoperfusion due to elevated lactate. 
At each stage, the covariates consist of 40 clinical state variables, including demographics and context, vital signs, ventilation and oxygenation measures, blood gas measurements, chemistry panels, liver function markers, hematology and coagulation measures, and shock index. All variables used in this application are listed in Table~\ref{tab:mimic_variable_summary}.

Table~\ref{tab:mimic_results} summarizes the results over 100 independent random splits, each using 2,000 training samples and 5,000 testing samples. Across both upper-tail quantile levels, the proposed RQDTR-C achieves the highest efficacy-based and utility-based quantile values, indicating that the constraint-based formulation effectively identifies treatment sequences associated with stronger organ-function recovery. Compared with the strongest competing method at each quantile level, RQDTR-C improves $Q^Y_\tau$ and $Q^U_\tau$ by 16.85\% and 16.49\% at $\tau = 0.75$ over OVO-SCL, and by 7.98\% and 8.67\% at $\tau = 0.90$ over OVO-QIQ. The risk results differ across quantile levels. At $\tau=0.75$, RQDTR-E achieves the lowest average risk burden ($V_{\mathrm{risk}}=0.192$), whereas the higher efficacy and utility quantile values of RQDTR-C are accompanied by a higher risk burden ($V_{\mathrm{risk}}=0.215$). At $\tau=0.90$, RQDTR-C achieves the lowest average risk burden ($V_{\mathrm{risk}}=0.199$) while also attaining the highest efficacy-based and utility-based quantile values. These findings highlight the quantile-dependent benefit-risk trade-offs, with RQDTR-C achieving favorable performance across all three metrics at the more extreme upper-tail quantile.

Supplementary Material Section~S.14.1 summarizes the covariate contributions to the learned RQDTR-C regime. Across both quantile levels and treatment stages, the PaO$_2$/FiO$_2$ ratio and platelet count are among the most important variables. Glucose and PaO$_2$ also played significant roles, particularly for the second stage, where both baseline and updated measurements contributed to treatment recommendations. In contrast, sodium, chloride, and systolic blood pressure were more influential for the stage-1 recommendation. These findings indicate that the learned RQDTR-C regime primarily used information on oxygenation status, hematologic function, and metabolic burden, which is clinically consistent with sepsis management and resuscitation decision-making.

\begin{table}[tb]
\centering
\caption{Summary of variables used in the MIMIC-III sepsis application.}
\label{tab:mimic_variable_summary}
\footnotesize
\begin{tabular}{@{}lp{11cm}@{}}
\toprule
Component & Variables \\
\midrule
Baseline covariates $X_1$
& Forty clinical state variables measured during the initial 4-hour block, comprising demographics and context (gender, age, readmission status, weight), vital signs (GCS, heart rate, systolic/mean/diastolic blood pressure, respiratory rate, SpO$_2$, temperature), ventilation and oxygenation (FiO$_2$, PaO$_2$/FiO$_2$, mechanical ventilation), blood gas indices (arterial pH, PaO$_2$, PaCO$_2$, bicarbonate, base excess, arterial lactate), chemistry panel (potassium, sodium, chloride, glucose, BUN, creatinine, magnesium, calcium, CO$_2$), liver function markers (SGOT, SGPT, total bilirubin, albumin), hematology and coagulation measures (hemoglobin, white blood cell count, platelet count, PTT, INR), and shock index. \\
Treatment $A_t$
& Four categories of resuscitation intensity derived from discretized intravenous fluid and vasopressor exposure: no treatment, low intensity, medium intensity, and high intensity. \\
Updated covariates $X_2$
& The same forty clinical state variables measured at the second 4-hour block, together with accumulated stage-1 history. \\
Efficacy outcome $Y_t$
& Stagewise reduction in SOFA score; larger values indicate greater short-term organ-function recovery. \\
Risk outcome $Z_t$
& Two continuous post-treatment physiologic burden measures derived from the subsequent 4-hour state: hemodynamic risk burden $Z_t^{(1)}$ and tissue-perfusion risk burden $Z_t^{(2)}$; larger values indicate greater physiologic risk burden. \\
\bottomrule
\end{tabular}
\end{table}

\begin{table}[tb]
\centering
\caption{Results for the MIMIC-III sepsis dataset. Means of the efficacy-based value $Q_\tau^Y$, utility-based value $Q_\tau^U$, and average observed risk burden $V_{\mathrm{risk}}$, with standard deviations in parentheses, are reported at quantile levels $\tau \in \{0.75, 0.90\}$. The best result in each scenario is highlighted in bold.}
\label{tab:mimic_results}
\footnotesize
\setlength{\tabcolsep}{3pt}
\begin{tabular}{@{}lcccccc@{}}
\toprule
& \multicolumn{3}{c}{$\tau = 0.75$} & \multicolumn{3}{c}{$\tau = 0.90$} \\
\cmidrule(lr){2-4} \cmidrule(lr){5-7}
Method & $Q_\tau^Y$ (SD) $\uparrow$ & $Q_\tau^U$ (SD) $\uparrow$ & $V_{\text{risk}}$ (SD) $\downarrow$ & $Q_\tau^Y$ (SD) $\uparrow$ & $Q_\tau^U$ (SD) $\uparrow$ & $V_{\text{risk}}$ (SD) $\downarrow$ \\
\midrule
OVO-SCL  & 3.075 (1.663)          & 3.002 (1.674)          & 0.212 (0.079)          & 4.560 (1.166)          & 4.515 (1.166)          & 0.204 (0.091)          \\
OVO-QIQ  & 3.067 (1.243)          & 2.988 (1.219)          & 0.206 (0.083)          & 4.686 (0.941)          & 4.589 (0.918)          & 0.210 (0.080)          \\
OVO-Wang & 2.400 (1.393)          & 2.320 (1.359)          & 0.209 (0.082)          & 4.313 (1.616)          & 4.265 (1.622)          & 0.203 (0.102)          \\
RQDTR-E  & 3.215 (1.229)          & 3.172 (1.242)          & \textbf{0.192 (0.052)} & 4.950 (1.192)          & 4.890 (1.163)          & 0.215 (0.070)          \\
RQDTR-U  & 3.062 (1.115)          & 2.960 (1.114)          & 0.202 (0.073)          & 4.890 (1.113)          & 4.825 (1.090)          & 0.208 (0.065)          \\
RQDTR-C  & \textbf{3.593 (1.428)} & \textbf{3.497 (1.430)} & 0.215 (0.070)          & \textbf{5.060 (1.609)} & \textbf{4.987 (1.610)} & \textbf{0.199 (0.064)} \\
\bottomrule
\end{tabular}
\end{table}

\section{Discussion}
\label{sc:Discussion}
In this article we propose a risk-aware quantile learning framework for estimating personalized dynamic treatment regimes with multicategory treatment options. The proposed RQDTR framework extends existing quantile DTR methods in two clinically important directions. First, it moves beyond efficacy-only quantile optimization by incorporating treatment-related risk through either explicit population-level constraints or a composite utility outcome. Second, it adopts angle-based learning to accommodate more than two treatment options, allowing all treatment categories to be learned jointly within a single optimization problem.

The proposed framework encompasses three interpretable subclasses with complementary roles. RQDTR-E focuses on efficacy-oriented quantile optimization and serves as a benchmark when risk is not part of the decision criterion. RQDTR-U incorporates risk into the utility outcome to provide a flexible benefit-risk trade-off, which can be useful when risk measures serve as imperfect proxies for latent treatment-related harm. RQDTR-C imposes explicit population-level risk constraints and is particularly relevant when risk measures are clinically well defined and explicit safety constraints are required. The general hybrid formulation further allows utility-based quantile optimization subject to risk constraints. Simulation studies and real-data applications illustrate these complementary roles, indicating that the choice of formulation should account for the clinical objectives and the interpretation of the available risk measures.

The theoretical guarantees are established for a restricted score class under vanishing smoothing, compatible increasing penalties, and approximate global optimization. The numerical studies use a broader linear score class, the default smoothing parameters, capped penalty updates, and alternating optimization. Thus, the theoretical results characterize the restricted estimator under the stated conditions, whereas the simulations and applications assess the finite-sample performance of the practical implementation.

Several directions for future research remain. As in most existing DTR methods, identification in our framework relies on the no-unmeasured confounding assumption. This assumption may be violated in observational studies when latent factors influence both treatment assignment and outcomes. Incorporating sensitivity analysis tools to assess robustness to unmeasured confounding is a natural extension \citep{jin2026sensitivity}. In addition, the current framework represents treatment decisions through a single categorical action at each stage, whereas clinical practice may involve combination therapies or continuous dosages. Extending RQDTR to continuous, mixture, or dosage-based treatment spaces \citep{kallus2018policy, schweisthal2023reliable} is another valuable direction. Moreover, the proposed constraints bound population-average cumulative risk rather than imposing separate risk bounds within each patient subgroup. Extensions to subgroup-specific or covariate-conditional expected-risk constraints could provide more targeted risk control (see Supplementary Material Section~S.16). Finally, although the three-stage simulation in Supplementary Material Section~S.7 illustrates the feasibility of applying RQDTR beyond two stages, the theoretical guarantees are established for the two-stage setting. Extending these guarantees to general finite-horizon settings remains an important direction for future research.

\section{Significance Statement}
Treatment decisions often involve several options over time and must balance potential benefits against possible harms. Many existing approaches emphasize average improvement, potentially overlooking poor treatment responses, while methods focused on poorer outcomes rarely address treatment risks directly. We develop a new method that can target poor responses or stronger recovery while accounting for treatment-related risk. It supports multiple treatment options across successive decisions and offers two ways to consider risk: setting limits on average harm across patients or weighing harm against benefit. Analyses based on simulated data and real-world patient records for major depressive disorder and sepsis illustrate how the proposed method can support research on personalized treatment strategies when both effectiveness and safety matter.



\begin{supplement}
\textbf{Supplementary Material to ``Risk-Aware Quantile Learning for Personalized Dynamic Treatment Regimes''.} The supplementary material contains technical proofs, additional theoretical results, implementation details, additional simulation studies, and extended real-data analyses.
\end{supplement}

\bibliographystyle{imsart-nameyear} 
\bibliography{references}

\end{document}


\title{Supplement to ``Risk-Aware Quantile Learning for Personalized Dynamic Treatment Regimes''}

\begin{aug}
\author[A]{\fnms{Chunyin}~\snm{Lei}\ead[label=e1]{chunyin\_lei@ucsb.edu}}
\author[A]{\fnms{Annie}~\snm{Qu}\ead[label=e2]{aqu2@ucsb.edu}}

\address[A]{Department of Statistics and Applied Probability, University of California, Santa Barbara\printead[presep={,\ }]{e1,e2}}
\end{aug}

\setcounter{tocdepth}{2}
\tableofcontents
\newpage

\section{Methodological Novelty}
To better position the proposed method relative to prior work, we summarize in Table~\ref{tab:novelty} whether each method supports lower-tail protection through quantile optimization, explicit treatment-related risk control, multicategory treatment assignment, and multistage decision-making. 

\begin{table}[tb]
\centering
\caption{Methodological positioning of RQDTR relative to representative treatment-regime learning methods. ``Quantile'' indicates whether the method directly targets a tail quantile of the outcome distribution, rather than optimizing the mean potential outcome. ``Risk control'' refers to whether treatment-related risks are explicitly considered. ``Multi-cat.'' indicates support for $K \ge 3$ treatment options. ``Stage'' indicates the number of decision stages that each method is designed to handle.}
\label{tab:novelty}
\footnotesize
\begin{tabular}{@{}lcccc@{}}
\toprule
Method & Quantile & Risk control & Multi-cat. & Stage \\
\midrule
\cite{murphy2003optimal} & $\times$ & $\times$ & $\times$ & $\geq 2$ \\
\cite{zhao2015new} & $\times$ & $\times$ & $\times$ & $\geq 2$ \\
\cite{shi2018high} & $\times$ & $\times$ & $\times$ & $\geq 2$ \\ 
\cite{qi2020multi} & $\times$ & $\times$ & $\checkmark$ & 1 \\
\cite{zhang2020multicategory} & $\times$ & $\times$ & $\checkmark$ & 1 \\
\cite{xue2022multicategory} & $\times$ & $\times$ & $\checkmark$ & $\geq 2$ \\
\cite{wang2018learning} & $\times$ & $\checkmark$ & $\times$ & 1 \\
\cite{liu2024controlling,liu2024learning} & $\times$ & $\checkmark$ & $\times$ & $\geq 2$ \\ 
\cite{zhu2024risk} & $\times$ & $\checkmark$ & $\times$ & $\geq 2$ \\ 
\cite{linn2017interactive} & $\checkmark$ & $\times$ & $\times$ & 2 \\
\cite{wang2018quantile} & $\checkmark$ & $\times$ & $\times$ & 2 \\
\cite{xia2026successive} & $\checkmark$ & $\times$ & $\times$ & 2 \\ 
\textbf{Proposed method} & $\boldsymbol{\checkmark}$ & $\boldsymbol{\checkmark}$ & $\boldsymbol{\checkmark}$ & $\geq 2$\\
\bottomrule
\end{tabular}
\end{table}

Table~\ref{tab:novelty} shows that existing quantile DTR methods primarily focus on efficacy-only objectives and are largely restricted to binary treatment settings. In addition, existing multistage quantile DTR methods are restricted to binary treatments, while most multicategory treatment-rule methods focus on single-stage decisions. Moreover, existing risk-aware DTR methods are mainly mean-based and do not directly target lower-tail protection. Although multicategory angle-based learning methods provide an effective framework for multi-arm treatment assignment, they have not been developed for risk-aware quantile objectives. 

RQDTR addresses this methodological gap by integrating quantile optimization, explicit treatment-related risk control, multicategory treatment learning, and multistage decision-making within a unified DTR framework. In the main paper, we focus on the two-stage setting in both the simulation studies and the two real-data applications, which is the canonical setting in the DTR methodology literature. 
An extension of the simulation results to a three-stage study (Scenario~E) is provided in Section~\ref{app:three-stage}. Note that extending the theoretical analysis to general $T \ge 3$ stages introduces additional technical challenges: smoothing approximation and nuisance estimation errors can accumulate across stages, and stronger regularity conditions on the score classes and propensity estimators may be required to ensure consistency of the estimated regime. In addition to the methodological coverage, our theoretical results establish identification and oracle equivalence (Theorem~\ref{thm:Thm1}), convergence of population surrogate maximizers to the restricted oracle set as the smoothing parameters vanish and the quadratic penalty weights increase (Theorem~\ref{thm:Thm2}), and consistency and asymptotic risk feasibility of approximate global maximizers of the empirical criterion, with convergence of regime quantiles under an additional oracle-crossing condition (Theorem~\ref{thm:Thm3}). We further derive finite-sample bounds on threshold shortfall and constraint violations, together with rates for the distance to the oracle set and the quantile error under additional local regularity conditions (Theorem~\ref{thm:Thm4}). The surrogate and estimation results target the best feasible regime within a fixed restricted score class; recovery of the unrestricted optimum requires oracle representability, while comparison with that optimum otherwise includes an explicit class approximation gap.

\subsection{Connections to Related Literature}
Beyond the methods summarized in Table~\ref{tab:novelty}, we further clarify the relationship between RQDTR and three closely related strands of literature.

\begin{itemize}
    \item[(a)] \emph{Policy learning versus policy evaluation.} A complementary line of work, such as doubly robust quantile OPE \citep{xu2025doubly}, focuses on estimating the quantile value $Q_\tau\{U(\pi)\}$ of a given target policy $\pi$. In contrast, RQDTR is a policy-learning method that targets a prespecified quantile subject to population risk constraints. Its statistical analysis considers optimization over a fixed restricted score class, with the best feasible regime represented by that class as the primary target. Therefore, the two objectives are formally distinct. Quantile OPE evaluates an existing policy, whereas RQDTR learns an optimal policy under quantile and risk-aware criteria. More generally, many risk-sensitive or quantile RL methods are formulated under a Markov decision process, where decisions depend on a Markov state and values are characterized through Bellman-type recursions. In contrast, our work is formulated as causal DTR learning from observational clinical data, where regimes are functions of treatment histories and counterfactual quantities are identified under standard causal assumptions. We view the doubly robust ideas developed in the quantile-OPE literature as highly relevant for improving nuisance estimation and stabilizing value estimation in future extensions of RQDTR, but they address a different statistical task from the one studied in this paper.
    
    \item[(b)] \emph{Risk-sensitive and quantile RL methods.} Based on the MDP perspective mentioned above, two further distinctions between RQDTR and risk-sensitive/quantile RL methods are particularly relevant: (i) In terms of sample-size regime and identifiability, RQDTR is tailored to the small-to-moderate-sample, observational clinical decision-making setting typical of DTR studies (for example, our All of Us depression cohort contains 3,683 two-stage trajectories). In contrast, modern deep RL methods are usually developed in large-scale environments with millions or even billions of trajectories. RQDTR imposes structure on the policy class through RKHS-based stagewise score functions $f_t$, identifies counterfactual survival and risk-burden functionals under standard causal assumptions (SUTVA, sequential ignorability, and positivity), and provides statistical guarantees for the learned regime. In particular, Theorem~\ref{thm:Thm4} provides finite-sample bounds on threshold shortfall and constraint violations, together with performance error rates relative to the restricted-class oracle under additional local regularity conditions. Thus, we view RQDTR and deep RL as complementary approaches targeting different statistical regimes, rather than directly competing methods: RQDTR emphasizes causal identifiability, sample efficiency, and interpretable risk-aware policy learning, while deep RL is often most effective in large-scale sequential decision environments. (ii) In terms of objective type, our objective is VaR/quantile-style rather than CVaR-style. The CVaR and risk-sensitive RL literatures are relevant to our work because they highlight important dynamic-consistency and tractability issues that arise when optimizing tail functionals of cumulative rewards. In particular, \citet{hau2023dynamic} show that the state-augmentation dynamic-programming decompositions widely used for CVaR and EVaR in risk-sensitive RL are inherently suboptimal regardless of discretization level, due to a saddle-point property violation; in contrast, VaR admits a valid primal dynamic-programming decomposition. A VaR/quantile-style objective is therefore structurally better aligned with dynamic decomposition than a CVaR-style objective for multistage decision-making. In RQDTR, the quantile-based efficacy objective and the population-level risk control act as complementary mechanisms: the former targets lower-tail performance, while the latter controls clinically meaningful risk burdens through either hard constraints or soft utility penalization.
    
    \item[(c)] \emph{Robust OPE and sequential unmeasured confounding.} In addition, RQDTR is related to the recent literature on robust off-policy evaluation and sensitivity analysis under unmeasured confounding \citep{bennett2024efficient, bruns2023robust, tan2025sensitivity}. The present paper develops RQDTR under sequential ignorability, which is the standard identification assumption in the DTR literature. In contrast, the robust-OPE and sensitivity-analysis literatures develop tools for evaluating policy value or bounding causal effects when this assumption may fail. Thus, the two lines of work address related but distinct parts of the problem: RQDTR focuses on optimal risk-aware quantile policy learning under identification, whereas robust-OPE and sensitivity-analysis methods quantify robustness to hidden confounding or model uncertainty. Combining the RQDTR framework with these tools to obtain risk-aware quantile DTRs that remain valid under partial violations of sequential ignorability is a particularly promising direction for future work.
\end{itemize}

\section{Pointwise Convergence of the Softmax Treatment Smoother}
\label{pf:Lemma1}

\begin{lemma}
\label{lem:softmax}
Fix a stage $t\in\{1,2\}$ and a measurable score function $f_t$.
Suppose that the maximizer
$\argmax_{a\in\mathcal A}\langle V_a,f_t(\H_t)\rangle$
is unique almost surely.
Then, for every $a\in\mathcal A$,
\[
p_{t,a}(\H_t;f_t,h)
\to I\{a=d_{f_t}(\H_t)\}
\ \text{a.s. as }h\downarrow0,
\]
where
$d_{f_t}(\H_t)=\argmax_{a\in\mathcal A}
\langle V_a,f_t(\H_t)\rangle$.
\end{lemma}

\begin{proof}
Fix a history $\H_t$ for which the maximizer is unique, and let $a^*=d_{f_t}(\H_t)$. By the uniqueness assumption, $\langle V_a,f_t(\H_t)\rangle<\langle V_{a^*},f_t(\H_t)\rangle$ for every $a\ne a^*$. 

Dividing the numerator and denominator of the softmax probability by $\exp\{\langle V_{a^*},f_t(\H_t)\rangle/h\}$ yields
\[
p_{t,a}(\H_t;f_t,h)=\frac{\exp\left\{\left(\langle V_a,f_t(\H_t)\rangle-\langle V_{a^*},f_t(\H_t)\rangle\right)/h\right\}}{1+\sum_{k\ne a^*}\exp\left\{\left(\langle V_k,f_t(\H_t)\rangle-\langle V_{a^*},f_t(\H_t)\rangle\right)/h\right\}}.
\]
As $h\downarrow0$, each exponential term corresponding to $k\ne a^*$ converges to zero, so the denominator converges to one. The numerator equals one when $a=a^*$ and converges to zero otherwise. Therefore,
\[
\lim_{h\downarrow0}p_{t,a}(\H_t;f_t,h)=I\{a=a^*\}=I\{a=d_{f_t}(\H_t)\}.
\]
Since the maximizer is unique almost surely, the convergence holds almost surely. The proof of Lemma~\ref{lem:softmax} is complete.
\end{proof}

\section{Pointwise Convergence of the Logistic Tail Smoother}
\label{pf:Lemma2}

\begin{lemma}
\label{lem:logistic}
For any fixed $q\in\mathbb R$, $g_\delta(U-q)\to I(U>q)+\frac12 I(U=q)$ almost surely as $\delta\downarrow0$, where $U$ denotes the observed cumulative utility. In particular, if $\PP(U=q)=0$, then $g_\delta(U-q)\to I(U>q)$ almost surely.
\end{lemma}

\begin{proof}
Fix $q\in\mathbb R$ and consider any realization $u$ of $U$. By the definition of the logistic smoother,
\[
g_\delta(u-q)=\frac{1}{1+\exp\{-(u-q)/\delta\}}.
\]
If $u>q$, then $\exp\{-(u-q)/\delta\}\to0$ as $\delta\downarrow0$, and hence $g_\delta(u-q)\to1$. Similarly, if $u<q$, then $\exp\{-(u-q)/\delta\}\to\infty$, so $g_\delta(u-q)\to0$. If $u=q$, then $g_\delta(u-q)=1/2$ for every $\delta>0$. Combining these three cases gives
\[
\lim_{\delta\downarrow0}g_\delta(u-q)=I(u>q)+\frac12 I(u=q).
\]
Since this limit holds for every $u\in\mathbb R$, it also holds almost surely when $u$ is replaced by $U$. 

If $\PP(U=q)=0$, then $\frac12 I(U=q)$ vanishes almost surely,
yielding $g_\delta(U-q)\to I(U>q)$. The proof of Lemma~\ref{lem:logistic} is complete.
\end{proof}

\section{Theoretical Properties}
\label{sec:restricted_theory}
In this section, we study the theoretical properties of the proposed RQDTR method. We first establish identification of the counterfactual survival function and population risk burdens, together with the equivalence between quantile optimization and its constrained formulation. We then study the smoothed penalized estimator over a fixed restricted score class, establishing uniform approximation, consistency, and finite-sample performance guarantees relative to the best feasible regime represented by this class. A separate result relates the restricted-class target to the unrestricted oracle.

Throughout the theoretical analysis, the clinical utility parameters $(\lambda_\mathcal{R},\rho,\xi)$ and the population risk tolerances $b_m$, $m=1,\ldots,M$, are treated as fixed, while the quadratic penalty weights $\mu_U,\mu_1,\ldots,\mu_M$ are allowed to increase with the sample size. Estimation of the risk tolerances is explicitly accommodated in the statistical error bounds.

\subsection{Identification and Oracle Equivalence}
Let $S(q,d)=\PP\{U(d)>q\}$ denote the counterfactual survival function under regime $d$. Throughout, we adopt the survival-based quantile convention $Q_\tau\{U(d)\}=\sup\{q\in\mathbb R:S(q,d)\ge 1-\tau\}$. The following assumption ensures that the quantile is attainable under the corresponding survival constraint.

\begin{assumption}[Quantile attainment]
\label{ass:quantile}
For every $d\in\mathcal D_b$, the $\tau$th quantile $Q_\tau\{U(d)\}$ is finite and satisfies $S(Q_\tau\{U(d)\},d)\ge 1-\tau$. A sufficient condition is that the distribution of $U(d)$ is continuous at its finite $\tau$th quantile.
\end{assumption}

\begin{condition}[Feasibility and existence of an oracle regime]
\label{con:feasibility}
The feasible regime class $\mathcal D_b$ is nonempty, and the optimal quantile $Q_\tau^*=\max_{d\in\mathcal D_b}Q_\tau\{U(d)\}$ is finite and attained by at least one regime $d^*\in\mathcal D_b$.
\end{condition}

The following theorem formalizes the population target and justifies the reformulation presented in Section~2 of the main manuscript.

\begin{theorem}
\label{thm:Thm1}
Under Assumptions~1--3 of the main manuscript, part (i) holds. If Assumption~\ref{ass:quantile} and Condition~\ref{con:feasibility} also hold, then parts (ii) and (iii) follow.
\begin{enumerate}
\item[(i)] \textbf{Identification.} For any measurable regime $d=(d_1,d_2)\in\mathcal D$ and any $q\in\mathbb R$, the counterfactual survival function $S(q,d)$ and the population risk burdens $\mathcal{B}_m(d)$ admit the inverse-probability-weighted representations:
\[
S(q,d)=\mathbb E\left[\frac{I(U>q)I\{A_1=d_1(\H_1)\}I\{A_2=d_2(\H_2)\}}{\pi_1(A_1|\H_1)\pi_2(A_2|\H_2)}\right],
\]
\[
\mathcal B_m(d)=\mathbb E\left[\frac{\psi_{\mathrm{agg}}(Z_1^{(m)},Z_2^{(m)})I\{A_1=d_1(\H_1)\}I\{A_2=d_2(\H_2)\}}{\pi_1(A_1|\H_1)\pi_2(A_2|\H_2)}\right],\ m=1,\ldots,M.
\]
\item[(ii)] \textbf{Optimization equivalence.} Maximizing $Q_\tau\{U(d)\}$ over $\mathcal D_b$ is equivalent to solving
\[
\begin{aligned}
\max_{q\in\mathbb R,\ d\in\mathcal D}\quad &q\\
\text{subject to}\quad &S(q,d)\ge 1-\tau,\\
&\mathcal B_m(d)\le b_m,\ m=1,\ldots,M.
\end{aligned}
\]
In particular, both problems have the same optimal value $Q_\tau^*$.
\item[(iii)] \textbf{Optimal solution equivalence.} If $d^*\in\argmax_{d\in\mathcal D_b}Q_\tau\{U(d)\}$, then $(Q_\tau\{U(d^*)\},d^*)$ solves the constrained problem in (ii). Conversely, if $(q^*,d^*)$ solves the constrained problem in (ii), then $d^*$ is a quantile-optimal feasible regime and $q^*=Q_\tau\{U(d^*)\}$. 
\end{enumerate}
\end{theorem}

Theorem~\ref{thm:Thm1} establishes the population-level foundation of the proposed framework. Part (i) identifies the counterfactual quantities through observed-data representations. Parts (ii) and (iii) justify the constrained optimization formulation and characterize its optimal solutions. These results hold over the unrestricted regime class and do not depend on the choice of score functions or smoothing approximations. The proof is given in Section~\ref{pf:Thm1}.

As a direct consequence of Theorem~\ref{thm:Thm1}, the unified formulation recovers several practically useful subclasses through different choices of the risk-control parameters.

\begin{corollary}[Special cases of the unified formulation]
\label{cor:special_cases}
Under the conditions of Theorem~\ref{thm:Thm1} for the corresponding feasible regime class, the proposed framework recovers three particularly interpretable special cases: (i) If $\lambda_\mathcal{R}=0$ and $b_m=+\infty$ for all $m$, or equivalently $\mathcal D_b=\mathcal D$, it reduces to an efficacy-only quantile DTR method. (ii) If $\lambda_\mathcal{R}=0$ and finite risk bounds $b_m$ are imposed, it reduces to a constraint-based quantile DTR method. (iii)  If $\lambda_\mathcal{R}>0$ and $b_m=+\infty$ for all $m$, equivalently $\mathcal D_b=\mathcal D$, it reduces to a utility-based quantile DTR method. In addition, allowing $\lambda_{\mathcal R}>0$ together with population risk constraints yields the hybrid formulation.
\end{corollary}

\subsection{Population Surrogate Consistency}
\label{sec:surrogate_consistency}
We next examine whether maximizing the smoothed population criterion recovers a quantile-optimal feasible regime as the smoothing parameters vanish and the quadratic penalty weights increase. We first establish recovery of an optimal feasible regime within a fixed score class. Recovery of an unrestricted oracle further requires that the score class contain a representation of such an oracle.

Let $\mathcal F_t^\dagger\subseteq\mathcal F_t(C_{\mathcal F})$, $t=1,2$, be fixed, nonempty, deterministic score classes, and write $\mathcal F=\mathcal F_1^\dagger\times\mathcal F_2^\dagger$. All score optimizations in this subsection are over $\mathcal F$. For $f=(f_1,f_2)\in\mathcal F$, let $d_f=(d_{f_1},d_{f_2})$, where $d_{f_t}(h_t)=\argmax_{a\in\mathcal A}\langle V_a,f_t(h_t)\rangle$, with ties resolved by a fixed measurable rule. Define $\rho_{\mathcal F}(f,f')=\max_{t=1,2}\sup_{h_t\in\mathcal H_t}\|f_t(h_t)-f_t'(h_t)\|$.

Let $c_\tau=1-\tau$ and take $\mathcal Q=[q_L,q_U]$ to be a fixed, sufficiently wide interval satisfying $\inf_{f\in\mathcal F}S(q_L,d_f)>c_\tau$ and $\sup_{f\in\mathcal F}S(q_U,d_f)<c_\tau$. These inequalities ensure that the quantiles of the represented regimes lie in the interior of $\mathcal Q$. Under uniform positivity and almost-sure finiteness of $U$, such finite endpoints exist because $1-S(q_L,d_f)\le\pi_{\min}^{-2}\PP(U\le q_L)$ and $S(q_U,d_f)\le\pi_{\min}^{-2}\PP(U>q_U)$ uniformly over $f\in\mathcal F$.

Let $\Theta=\mathcal Q\times\mathcal F$ and define $\rho_\Theta\{(q,f),(q',f')\}=|q-q'|+\rho_{\mathcal F}(f,f')$. Distances between elements of $\Theta$ are measured using $\rho_\Theta$. Let $\mathcal I\subseteq\{1,\ldots,M\}$ index the imposed risk constraints, and put $\mathcal J=\{0\}\cup\mathcal I$, where index $0$ denotes the survival constraint. Define $\mathcal F_b=\{f\in\mathcal F:\mathcal B_m(d_f)\le b_m,\ m\in\mathcal I\}$ and $\psi_m=\psi_{\mathrm{agg}}(Z_1^{(m)},Z_2^{(m)})$. When $\mathcal I=\varnothing$, the risk constraints and their penalties are omitted, and maxima of nonnegative quantities over $\mathcal I$ are defined as zero.

\begin{assumption}[Restricted-class feasibility and regularity]
\label{ass:restricted_regular}
The score class $\mathcal F$ is a nonempty compact subset of bounded measurable score functions under $\rho_{\mathcal F}$, and the restricted feasible class $\mathcal F_b$ is nonempty. The observed cumulative utility $U$ is finite almost surely, and for every $m\in\mathcal I$, the risk aggregate satisfies $|\psi_m|\le C_\psi$ almost surely for some finite constant $C_\psi$.
\end{assumption}

Since $\mathcal Q$ is compact, Assumption~\ref{ass:restricted_regular} also implies that $\Theta$ is compact under $\rho_\Theta$.

\begin{assumption}[Uniform boundary regularity]
\label{ass:boundary}
The following conditions hold:
\begin{enumerate}
\item[(i)] For each $t=1,2$, let $\Delta_t(\H_t;f_t)$ denote the difference between the largest and second-largest treatment scores $\langle V_a,f_t(\H_t)\rangle$, $a\in\mathcal A$. There exist constants $C_t,\alpha_t,r_t>0$ such that
\[
\sup_{f_t\in\mathcal F_t^\dagger}\PP\{\Delta_t(\H_t;f_t)\le r\}\le C_tr^{\alpha_t},\ 0<r\le r_t.
\]
\item[(ii)] There exist constants $C_U,r_U>0$ such that
\[
\sup_{q\in\mathcal Q}\PP(|U-q|\le r)\le C_Ur,\ 0<r\le r_U.
\]
\end{enumerate}
\end{assumption}

The probabilities in Assumption~\ref{ass:boundary} are taken under the observed-data law. A bounded density of $U$ on a neighborhood of $\mathcal Q$ is sufficient for part (ii).
For example, consider binary treatments with simplex vertices $V_1=-1$ and $V_2=1$, and suppose that a history coordinate $X_t$ follows $\operatorname{Unif}[-1,1]$. The restricted class $\mathcal F_t^\dagger=\{f_{t,b}(H_t)=X_t-b:|b|\le b_0\}$, where $0<b_0<1$, is compact under the supremum norm and consists of bounded scores. Since $\Delta_t(H_t;f_{t,b})=2|X_t-b|$, we have $\sup_{|b|\le b_0}P\{\Delta_t(H_t;f_{t,b})\le r\}\le r/2$ for sufficiently small $r>0$, verifying Assumption~\ref{ass:boundary}(i) with $\alpha_t=1$.

For $\theta=(q,f)\in\Theta$, define the hard constraint violations and feasible parameter set by
\[
v_0(q,f)=[c_\tau-S(q,d_f)]_+,\ v_m(q,f)=[\mathcal B_m(d_f)-b_m]_+,\ m\in\mathcal I,
\]
\[
\Theta_b=\{(q,f)\in\Theta:v_j(q,f)=0,\ j\in\mathcal J\}.
\]
The restricted oracle value and its solution set are
\[
q_\dagger=Q_{\tau,\dagger}^*=\sup_{(q,f)\in\Theta_b}q,\ \mathcal A_\dagger=\{(q,f)\in\Theta_b:q=q_\dagger\}.
\]
The theorem below establishes that this supremum is attained and agrees with the best feasible quantile represented by $\mathcal F$.

Write $\Pi=\prod_{t=1}^2\pi_t(A_t|\H_t)$. The smoothed population quantities are
\[
\begin{aligned}
S_{h,\delta}(q,f)&=\mathbb E\left[\frac{g_\delta(U-q)\prod_{t=1}^2p_{t,A_t}(\H_t;f_t,h)}{\Pi}\right],\\
\mathcal B_{m,h}(f)&=\mathbb E\left[\frac{\psi_m\prod_{t=1}^2p_{t,A_t}(\H_t;f_t,h)}{\Pi}\right].
\end{aligned}
\]
Let $v_{0,h,\delta}(q,f)=[c_\tau-S_{h,\delta}(q,f)]_+$ and $v_{m,h,\delta}(q,f)=[\mathcal B_{m,h}(f)-b_m]_+$ for $m\in\mathcal I$. With $\mu_{0,n}=\mu_{U,n}$, define
\[
\begin{gathered}
\mathcal M_{n,h,\delta}(q,f)=q-\sum_{j\in\mathcal J}\mu_{j,n}v_{j,h,\delta}(q,f)^2,\\
\underline\mu_n=\min_{j\in\mathcal J}\mu_{j,n},\ \overline\mu_n=\max_{j\in\mathcal J}\mu_{j,n}.
\end{gathered}
\]
Here $n$ indexes a deterministic population approximation sequence; in the subsequent statistical analysis, it also indexes the sample size. For $0<h,\delta<e^{-1}$, define
\begin{equation*}
\label{eq:smoothing_rate}
a_{h,\delta}=\delta\log(1/\delta)+h+\sum_{t=1}^2\{h\log(1/h)\}^{\alpha_t}.
\end{equation*}

\begin{theorem}[Surrogate consistency with increasing penalties]
\label{thm:Thm2}
Suppose Assumptions~1--3 of the main manuscript and Assumptions~\ref{ass:restricted_regular} and \ref{ass:boundary} hold. In addition, suppose positivity holds uniformly, with $\pi_t(a|\H_t)\ge\pi_{\min}>0$ almost surely for each $t=1,2$ and $a\in\mathcal A$. Then:
\begin{enumerate}
\item[(i)] \textbf{Uniform approximation and the restricted oracle.} There is a constant $C<\infty$, independent of $h$, $\delta$, and the penalty weights, such that, for all sufficiently small $h,\delta>0$,
\begin{equation}
\label{eq:uniform_primitive_approximation}
\sup_{(q,f)\in\Theta}|S_{h,\delta}(q,f)-S(q,d_f)|+\max_{m\in\mathcal I}\sup_{f\in\mathcal F}|\mathcal B_{m,h}(f)-\mathcal B_m(d_f)|\le Ca_{h,\delta}.
\end{equation}
The hard constraint violations are continuous on $\Theta$, and $\Theta_b$ and $\mathcal A_\dagger$ are nonempty and compact. Moreover,
\begin{equation*}
\label{eq:restricted_target}
q_\dagger=\max_{(q,f)\in\Theta_b}q=\max_{f\in\mathcal F_b}Q_\tau\{U(d_f)\}\in(q_L,q_U).
\end{equation*}
\item[(ii)] \textbf{Consistency of population surrogate maximizers.} Suppose further that the penalty weights $\mu_{j,n}$, $j\in\mathcal J$, are finite, positive, and deterministic for every $n$, and that $\underline\mu_n\to\infty$. Let $h_n,\delta_n\in(0,e^{-1})$ tend to zero and satisfy $\overline\mu_n a_{h_n,\delta_n}^2\to0$.
For every $n$, the population criterion $\mathcal M_{n,h_n,\delta_n}$ attains its maximum on $\Theta$. Any sequence of maximizers $\theta_n^\circ=(q_n^\circ,f_n^\circ)$ satisfies
\[
\operatorname{dist}(\theta_n^\circ,\mathcal A_\dagger)\to0,\ q_n^\circ\to q_\dagger,\ \max_{j\in\mathcal J}v_j(\theta_n^\circ)\to0.
\]
Every accumulation point $(\bar q,\bar f)$ therefore satisfies $\bar f\in\mathcal F_b$ and $\bar q=Q_\tau\{U(d_{\bar f})\}=q_\dagger$.
\item[(iii)] \textbf{Recovery of an unrestricted oracle.} Let $\mathcal D_{b,\mathcal I}=\{d\in\mathcal D:\mathcal B_m(d)\le b_m,\ m\in\mathcal I\}$ denote the unrestricted feasible regime class, and let $Q_\tau^*=\sup_{d\in\mathcal D_{b,\mathcal I}}Q_\tau\{U(d)\}$ be its optimal quantile value. If some $f^*\in\mathcal F_b$ satisfies $Q_\tau\{U(d_{f^*})\}=Q_\tau^*$, then $q_\dagger=Q_\tau^*$, and every accumulation point $(\bar q,\bar f)$ in (ii) induces a globally quantile-optimal feasible regime:
\[
d_{\bar f}\in\argmax_{d\in\mathcal D_{b,\mathcal I}}Q_\tau\{U(d)\},\ \bar q=Q_\tau^*.
\]
For $\mathcal I=\{1,\ldots,M\}$, this is the unrestricted oracle characterized in Theorem~\ref{thm:Thm1}.
\end{enumerate}
\end{theorem}

Theorem~\ref{thm:Thm2} establishes that the smoothed penalized population criterion asymptotically recovers the best feasible regime within the restricted score class as the smoothing parameters vanish and the penalty weights increase. Importantly, this guarantee does not require the restricted class to contain an unrestricted oracle; such representability is needed only to extend the result to global optimality.  The proof is given in Section~\ref{pf:Thm2}.

\subsection{Consistency of the Estimated Treatment Regime}
\label{sec:thm-3}
We next extend the population-level analysis to an empirical estimator based on $n$ observed trajectories. The empirical criterion accommodates estimated propensity scores and risk tolerances, regularization, and approximate optimization. The target remains the restricted oracle characterized in Theorem~\ref{thm:Thm2}, with the regularization parameter required to vanish asymptotically to preserve this target.

Let $\mathcal T_1,\ldots,\mathcal T_n$ denote the observed trajectories, and retain $\Theta$, $\mathcal I$, $\mathcal J$, $q_\dagger$, and $\mathcal A_\dagger$ from Section~\ref{sec:surrogate_consistency}. Write $\Pi_i=\prod_{t=1}^2\pi_t(A_{ti}|\H_{ti})$, $\widetilde\Pi_{n,i}=\prod_{t=1}^2\widetilde\pi_{n,t}(A_{ti}|\H_{ti})$, and $\psi_{m,i}=\psi_{\mathrm{agg}}(Z_{1i}^{(m)},Z_{2i}^{(m)})$. The empirical smoothed survival function and risk burdens are
\[
\begin{aligned}
\widehat S_{h_n,\delta_n}(q,f)&=\frac1n\sum_{i=1}^n\frac{g_{\delta_n}(U_i-q)\prod_{t=1}^2p_{t,A_{ti}}(\H_{ti};f_t,h_n)}{\widetilde\Pi_{n,i}},\\
\widehat{\mathcal B}_{m,h_n}(f)&=\frac1n\sum_{i=1}^n\frac{\psi_{m,i}\prod_{t=1}^2p_{t,A_{ti}}(\H_{ti};f_t,h_n)}{\widetilde\Pi_{n,i}},\ m\in\mathcal I.
\end{aligned}
\]
For risk-tolerance estimates $\widehat b_{m,n}$, define
\[
\widehat v_{0,n}(q,f)=[c_\tau-\widehat S_{h_n,\delta_n}(q,f)]_+,\ \widehat v_{m,n}(q,f)=[\widehat{\mathcal B}_{m,h_n}(f)-\widehat b_{m,n}]_+,\ m\in\mathcal I.
\]
Known propensities and fixed risk tolerances are included by setting $\widetilde\pi_{n,t}=\pi_t$ and $\widehat b_{m,n}=b_m$, respectively. Let $J(f)=\|f_1\|^2+\|f_2\|^2$ be the regularizer from the main manuscript, with $J_{\max}=\sup_{f\in\mathcal F}J(f)<\infty$ under the fixed RKHS norm bounds. The empirical regularized objective is
\[
\widehat{\mathcal L}_n(q,f)=q-\sum_{j\in\mathcal J}\mu_{j,n}\widehat v_{j,n}(q,f)^2-\lambda_nJ(f).
\]
Here $\lambda_n$ is the regularization parameter, distinct from the fixed clinical utility parameter $\lambda_{\mathcal R}$. The estimator is optimized over the same restricted domain $\Theta$ used in Theorem~\ref{thm:Thm2}.

\begin{assumption}[Sampling, entropy, and nuisance regularity]
\label{ass:primitive_entropy}
The following conditions hold:
\begin{enumerate}
\item[(i)] The observed trajectories are independent and identically distributed. The bandwidths $h_n,\delta_n\in(0,e^{-1})$ are deterministic and tend to zero. There exist a deterministic sequence $u_n\in(0,1)$ tending to zero and a sufficiently small fixed constant $c>0$ such that
\[
\mathcal E_n=1+\log\{1/(u_n\delta_n)\}+\log\mathcal N(\mathcal F,\rho_{\mathcal F},cu_nh_n)=o(n),\ r_{n,\mathrm{emp}}=u_n+\sqrt{\mathcal E_n/n}.
\]
Here $\mathcal N(\mathcal F,\rho_{\mathcal F},r)$ is the covering number at radius $r$, and $u_n$ is an auxiliary covering radius used in the proof, not an algorithm parameter.
\item[(ii)] The fitted propensities are positive and have uniformly bounded inverses. Moreover,
\[
\Xi_{\pi,n}=\frac1n\sum_{i=1}^n|\widetilde\Pi_{n,i}^{-1}-\Pi_i^{-1}|=o_p(1),\ \Delta_{b,n}=\max_{m\in\mathcal I}|\widehat b_{m,n}-b_m|=o_p(1).
\]
The corresponding errors are zero when propensities are known or risk tolerances are fixed. For $\mathcal I=\varnothing$, set $\Delta_{b,n}=0$.
\item[(iii)] The regularization parameters are deterministic and nonnegative; $r_{n,\mathrm{reg}}=\lambda_nJ_{\max}\to0$.
\end{enumerate}
\end{assumption}

Uniform consistency of the bounded inverse propensity estimates is sufficient for part (ii), but the proof only requires the displayed average denominator error. The propensity estimates may be fitted using the same observations as the empirical criterion.

\begin{assumption}[Approximate empirical optimization]
\label{ass:approx_opt}
The estimator $\widehat\theta_n=(\widehat q_n,\widehat f_n)$ is a measurable element of $\Theta$ satisfying $\widehat{\mathcal L}_n(\widehat\theta_n)\ge\sup_{\theta\in\Theta}\widehat{\mathcal L}_n(\theta)-\varepsilon_n$, where $\varepsilon_n\ge0$ and $\varepsilon_n=o_p(1)$.
\end{assumption}

\begin{assumption}[Uniform crossing at restricted oracle quantiles]
\label{ass:oracle_crossing}
There exists $r_0>0$ such that $[q_\dagger-r_0,q_\dagger+r_0]\subset\mathcal Q$ and, for every $0<r\le r_0$, there is an $\eta(r)>0$ satisfying
\[
\inf_{(q_\dagger,f^*)\in\mathcal A_\dagger}S(q_\dagger-r,d_{f^*}) \ge c_\tau+\eta(r),\ \sup_{(q_\dagger,f^*)\in\mathcal A_\dagger}S(q_\dagger+r,d_{f^*}) \le c_\tau-\eta(r).
\]
\end{assumption}

This condition is used to translate convergence near the oracle set into convergence of the actual regime quantiles. It is not needed for asymptotic risk feasibility.

\begin{lemma}[Uniform empirical approximation]
\label{lem:empirical_uniform}
Suppose the assumptions of Theorem~\ref{thm:Thm2}(i) and Assumption~\ref{ass:primitive_entropy} hold. Let $\widehat S^{\pi}_{h_n,\delta_n}$ and $\widehat{\mathcal B}^{\pi}_{m,h_n}$ denote the empirical quantities with $\widetilde\Pi_{n,i}$ replaced by $\Pi_i$. Then
\[
E_n^0:=\max\Bigg\{\sup_{(q,f)\in\Theta}|\widehat S^{\pi}_{h_n,\delta_n}(q,f)-S_{h_n,\delta_n}(q,f)|,\ \max_{m\in\mathcal I}\sup_{f\in\mathcal F}|\widehat{\mathcal B}^{\pi}_{m,h_n}(f)-\mathcal B_{m,h_n}(f)|\Bigg\}=O_p(r_{n,\mathrm{emp}}).
\]
Consequently, for a constant $C<\infty$ independent of $n$, the empirical constraint violations satisfy, for all sufficiently large $n$,
\begin{equation}
\label{eq:primitive_error}
e_n:=\max_{j\in\mathcal J}\sup_{\theta\in\Theta}|\widehat v_{j,n}(\theta)-v_j(\theta)|\le E_n^0+C\Xi_{\pi,n}+Ca_{h_n,\delta_n}+\Delta_{b,n}.
\end{equation}
\end{lemma}
\begin{lemma}[Continuity near the restricted oracle set]
\label{lem:oracle_continuity}
Suppose the assumptions of Theorem~\ref{thm:Thm2}(i) hold. If a deterministic sequence $\theta_n=(q_n,f_n)\in\Theta$ satisfies $\operatorname{dist}(\theta_n,\mathcal A_\dagger)\to0$, then $\max_{m\in\mathcal I}[\mathcal B_m(d_{f_n})-b_m]_+\to0$. If Assumption~\ref{ass:oracle_crossing} also holds, then $Q_\tau\{U(d_{f_n})\}\to q_\dagger$.
\end{lemma}
\begin{theorem}[Consistency of the estimated treatment regime]
\label{thm:Thm3}
Suppose the assumptions of Theorem~\ref{thm:Thm2}(i) and Assumptions~\ref{ass:primitive_entropy}--\ref{ass:approx_opt} hold. Let the penalty weights $\mu_{j,n}$, $j\in\mathcal J$, be finite, positive, and deterministic for every $n$, with $\underline\mu_n\to\infty$, and suppose
\begin{equation}
\label{eq:qualitative_compatibility}
\overline\mu_n\{a_{h_n,\delta_n}^2+r_{n,\mathrm{emp}}^2+\Xi_{\pi,n}^2+\Delta_{b,n}^2\}\xrightarrow{p}0.
\end{equation}
Then, $\operatorname{dist}(\widehat\theta_n,\mathcal A_\dagger)\xrightarrow{p}0,\ \widehat q_n\xrightarrow{p}q_\dagger,\ \max_{j\in\mathcal J}v_j(\widehat\theta_n)\xrightarrow{p}0$. In particular, the estimated regime is asymptotically risk-feasible, i.e., $\max_{m\in\mathcal I}[\mathcal B_m(d_{\widehat f_n})-b_m]_+\xrightarrow{p}0$. If Assumption~\ref{ass:oracle_crossing} also holds, then
\[
Q_\tau\{U(d_{\widehat f_n})\}\xrightarrow{p}q_\dagger=Q_{\tau,\dagger}^*.
\]
Under the additional representability condition in Theorem~\ref{thm:Thm2}(iii), this limiting quantile equals the unrestricted optimal value $Q_\tau^*$.
\end{theorem}

Theorem~\ref{thm:Thm3} extends the population surrogate consistency result to the empirical estimator, establishing convergence to the restricted oracle set and asymptotic satisfaction of the population risk constraints. Under the additional oracle-crossing condition, the learned treatment regime also achieves the restricted optimal quantile asymptotically, and this guarantee extends to the unrestricted optimum when an unrestricted oracle is representable in the score class. Note that asymptotic risk feasibility concerns vanishing constraint violation magnitudes and does not, by itself, imply that the probability of exact constraint satisfaction converges to one.

The proof is given in Section~\ref{pf:Thm3}.

\subsection{Finite-Sample Performance Error Bounds}
\label{sec:restricted_rates}
We next derive estimation error bounds relative to the restricted oracle. We first establish bounds on the threshold shortfall and population constraint violations, and then impose additional local regularity conditions to obtain convergence rates for the distance to the oracle set and the quantile-value error relative to the restricted oracle.

We retain the notation from Section~\ref{sec:thm-3}, including $\widehat\theta_n=(\widehat q_n,\widehat f_n)$, $e_n$ defined in \eqref{eq:primitive_error}, $r_{n,\mathrm{emp}}$ from Assumption~\ref{ass:primitive_entropy}(i), and $r_{n,\mathrm{reg}}=\lambda_nJ_{\max}$. The following assumption specifies convergence rates for the remaining error terms.

\begin{assumption}[Nuisance, tolerance, and optimization rates]
\label{ass:emp_rate}
There exist deterministic nonnegative sequences $r_{\pi,n}$, $r_{b,n}$, and $r_{n,\mathrm{opt}}$ tending to zero such that
\[
\Xi_{\pi,n}=O_p(r_{\pi,n}),\ \Delta_{b,n}=O_p(r_{b,n}),\ \varepsilon_n=O_p(r_{n,\mathrm{opt}}).
\]
The corresponding rate terms are set to zero when the errors vanish identically.
\end{assumption}

Define the deterministic rates: $s_n =r_{n,\mathrm{emp}} +r_{\pi,n}+r_{b,n}+a_{h_n,\delta_n}$, $d_n =\overline\mu_ns_n^2 +r_{n,\mathrm{reg}}+r_{n,\mathrm{opt}}$, $t_n=s_n +\underline\mu_n^{-1/2}$, and $\Delta_n=d_n+t_n$. Here $s_n$ collects the errors in approximating the hard population constraints, $d_n$ controls the oracle comparison loss, $t_n$ controls constraint violations, and $\Delta_n$ controls the combined optimality and feasibility residual. In this subsection, the scalar $d_n$ denotes an error rate, not a treatment regime.
For $\theta=(q,f)\in\Theta$, define
\[
G_\dagger(\theta)=(q_\dagger-q)_++\max_{j\in\mathcal J}v_j(\theta).
\]
This nonnegative residual vanishes if and only if $\theta\in\mathcal A_\dagger$. Indeed, a zero residual implies feasibility and $q\ge q_\dagger$, which, by restricted optimality, yields $q=q_\dagger$.

\begin{assumption}[Quantitative oracle regularity]
\label{ass:quant_oracle}
The following conditions hold:
\begin{enumerate}
\item[(i)] \textbf{Local constrained error bound.} There exist constants $c_{\mathrm{sep}},\gamma,r_{\mathrm{sep}}>0$ such that
\[
G_\dagger(\theta)\ge c_{\mathrm{sep}}\operatorname{dist}(\theta,\mathcal A_\dagger)^\gamma
\]
for every $\theta\in\Theta$ with $\operatorname{dist}(\theta,\mathcal A_\dagger)\le r_{\mathrm{sep}}$.
\item[(ii)] \textbf{Uniform linear oracle crossing.} There exist constants $c_{\mathrm{cross}},r_0>0$ with $[q_\dagger-r_0,q_\dagger+r_0]\subset\mathcal Q$ such that, for every $(q_\dagger,f^*)\in\mathcal A_\dagger$ and every $0<r\le r_0$,
\[
S(q_\dagger-r,d_{f^*})\ge c_\tau+c_{\mathrm{cross}}r,\ S(q_\dagger+r,d_{f^*})\le c_\tau-c_{\mathrm{cross}}r.
\]
\end{enumerate}
\end{assumption}

In Assumption~\ref{ass:quant_oracle}, part (i) strengthens the qualitative separation argument in the proof of Theorem~\ref{thm:Thm3} by imposing a local polynomial lower bound on the constrained residual, which does not follow from compactness alone. Part (ii) strengthens Assumption~\ref{ass:oracle_crossing} by requiring a uniform linear crossing at the restricted oracle quantile. A sufficient condition is that the counterfactual utility distributions under all restricted oracle regimes have densities uniformly bounded away from zero on a common neighborhood of $q_\dagger$. This lower-density condition differs from the upper probability-mass bound on the observed utility in Assumption~\ref{ass:boundary}(ii), which controls the logistic smoothing error.

\begin{theorem}[Finite-sample bounds and performance error rates]
\label{thm:Thm4}
Suppose the assumptions of Theorem~\ref{thm:Thm2}(i) and Assumptions~\ref{ass:primitive_entropy}, \ref{ass:approx_opt}, and \ref{ass:emp_rate} hold. Let the penalty weights $\mu_{j,n}$, $j\in\mathcal J$, be finite, positive, and deterministic for every $n$, with $\underline\mu_n\to\infty$, and suppose $\overline\mu_ns_n^2\to0$. Let $\alpha=\min(\alpha_1,\alpha_2)$, where $\alpha_1, \alpha_2$ are the treatment-score margin exponents in Assumption~\ref{ass:boundary}(i). Then:
\begin{enumerate}
\item[(i)] \textbf{Direct threshold and feasibility bounds.} Define $\widehat D_n=|\mathcal J|\overline\mu_ne_n^2+r_{n,\mathrm{reg}}+\varepsilon_n$. For every $n$, the following inequalities hold almost surely:
$$(q_\dagger-\widehat q_n)_+ \le\widehat D_n,\ v_j(\widehat\theta_n) \le e_n+\left(\frac{q_U-q_\dagger+\widehat D_n}{\mu_{j,n}}\right)^{1/2},\ j\in\mathcal J.$$
Consequently, $(q_\dagger-\widehat q_n)_+=O_p(d_n)$, $\max_{j\in\mathcal J}v_j(\widehat\theta_n)=O_p(t_n)$, $G_\dagger(\widehat\theta_n) =O_p(\Delta_n)$.
These conclusions do not require Assumption~\ref{ass:quant_oracle} or Assumption~\ref{ass:oracle_crossing}.
\item[(ii)] \textbf{Oracle-distance and risk rates.} If Assumption~\ref{ass:quant_oracle}(i) also holds, then
\[
\operatorname{dist}(\widehat\theta_n,\mathcal A_\dagger)=O_p(\Delta_n^{1/\gamma}),\ |\widehat q_n-q_\dagger|=O_p(\Delta_n^{1/\gamma}),\ \max_{m\in\mathcal I}[\mathcal B_m(d_{\widehat f_n})-b_m]_+=O_p(\Delta_n^{\alpha/\gamma}).
\]
The direct risk bound $O_p(t_n)$ from part (i) remains valid without this additional assumption.
\item[(iii)] \textbf{Actual quantile performance.} If both parts of Assumption~\ref{ass:quant_oracle} hold, then
\[
|Q_\tau\{U(d_{\widehat f_n})\}-q_\dagger|=O_p(\Delta_n^{\alpha/\gamma}).
\]
Under the representability condition in Theorem~\ref{thm:Thm2}(iii), $q_\dagger=Q_\tau^*$, so the same rate holds relative to the unrestricted optimal quantile. More generally, if $Q_\tau^*=\sup_{d\in\mathcal D_{b,\mathcal I}}Q_\tau\{U(d)\}$ is finite, then
\[
[Q_\tau^*-Q_\tau\{U(d_{\widehat f_n})\}]_+ \le (Q_\tau^*-q_\dagger)
+[q_\dagger-Q_\tau\{U(d_{\widehat f_n})\}]_+,
\]
where $Q_\tau^*-q_\dagger\ge0$ is the class approximation
gap, and $[q_\dagger-Q_\tau\{U(d_{\widehat f_n})\}]_+=O_p(\Delta_n^{\alpha/\gamma})$.
\end{enumerate}
\end{theorem}

Theorem~\ref{thm:Thm4} reveals a trade-off in the penalty weights: increasing $\underline\mu_n$ reduces the direct feasibility error term $\underline\mu_n^{-1/2}$, whereas increasing $\overline\mu_n$ amplifies the squared constraint approximation error through $\overline\mu_ns_n^2$. The exponent $\gamma$ governs the conversion of the constrained residual into distance to the restricted oracle set, while $\alpha$ controls the effect of score perturbations on treatment decisions. Under uniform linear oracle crossing (Assumption~\ref{ass:quant_oracle}(ii)), the resulting survival approximation bound further yields the actual quantile error rate. The resulting performance rates account for empirical fluctuations, propensity and risk-tolerance estimation, smoothing approximation, regularization, numerical optimization, and finite-penalty effects. The proof is given in Section~\ref{pf:Thm4}.

\subsection{Proof of Theorem~\ref{thm:Thm1}}
\label{pf:Thm1}
This proof is standard but included for completeness, since the present formulation combines a quantile-based utility target with multiple attribute-level cumulative risk constraints.

\begin{proof}
For simplicity, write
$$I_d = I \{A_1=d_1(\bm H_1)\} I \{A_2=d_2(\bm H_2)\}, \ \Pi_d = \pi_1(A_1 | \bm H_1) \pi_2(A_2 | \bm H_2).$$

We prove the three parts in turn.

\medskip
\noindent
\textit{Proof of (i).}
By consistency (Assumption~1 of the main manuscript), on the event $\{I_d=1\}$, the observed utility coincides with its counterfactual version under regime $d$, i.e., $U=U(d)$. Therefore,
$$\mathbb{E} \left[\frac{I \{U>q\}I_d}{\Pi_d}\right] = \mathbb{E} \left[\frac{I \{U(d)>q\}I_d}{\Pi_d}\right].$$

Using sequential ignorability (Assumption~2 of the main manuscript) at stage 2, on the event $\{A_1=d_1(\H_1)\}$, we have
\begin{align*}
&\mathbb E\left[\frac{I\{U(d)>q\}I\{A_2=d_2(\H_2)\}}{\pi_2(A_2|\H_2)}\bigg|\H_2\right]\\
&=\mathbb E\left[I\{U(d)>q\}|\H_2,A_2=d_2(\H_2)\right] =\mathbb E\left[I\{U(d)>q\}|\H_2\right],
\end{align*}
where the first equality follows from the definition of the propensity score and positivity (Assumption~3 of the main manuscript), and the second follows from sequential ignorability. Since $I\{A_1=d_1(\H_1)\}/\pi_1(A_1|\H_1)$ is measurable with respect to $\H_2$, the law of total expectations yields
\begin{align*}
\mathbb E\left[\frac{I\{U(d)>q\}I_d}{\Pi_d}\right]
&= \mathbb{E} \left[ \mathbb{E} \left\{
\frac{I \{U(d)>q\} I \{A_1=d_1(\bm H_1)\} I \{A_2=d_2(\bm H_2)\}}{\pi_1(A_1| \bm H_1)\pi_2(A_2| \bm H_2)}
\,\Bigg|\, \bm H_2 \right\} \right] \\
&=\mathbb E\left[\frac{I\{A_1=d_1(\H_1)\}}{\pi_1(A_1|\H_1)}\mathbb E\left\{\frac{I\{U(d)>q\}I\{A_2=d_2(\H_2)\}}{\pi_2(A_2|\H_2)}\bigg|\H_2\right\}\right]\\
&=\mathbb E\left[\frac{I\{A_1=d_1(\H_1)\}}{\pi_1(A_1|\H_1)}\mathbb E\{I\{U(d)>q\}|\H_2\}\right]\\
&=\mathbb E\left[\frac{I\{U(d)>q\}I\{A_1=d_1(\H_1)\}}{\pi_1(A_1|\H_1)}\right].
\end{align*}

Similarly, sequential ignorability at stage 1, together with positivity, implies
\begin{align*}
&\mathbb E\left[\frac{I\{U(d)>q\}I\{A_1=d_1(\H_1)\}}{\pi_1(A_1|\H_1)}\bigg|\H_1\right]\\
&=\mathbb E\left[I\{U(d)>q\}|\H_1,A_1=d_1(\H_1)\right] =\mathbb E\left[I\{U(d)>q\}|\H_1\right].
\end{align*}
Taking expectations on both sides gives
\begin{align*}
\mathbb E\left[\frac{I\{U(d)>q\}I\{A_1=d_1(\H_1)\}}{\pi_1(A_1|\H_1)}\right]
&=\mathbb E\left[\mathbb E\{I\{U(d)>q\}|\H_1\}\right]\\
&=\PP\{U(d)>q\}=S(q,d).
\end{align*}
This establishes the identification formula for $S(q,d)$.

For the population risk burden, consistency implies
$$\psi_{\mathrm{agg}}(Z_1^{(m)},Z_2^{(m)})I_d=\psi_{\mathrm{agg}}(Z_1^{(m)}(d),Z_2^{(m)}(d))I_d.$$
Applying the same two-stage iterated-expectation argument to the integrable counterfactual risk aggregate yields
$$\mathbb E\left[\frac{\psi_{\mathrm{agg}}(Z_1^{(m)},Z_2^{(m)})I_d}{\Pi_d}\right]=\mathbb E\left[\psi_{\mathrm{agg}}(Z_1^{(m)}(d),Z_2^{(m)}(d))\right]=\mathcal B_m(d),\ m=1,\ldots,M.$$

\noindent
\textit{Proof of (ii).}
Under Assumption~\ref{ass:quantile}, for any fixed regime $d\in\mathcal D_b$, the $\tau$th quantile is
$$Q_\tau\{U(d)\}=\max\{q:S(q,d)\ge 1-\tau\}.$$
Hence, maximizing $Q_\tau\{U(d)\}$ over the feasible set $\mathcal D_b$ is equivalent to maximizing $q$ over all pairs $(q,d)$ such that
$S(q,d)\ge 1-\tau$ and $\mathcal B_m(d)\le b_m,\ m=1,\dots,M$.

Thus, $\max_{d\in\mathcal D_b}Q_\tau\{U(d)\}$ is equivalent to
\begin{align*}
    &\max_{q\in\mathbb R,\ d\in\mathcal D} \ q \\
    & \text{subject to} \ S(q,d)\ge 1-\tau,\\
    &\hspace{18mm} \mathcal{B}_m(d)\le b_m,\ m=1,\dots,M.
\end{align*}

\noindent
\textit{Proof of (iii).}
Let $d^* \in \argmax_{d\in\mathcal D_b}Q_\tau\{U(d)\}$ and define $q^*=Q_\tau\{U(d^*)\}$. Since $d^*\in\mathcal D_b$, we have $\mathcal B_m(d^*)\le b_m,\ m=1,\dots,M$. By the definition of $Q_\tau\{U(d^*)\}$ and Assumption~\ref{ass:quantile}, $S(q^*,d^*)\ge 1-\tau$. Thus, $(q^*,d^*)$ is feasible for the constrained problem in (ii).

Let $(q,d)$ be any feasible pair. Then $S(q,d)\ge 1-\tau, \ \mathcal B_m(d)\le b_m, \ m=1,\dots,M$, so $d\in\mathcal D_b$. By the definition of $Q_\tau\{U(d)\}$, we know that $q\le Q_\tau\{U(d)\}\le Q_\tau\{U(d^*)\}=q^*$. Therefore, $(q^*,d^*)$ is an optimizer of the constrained problem.

Conversely, suppose that $(q^*,d^*)$ solves the constrained problem in (ii). Since $(q^*,d^*)$ is feasible for the constrained problem, we know that $\mathcal B_m(d^*)\le b_m,\ m=1,\dots,M$. Thus, $d^*\in\mathcal D_b$. 

Since $S(q^*,d^*)\ge 1-\tau$, the definition of $Q_\tau\{U(d^*)\}$ implies $q^*\le Q_\tau\{U(d^*)\}$. If the inequality here were strict, then $(Q_\tau\{U(d^*)\},d^*)$ would also be feasible and would achieve a strictly larger objective value, contradicting the optimality of $(q^*,d^*)$. Hence, $q^* = Q_\tau\{U(d^*)\}$. By Assumption~\ref{ass:quantile}, for any regime $d\in\mathcal D_b$, the pair $(Q_\tau\{U(d)\},d)$ is feasible. The optimality of $(q^*,d^*)$ implies
$$Q_\tau\{U(d)\}\le q^*=Q_\tau\{U(d^*)\}.$$
Therefore, $d^*$ is a quantile-optimal feasible regime and $d^*\in\argmax_{d\in\mathcal D_b}Q_\tau\{U(d)\}$.
The proof of Theorem~\ref{thm:Thm1} is complete.
\end{proof}

\subsection{Proof of Theorem~\ref{thm:Thm2}}
\label{pf:Thm2}
\begin{proof}
We prove the three parts in turn.

\medskip\noindent\textit{Proof of (i).} We first establish the uniform approximation bounds for the logistic and softmax smoothers, and then prove the existence and characterization of the restricted oracle.

\medskip\noindent\textit{Step 1: Uniform control of the logistic smoothing error.} By the identification formula in Theorem~\ref{thm:Thm1}(i), $S(q,d_f)$ admits its observed-data IPW representation. For any $(q,f)\in\Theta$, write $S_{h,\delta}(q,f)-S(q,d_f)=I_1(q,f)+I_2(q,f)$, where
\[
I_1(q,f)=\mathbb E\left[\{g_\delta(U-q)-I(U>q)\}\frac{\prod_{t=1}^2p_{t,A_t}(\H_t;f_t,h)}{\Pi}\right],
\]
\[
I_2(q,f)=\mathbb E\left[I(U>q)\frac{\prod_{t=1}^2p_{t,A_t}(\H_t;f_t,h)-\prod_{t=1}^2I\{A_t=d_{f_t}(\H_t)\}}{\Pi}\right].
\]
Since $0\le p_{t,a}(\cdot)\le1$ and $\Pi^{-1}\le\pi_{\min}^{-2}$ by uniform positivity, we have
\[
\sup_{(q,f)\in\Theta}|I_1(q,f)|\le\pi_{\min}^{-2}\sup_{q\in\mathcal Q}\mathbb E|g_\delta(U-q)-I(U>q)|.
\]
Let $r_\delta=\delta\log(1/\delta)$. 

On the event $\{|U-q|>r_\delta\}$, if $U>q$, then
\[
|g_\delta(U-q)-I(U>q)|=1-g_\delta(U-q)=\frac{\exp\{-(U-q)/\delta\}}{1+\exp\{-(U-q)/\delta\}}\le\exp(-r_\delta/\delta)=\delta.
\]
If $U<q$, then
\[
|g_\delta(U-q)-I(U>q)|=g_\delta(U-q)=\frac{1}{1+\exp\{(q-U)/\delta\}}\le\exp(-r_\delta/\delta)=\delta.
\]
Thus, $|g_\delta(U-q)-I(U>q)| \le \exp(-r_\delta/\delta) = \delta$. 

On the event $\{|U-q|\le r_\delta\}$, if $U>q$, then $|g_\delta(U-q)-I(U>q)| = 1-g_\delta(U-q) \le 1$. If $U<q$, then $|g_\delta(U-q)-I(U>q)| = g_\delta(U-q) \le 1$. Thus, the absolute difference is bounded by $1$.

We have
$$\begin{aligned}
\mathbb{E}|g_\delta(U-q)-I(U>q)|
&= \mathbb{E}\left[ |g_\delta(U-q)-I(U>q)| I(|U-q|\le r_\delta) \right] \\
&\quad + \mathbb{E}\left[ |g_\delta(U-q)-I(U>q)| I(|U-q|> r_\delta) \right] \\
&\le \mathbb{E}\left[ I(|U-q|\le r_\delta) \right] + \delta = \PP(|U-q|\le r_\delta) + \delta.
\end{aligned}$$
By Assumption~\ref{ass:boundary}(ii), for sufficiently small $\delta$,
\[
\sup_{q\in\mathcal Q}\mathbb E|g_\delta(U-q)-I(U>q)|\le C_Ur_\delta+\delta=C_U\delta\log(1/\delta)+\delta.
\]
Therefore,
\begin{equation}
\label{eq:logistic_uniform_bound}
\sup_{(q,f)\in\Theta}|I_1(q,f)|\le\pi_{\min}^{-2}\{C_U\delta\log(1/\delta)+\delta\}.
\end{equation}

\medskip\noindent\textit{Step 2: Uniform control of the softmax smoothing error.} For each stage $t=1,2$, let $\Delta_t(\H_t;f_t)$ denote the difference between the largest and second-largest treatment scores, as defined in Assumption~\ref{ass:boundary}(i). This assumption implies $\PP\{\Delta_t(\H_t;f_t)=0\}=0$ for every $f_t\in\mathcal F_t^\dagger$, so the maximal treatment is unique almost surely.

Let $r_h=h\log(1/h)$ and write $a_t^*=d_{f_t}(\H_t)$. On the event $\{\Delta_t(\H_t;f_t)>r_h\}$, the maximum treatment score exceeds every other treatment score by at least $r_h$. For any nonmaximal treatment $a\ne a_t^*$,
\[
\begin{aligned}
p_{t,a}(\H_t;f_t,h)
&=\frac{\exp\{\langle V_a,f_t(\H_t)\rangle/h\}}{\sum_{k\in\mathcal A}\exp\{\langle V_k,f_t(\H_t)\rangle/h\}}\\
&=\frac{\exp\{(\langle V_a,f_t(\H_t)\rangle-\langle V_{a_t^*},f_t(\H_t)\rangle)/h\}}{1+\sum_{k\ne a_t^*}\exp\{(\langle V_k,f_t(\H_t)\rangle-\langle V_{a_t^*},f_t(\H_t)\rangle)/h\}}\\
&\le\exp\{(\langle V_a,f_t(\H_t)\rangle-\langle V_{a_t^*},f_t(\H_t)\rangle)/h\}\\
&\le\exp(-r_h/h)=h.
\end{aligned}
\]
Thus, the total softmax probability assigned to nonmaximal treatments is at most $(K-1)h$. Since the difference for the maximal treatment equals this total probability, we obtain
\[
\max_{a\in\mathcal A}|p_{t,a}(\H_t;f_t,h)-I\{a=d_{f_t}(\H_t)\}|\le(K-1)h.
\]
On the complementary event $\{\Delta_t(\H_t;f_t)\le r_h\}$, both $p_{t,a}(\H_t;f_t,h)$ and $I\{a=d_{f_t}(\H_t)\}$ lie in $[0,1]$, so their absolute difference is bounded by $1$. Therefore,
$$\mathbb E\left[\max_{a\in\mathcal A}|p_{t,a}(\H_t;f_t,h)-I\{a=d_{f_t}(\H_t)\}|\right] \le(K-1)h+\PP\{\Delta_t(\H_t;f_t)\le r_h\}.$$
Taking the supremum over $f_t\in\mathcal F_t^\dagger$ and applying Assumption~\ref{ass:boundary}(i), we obtain, for sufficiently small $h$,
\begin{equation}
\label{eq:softmax_uniform_bound}
\begin{aligned}
E_t(h)&:=\sup_{f_t\in\mathcal F_t^\dagger}\mathbb E\left[\max_{a\in\mathcal A}|p_{t,a}(\H_t;f_t,h)-I\{a=d_{f_t}(\H_t)\}|\right]\\
&\le(K-1)h+C_t\{h\log(1/h)\}^{\alpha_t}.
\end{aligned}
\end{equation}
Using the inequality $|xy-x'y'|\le|x-x'|+|y-y'|$ for $x,y,x',y'\in[0,1]$, we have
\[
\left|\prod_{t=1}^2p_{t,A_t}(\H_t;f_t,h)-\prod_{t=1}^2I\{A_t=d_{f_t}(\H_t)\}\right| \le\sum_{t=1}^2|p_{t,A_t}(\H_t;f_t,h)-I\{A_t=d_{f_t}(\H_t)\}|.
\]
Consequently, by uniform positivity,
\[
\begin{aligned}
|I_2(q,f)|
&\le\pi_{\min}^{-2}\mathbb E\left[\left|\prod_{t=1}^2p_{t,A_t}(\H_t;f_t,h)-\prod_{t=1}^2I\{A_t=d_{f_t}(\H_t)\}\right|\right]\\
&\le\pi_{\min}^{-2}\sum_{t=1}^2\mathbb E\left[\max_{a\in\mathcal A}|p_{t,a}(\H_t;f_t,h)-I\{a=d_{f_t}(\H_t)\}|\right].
\end{aligned}
\]
It follows from \eqref{eq:softmax_uniform_bound} that
\begin{equation}
\label{eq:survival_softmax_bound}
\sup_{(q,f)\in\Theta}|I_2(q,f)|\le\pi_{\min}^{-2}\sum_{t=1}^2E_t(h)\le\pi_{\min}^{-2}\sum_{t=1}^2\left[(K-1)h+C_t\{h\log(1/h)\}^{\alpha_t}\right].
\end{equation}
Combining \eqref{eq:logistic_uniform_bound} and \eqref{eq:survival_softmax_bound} yields
\[
\sup_{(q,f)\in\Theta}|S_{h,\delta}(q,f)-S(q,d_f)| \le\pi_{\min}^{-2}\left[C_U\delta\log(1/\delta)+\delta+2(K-1)h+\sum_{t=1}^2C_t\{h\log(1/h)\}^{\alpha_t}\right].
\]

\medskip\noindent\textit{Step 3: Uniform control of the smoothed risk burdens.} For each $m\in\mathcal I$, the IPW identification formula gives
\[
\mathcal B_{m,h}(f)-\mathcal B_m(d_f) =\mathbb E\left[\psi_m\frac{\prod_{t=1}^2p_{t,A_t}(\H_t;f_t,h)-\prod_{t=1}^2I\{A_t=d_{f_t}(\H_t)\}}{\Pi}\right].
\]
By Assumption~\ref{ass:restricted_regular}, $|\psi_m|\le C_\psi$ almost surely, and uniform positivity gives $\Pi^{-1}\le\pi_{\min}^{-2}$. Applying the same product inequality as in Step 2, we obtain
\[
\begin{aligned}
|\mathcal B_{m,h}(f)-\mathcal B_m(d_f)|
&\le C_\psi\pi_{\min}^{-2}\mathbb E\left[\left|\prod_{t=1}^2p_{t,A_t}(\H_t;f_t,h)-\prod_{t=1}^2I\{A_t=d_{f_t}(\H_t)\}\right|\right]\\
&\le C_\psi\pi_{\min}^{-2}\sum_{t=1}^2\mathbb E\left[\max_{a\in\mathcal A}|p_{t,a}(\H_t;f_t,h)-I\{a=d_{f_t}(\H_t)\}|\right].
\end{aligned}
\]
Therefore,
\[
\max_{m\in\mathcal I}\sup_{f\in\mathcal F}|\mathcal B_{m,h}(f)-\mathcal B_m(d_f)|\le C_\psi\pi_{\min}^{-2}\sum_{t=1}^2E_t(h).
\]

Combining Steps 1--3 and using $\delta\le\delta\log(1/\delta)$ for $\delta<e^{-1}$, there exists a constant $C<\infty$, independent of $h$, $\delta$, and the penalty weights, such that as $h,\delta\downarrow0$,
\[
\sup_{(q,f)\in\Theta}|S_{h,\delta}(q,f)-S(q,d_f)|+\max_{m\in\mathcal I}\sup_{f\in\mathcal F}|\mathcal B_{m,h}(f)-\mathcal B_m(d_f)|\le Ca_{h,\delta}\to0,
\]
where $a_{h,\delta}=\delta\log(1/\delta)+h+\sum_{t=1}^2\{h\log(1/h)\}^{\alpha_t}$. This establishes the uniform convergence of the smoothed population quantities to their hard counterparts.

\medskip\noindent\textit{Step 4: Continuity and characterization of the restricted oracle.} For fixed $h,\delta>0$, the smoothed functions $(q,f)\mapsto S_{h,\delta}(q,f)$ and $f\mapsto\mathcal B_{m,h}(f)$ are continuous under the stated metrics. Indeed, convergence under $\rho_\Theta$ implies pointwise convergence of the corresponding scores and thresholds. Continuity of the logistic and softmax maps, together with dominated convergence and the bounds $\Pi^{-1}\le\pi_{\min}^{-2}$ and $|\psi_m|\Pi^{-1}\le C_\psi\pi_{\min}^{-2}$, establishes the claimed continuity. By the uniform approximation obtained in Steps 1--3, the hard population functions $(q,f)\mapsto S(q,d_f)$ and $f\mapsto\mathcal B_m(d_f)$ are uniform limits of continuous functions and are therefore continuous. Consequently, all hard constraint violations $v_j$ are continuous on $\Theta$.

By Assumption~\ref{ass:restricted_regular}, $\mathcal F_b$ is nonempty. Since $S(q_L,d_f)>c_\tau$ for every $f\in\mathcal F$, the pair $(q_L,f)$ belongs to $\Theta_b$ for every $f\in\mathcal F_b$. Hence $\Theta_b$ is nonempty. Moreover, $\Theta_b$ is closed in compact $\Theta$ because all hard constraint violations are continuous. Thus, $\Theta_b$ is compact, and the continuous map $(q,f)\mapsto q$ attains its maximum on $\Theta_b$. Therefore, $\mathcal A_\dagger$ is nonempty and compact.

For any fixed $f\in\mathcal F$, define $q_f=\max\{q\in\mathcal Q:S(q,d_f)\ge c_\tau\}$. By the continuity and monotonicity of $S(\cdot,d_f)$ and the endpoint conditions $S(q_L,d_f)>c_\tau$ and $S(q_U,d_f)<c_\tau$, we have $q_f\in(q_L,q_U)$ and $S(q_f,d_f)=c_\tau$. Since the survival function is nonincreasing, no $q>q_U$ satisfies the survival constraint, and thresholds below $q_L$ cannot increase its maximum. Thus,
\[
q_f=\sup\{q\in\mathbb R:S(q,d_f)\ge c_\tau\}=Q_\tau\{U(d_f)\}.
\]
Maximizing over $f\in\mathcal F_b$ yields
\[
q_\dagger=\max_{(q,f)\in\Theta_b}q=\max_{f\in\mathcal F_b}
\max_{\substack{q\in\mathcal Q\\S(q,d_f)\ge c_\tau}}q =\max_{f\in\mathcal F_b}Q_\tau\{U(d_f)\}\in(q_L,q_U).
\]
In particular, every $(q_\dagger,f)\in\mathcal A_\dagger$ satisfies $Q_\tau\{U(d_f)\}=q_\dagger$.

\medskip\noindent\textit{Proof of (ii).} We first establish the existence of population surrogate maximizers, then derive bounds on their constraint violations and objective values, and finally prove convergence to the restricted oracle set.

\medskip\noindent\textit{Step 1: Existence of population surrogate maximizers.} For each $n$, the smoothed population functions $S_{h_n,\delta_n}(q,f)$ and $\mathcal B_{m,h_n}(f)$ are continuous under the stated metrics, as established in the proof of (i). Since the penalty weights $\mu_{j,n}$ are finite and positive, the population criterion
\[
\mathcal M_{n,h_n,\delta_n}(q,f)=q-\sum_{j\in\mathcal J}\mu_{j,n}v_{j,h_n,\delta_n}(q,f)^2
\]
is continuous on $\Theta$. By Assumption~\ref{ass:restricted_regular}, $\Theta$ is compact. Thus, $\mathcal M_{n,h_n,\delta_n}$ attains its maximum on $\Theta$ for every $n$. Let $\theta_n^\circ=(q_n^\circ,f_n^\circ)$ denote any such maximizer.

\medskip\noindent\textit{Step 2: Oracle comparison and asymptotic feasibility.} For each $j\in\mathcal J$, define the uniform approximation error $e_{j,n}=\sup_{\theta\in\Theta}|v_{j,h_n,\delta_n}(\theta)-v_j(\theta)|$. By the uniform approximation bound in Eq.~\eqref{eq:uniform_primitive_approximation} and the fact that the positive-part map is $1$-Lipschitz, there exists a constant $C<\infty$, independent of $n$, such that
\[
\max_{j\in\mathcal J}e_{j,n}\le Ca_{h_n,\delta_n}.
\]
Define $D_n=\sum_{j\in\mathcal J}\mu_{j,n}e_{j,n}^2$. Since $\mu_{j,n}\le\overline\mu_n$ and $\overline\mu_n a_{h_n,\delta_n}^2\to0$, we have
\[
D_n\le C^2|\mathcal J|\overline\mu_n a_{h_n,\delta_n}^2\to0.
\]

By part (i), the restricted oracle set $\mathcal A_\dagger$ is nonempty. Choose any $\theta^\dagger=(q_\dagger,f^\dagger)\in\mathcal A_\dagger$. Since $\theta^\dagger\in\Theta_b$, we have $v_j(\theta^\dagger)=0$ for every $j\in\mathcal J$. Consequently,
\[
v_{j,h_n,\delta_n}(\theta^\dagger)\le |v_{j,h_n,\delta_n}(\theta^\dagger)-v_j(\theta^\dagger)|\le e_{j,n}.
\]
It follows that
\[
\mathcal M_{n,h_n,\delta_n}(\theta^\dagger) =q_\dagger-\sum_{j\in\mathcal J}\mu_{j,n}v_{j,h_n,\delta_n}(\theta^\dagger)^2 \ge q_\dagger-D_n.
\]
By the optimality of $\theta_n^\circ$, we obtain
\[
q_n^\circ-\sum_{j\in\mathcal J}\mu_{j,n}v_{j,h_n,\delta_n}(\theta_n^\circ)^2 =\mathcal M_{n,h_n,\delta_n}(\theta_n^\circ) \ge\mathcal M_{n,h_n,\delta_n}(\theta^\dagger) \ge q_\dagger-D_n.
\]
Since all penalty terms are nonnegative, this implies $q_n^\circ\ge q_\dagger-D_n$. Hence, $(q_\dagger-q_n^\circ)_+\le D_n\to0$.
Rearranging the preceding oracle comparison inequality and using $q_n^\circ\le q_U$ yields
\begin{equation*}
\label{eq:population_penalty_bound}
\sum_{j\in\mathcal J}\mu_{j,n}v_{j,h_n,\delta_n}(\theta_n^\circ)^2\le q_n^\circ-q_\dagger+D_n\le q_U-q_\dagger+D_n.
\end{equation*}
Since $\mu_{j,n}\ge\underline\mu_n>0$ for every $j\in\mathcal J$, we further obtain
\[
\max_{j\in\mathcal J}v_{j,h_n,\delta_n}(\theta_n^\circ)\le\left(\frac{q_U-q_\dagger+D_n}{\underline\mu_n}\right)^{1/2}.
\]
By the definition of $e_{j,n}$, $v_j(\theta_n^\circ)\le v_{j,h_n,\delta_n}(\theta_n^\circ)+e_{j,n}$. Therefore,
\begin{equation}
\label{eq:population_feasibility_bound}
\begin{aligned}
\max_{j\in\mathcal J}v_j(\theta_n^\circ)
&\le\max_{j\in\mathcal J}v_{j,h_n,\delta_n}(\theta_n^\circ)+\max_{j\in\mathcal J}e_{j,n}\\
&\le\left(\frac{q_U-q_\dagger+D_n}{\underline\mu_n}\right)^{1/2}+Ca_{h_n,\delta_n}\to0,
\end{aligned}
\end{equation}
where the convergence follows from $\underline\mu_n\to\infty$, $D_n\to0$, and $a_{h_n,\delta_n}\to0$. Thus, the hard survival and risk constraint violations of the population surrogate maximizers vanish asymptotically.

\medskip\noindent\textit{Step 3: Convergence to the restricted oracle set.} We first show that every accumulation point of $\{\theta_n^\circ\}$ belongs to $\mathcal A_\dagger$. Since $\Theta$ is compact, every subsequence of $\{\theta_n^\circ\}$ admits a convergent further subsequence. Let $\theta_{n_k}^\circ\to\bar\theta=(\bar q,\bar f)\in\Theta$ be any such convergent subsequence.

By part (i), each hard constraint violation $v_j$ is continuous on $\Theta$. Together with Eq.~\eqref{eq:population_feasibility_bound}, this gives
\[
v_j(\bar\theta)=v_j(\lim_{k\to\infty}\theta_{n_k}^\circ)=\lim_{k\to\infty}v_j(\theta_{n_k}^\circ)=0,\ j\in\mathcal J.
\]
Hence, $\bar\theta\in\Theta_b$. Moreover, by step 2, we have $q_n^\circ\ge q_\dagger-D_n$ and $D_n\to0$, so
\[
\bar q=\lim_{k\to\infty}q_{n_k}^\circ\ge q_\dagger.
\]
On the other hand, since $\bar\theta\in\Theta_b$ and $q_\dagger=\max_{(q,f)\in\Theta_b}q$, we have $\bar q\le q_\dagger$. Therefore,
\[
\bar q=q_\dagger,\ \bar\theta\in\mathcal A_\dagger.
\]
Thus, every accumulation point of the population surrogate maximizers belongs to the restricted oracle set.

We next establish convergence to the oracle set. By contradiction, suppose that $\operatorname{dist}(\theta_n^\circ,\mathcal A_\dagger)$ does not converge to zero. Then there exist $\varepsilon>0$ and a subsequence $\{\theta_{n_k}^\circ\}$ such that $\operatorname{dist}(\theta_{n_k}^\circ,\mathcal A_\dagger)\ge\varepsilon$ for all $k$. By compactness of $\Theta$, there exists a further subsequence $\{\theta_{n_{k_\ell}}^\circ\}$ converging to some $\bar\theta\in\Theta$. As shown above, every accumulation point belongs to $\mathcal A_\dagger$, so $\bar\theta\in\mathcal A_\dagger$. Consequently,
\[
\operatorname{dist}(\theta_{n_{k_\ell}}^\circ,\mathcal A_\dagger)
=\inf_{\theta^\dagger\in\mathcal A_\dagger}\rho_\Theta(\theta_{n_{k_\ell}}^\circ,\theta^\dagger)
\le\rho_\Theta(\theta_{n_{k_\ell}}^\circ,\bar\theta)\to0,
\]
contradicting $\operatorname{dist}(\theta_{n_{k_\ell}}^\circ,\mathcal A_\dagger)\ge\varepsilon$. Hence, $\operatorname{dist}(\theta_n^\circ,\mathcal A_\dagger)\to0$. Since every element of $\mathcal A_\dagger$ has threshold $q_\dagger$, we have $q_n^\circ\to q_\dagger$. Furthermore, by the characterization of $\mathcal A_\dagger$ in part (i), every accumulation point $(\bar q,\bar f)\in\mathcal A_\dagger$ satisfies
\[
\bar f\in\mathcal F_b,\ \bar q=Q_\tau\{U(d_{\bar f})\}=q_\dagger.
\]

\medskip\noindent\textit{Proof of (iii).} By definition, every $f\in\mathcal F_b$ induces a regime $d_f\in\mathcal D_{b,\mathcal I}$. Thus, the restricted oracle value cannot exceed the unrestricted oracle value. By part (i),
\[
q_\dagger=\max_{f\in\mathcal F_b}Q_\tau\{U(d_f)\}\le\sup_{d\in\mathcal D_{b,\mathcal I}}Q_\tau\{U(d)\}=Q_\tau^*.
\]
Suppose that there exists $f^*\in\mathcal F_b$ satisfying $Q_\tau\{U(d_{f^*})\}=Q_\tau^*$. Since $f^*$ belongs to the restricted feasible class, we also have
\[
q_\dagger=\max_{f\in\mathcal F_b}Q_\tau\{U(d_f)\}\ge Q_\tau\{U(d_{f^*})\}=Q_\tau^*.
\]
Combining the two inequalities yields $q_\dagger=Q_\tau^*$. By part (ii), every accumulation point $(\bar q,\bar f)$ of the population surrogate maximizers satisfies $\bar f\in\mathcal F_b$ and $\bar q=Q_\tau\{U(d_{\bar f})\}=q_\dagger$. Since $\mathcal F_b$ is the restricted feasible class, $d_{\bar f}\in\mathcal D_{b,\mathcal I}$. Hence, $Q_\tau\{U(d_{\bar f})\}=q_\dagger=Q_\tau^*$, which implies
\[
d_{\bar f}\in\argmax_{d\in\mathcal D_{b,\mathcal I}}Q_\tau\{U(d)\},\ \bar q=Q_\tau^*.
\]
Thus, every accumulation point induces an unrestricted quantile-optimal feasible regime. The proof of Theorem~\ref{thm:Thm2} is complete.
\end{proof}

\subsection{Proof of Lemma~\ref{lem:empirical_uniform}}
\label{pf:Lemma3}
\begin{proof}
Let $\mathcal T=(\X_1,A_1,Y_1,\Z_1,\X_2,A_2,Y_2,\Z_2)$ denote an observed trajectory. For any measurable function $\varphi$, write $\mathbb E\varphi=\mathbb E\{\varphi(\mathcal T)\}$ and $\mathbb E_n\varphi=n^{-1}\sum_{i=1}^n\varphi(\mathcal T_i)$. For each $(q,f)\in\Theta$ and $m\in\mathcal I$, define
\[
\varphi^S_{q,f,n}(\mathcal T)=\frac{g_{\delta_n}(U-q)\prod_{t=1}^2p_{t,A_t}(\H_t;f_t,h_n)}{\Pi},\
\varphi^{\mathcal B,m}_{f,n}(\mathcal T)=\frac{\psi_m\prod_{t=1}^2p_{t,A_t}(\H_t;f_t,h_n)}{\Pi}.
\]
Define the corresponding function classes by
$$\mathcal H_n^S=\{\varphi^S_{q,f,n}:(q,f)\in\Theta\},\
\mathcal H_n^{\mathcal B,m}=\{\varphi^{\mathcal B,m}_{f,n}:f\in\mathcal F\}.$$
Let $\mathcal H_n=\mathcal H_n^S\cup(\bigcup_{m\in\mathcal I}\mathcal H_n^{\mathcal B,m})$. By definition, $E_n^0=\sup_{\varphi\in\mathcal H_n}|(\mathbb E_n-\mathbb E)\varphi|$. Throughout the proof, $C$ denotes a generic finite constant independent of $n$, whose value may change from line to line.

\medskip\noindent\textit{Step 1: Uniform envelope.} Since $0\le g_{\delta_n}(\cdot)\le1$, $0\le p_{t,a}(\cdot)\le1$, $|\psi_m|\le C_\psi$ by Assumption~\ref{ass:restricted_regular}, and $\Pi^{-1}\le\pi_{\min}^{-2}$ by uniform positivity, we have
\[
|\varphi^S_{q,f,n}(\mathcal T)|\le\pi_{\min}^{-2},\
|\varphi^{\mathcal B,m}_{f,n}(\mathcal T)|\le C_\psi\pi_{\min}^{-2}.
\]
Thus, every function in $\mathcal H_n$ is bounded in absolute value by $B=\pi_{\min}^{-2}\max(1,C_\psi)$, uniformly in $n$. Therefore, both classes admit constant envelopes uniformly in $n$.

\medskip\noindent\textit{Step 2: Lipschitz continuity of the smoothing maps.} Let $\sigma(x)=(1+e^{-x})^{-1}$. Since $\sigma'(x)=\sigma(x)\{1-\sigma(x)\}$, we have $\sup_{x\in\mathbb R}|\sigma'(x)|=\tfrac{1}{4}$, and hence $\sup_{x\in\mathbb{R}}|g'_{\delta}(x)|=\sup_{x\in\mathbb{R}}|\tfrac{1}{\delta} \sigma'(\tfrac{x}{\delta})|=\tfrac{1}{4\delta}$. By the mean value theorem, $g_{\delta_n}(\cdot)$ is $\tfrac{1}{4\delta_n}$-Lipschitz on $\mathbb R$ and
\[
|g_{\delta_n}(U-q)-g_{\delta_n}(U-q')|\le\frac{|q-q'|}{4\delta_n}.
\]
For the softmax map, write $s_a(u):=p_{t,a}(u)=e^{u_a}/\sum_{k=1}^Ke^{u_k}$, $u\in\mathbb R^K$. Its partial derivatives satisfy
\[
\frac{\partial s_a(u)}{\partial u_k}=s_a(u)\{I(a=k)-s_k(u)\},\ k=1,\ldots,K.
\]
Hence, the $\ell^1$ norm of the corresponding row of the Jacobian is
\[
\sum_{k=1}^K\left|\frac{\partial s_a(u)}{\partial u_k}\right|
=s_a(u)\{1-s_a(u)\}+\sum_{k\ne a}s_a(u)s_k(u) =2s_a(u)\{1-s_a(u)\}\le\frac12.
\]
Therefore, $s_a$ is $\tfrac{1}{2}$-Lipschitz with respect to the $\ell^\infty$ norm: $|s_a(u)-s_a(u')|\le\frac12\|u-u'\|_\infty$. Let $C_V=\max_{a\in\mathcal A}\|V_a\|$. By the Cauchy--Schwarz inequality, for every $k\in\mathcal A$,
\[
|\langle V_k,f_t(\H_t)\rangle-\langle V_k,f_t'(\H_t)\rangle|
\le\|V_k\|\|f_t(\H_t)-f_t'(\H_t)\| \le C_V\rho_{\mathcal F}(f,f').
\]
Applying the softmax Lipschitz bound to treatment scores divided by $h_n$ yields
\[
|p_{t,A_t}(\H_t;f_t,h_n)-p_{t,A_t}(\H_t;f_t',h_n)|
\le\frac{C_V}{2h_n}\rho_{\mathcal F}(f,f').
\]
Using $|ab-cd|\le|a-c|+|b-d|$ for $a,b,c,d\in[0,1]$, together with Step 1, we obtain
\begin{equation}
\label{eq:empirical_summand_lipschitz}
\begin{aligned}
\|\varphi^S_{q,f,n}-\varphi^S_{q',f',n}\|_\infty
&\le C\left\{\frac{|q-q'|}{\delta_n}+\frac{\rho_{\mathcal F}(f,f')}{h_n}\right\},\\
\|\varphi^{\mathcal B,m}_{f,n}-\varphi^{\mathcal B,m}_{f',n}\|_\infty
&\le\frac{C}{h_n}\rho_{\mathcal F}(f,f').
\end{aligned}
\end{equation}
Here $\|\cdot\|_\infty$ denotes the essential supremum norm under the observed-data law.

\medskip\noindent\textit{Step 3: Construction of finite covering nets.} We now use Eq.~\eqref{eq:empirical_summand_lipschitz} to translate covering numbers of $\mathcal Q$ and $\mathcal F$ into a finite covering of the empirical summand classes.

Let $u_n$ be the auxiliary sequence in Assumption~\ref{ass:primitive_entropy}(i). Choose a fixed sufficiently small $c>0$. Construct a finite $cu_n\delta_n$-net of $\mathcal Q$ and a finite $cu_nh_n$-net of $\mathcal F$ under $\rho_{\mathcal F}$. Denote their cardinalities by $N_{q,n}$ and $N_{f,n}$, respectively.

For every $(q,f)\in\Theta$, there exist net points $(q',f')$ such that $|q-q'|\le cu_n\delta_n,\ \rho_{\mathcal F}(f,f')\le cu_nh_n$. By Eq.~\eqref{eq:empirical_summand_lipschitz}, $\|\varphi^S_{q,f,n}-\varphi^S_{q',f',n}\|_\infty\le 2Ccu_n$. Similarly, for every $f\in\mathcal F$ and $m\in\mathcal I$, there exists a net point $f'$ satisfying $\|\varphi^{\mathcal B,m}_{f,n}-\varphi^{\mathcal B,m}_{f',n}\|_\infty\le Ccu_n$. Choosing $c$ sufficiently small ensures that both approximation errors are at most $u_n$.

Let $\varphi_{1,n},\ldots,\varphi_{N_n,n}$ collect the resulting finite nets for the survival and risk summand classes. Their total cardinality satisfies $N_n\le N_{q,n}N_{f,n}+|\mathcal I|N_{f,n}$. Since $\mathcal Q=[q_L,q_U]$ is a fixed finite interval, a covering net with radius $cu_n\delta_n$ can be chosen with cardinality
\[
N_{q,n}\le 1+\frac{q_U-q_L}{cu_n\delta_n}=1+\frac{C_{\mathcal Q}}{u_n\delta_n},
\]
where $C_{\mathcal Q}=(q_U-q_L)/c<\infty$ is independent of $n$. Choosing a minimal covering net of $\mathcal F$ gives $N_{f,n}=\mathcal N(\mathcal F,\rho_{\mathcal F},cu_nh_n)$. Since $|\mathcal I|\le M$ is fixed, for all sufficiently large $n$,
\[
\begin{aligned}
\log(2N_n)
&\le\log\{2(N_{q,n}+M)N_{f,n}\}\\
&\le\log\{2(1+M+C_{\mathcal Q})\}+\log\{1/(u_n\delta_n)\}+\log N_{f,n}\\
&\le C\left[1+\log\{1/(u_n\delta_n)\}+\log\mathcal N(\mathcal F,\rho_{\mathcal F},cu_nh_n)\right]=C\mathcal E_n.
\end{aligned}
\]

\medskip\noindent\textit{Step 4: Uniform empirical convergence using concentration.} For any $\varphi\in\mathcal H_n$, choose a corresponding net element $\varphi_{k,n}$ satisfying $\|\varphi-\varphi_{k,n}\|_\infty\le u_n$. Then
\[
\begin{aligned}
|(\mathbb E_n-\mathbb E)\varphi|
&\le |(\mathbb E_n-\mathbb E)\varphi_{k,n}| + |(\mathbb E_n-\mathbb E)(\varphi-\varphi_{k,n})| \\
&\le|(\mathbb E_n-\mathbb E)\varphi_{k,n}|+\mathbb E_n|\varphi-\varphi_{k,n}|+\mathbb E|\varphi-\varphi_{k,n}|\\
&\le|(\mathbb E_n-\mathbb E)\varphi_{k,n}|+2u_n.
\end{aligned}
\]
Taking the supremum over $\varphi\in\mathcal H_n$ yields
\[
E_n^0\le2u_n+\max_{1\le k\le N_n}|(\mathbb E_n-\mathbb E)\varphi_{k,n}|.
\]
For each fixed net element $\varphi_{k,n}$, the random variables $\varphi_{k,n}(\mathcal T_1),\ldots,\varphi_{k,n}(\mathcal T_n)$ are i.i.d. and bounded in $[-B,B]$ by Step 1. Hoeffding's inequality therefore implies
\[
\PP\left\{|(\mathbb E_n-\mathbb E)\varphi_{k,n}|>t\right\}\le2\exp\left(-\frac{nt^2}{2B^2}\right).
\]
Applying the union bound over the finite net gives
\[
\PP\left\{\max_{1\le k\le N_n}|(\mathbb E_n-\mathbb E)\varphi_{k,n}|>t\right\}\le2N_n\exp\left(-\frac{nt^2}{2B^2}\right).
\]
For any $0<\beta<1$, choosing $t=B\sqrt{2\log(2N_n/\beta)/n}$ yields
\[
\PP\left\{E_n^0>2u_n+B\sqrt{\frac{2\log(2N_n/\beta)}{n}}\right\}\le\beta.
\]
Since $\log(2N_n)\le C\mathcal E_n$, we conclude that
\[
E_n^0=O_p\left(u_n+\sqrt{\frac{\mathcal E_n}{n}}\right)=O_p(r_{n,\mathrm{emp}}).
\]
Under the compactness assumptions, the summand classes are pointwise measurable through countable dense parameter subsets, so the displayed suprema are measurable.

\medskip\noindent\textit{Step 5: Estimated propensity scores and smoothing approximation.} We now bound the additional errors introduced by replacing true propensity scores with their estimates and by approximating the hard population constraints.

By definition,
\[
\begin{aligned}
\widehat S_{h_n,\delta_n}(q,f)-\widehat S^\pi_{h_n,\delta_n}(q,f)
&=\frac1n\sum_{i=1}^ng_{\delta_n}(U_i-q)\prod_{t=1}^2p_{t,A_{ti}}(\H_{ti};f_t,h_n)(\widetilde\Pi_{n,i}^{-1}-\Pi_i^{-1}).
\end{aligned}
\]
Since the logistic and softmax factors lie in $[0,1]$, we obtain
\[
\sup_{(q,f)\in\Theta}|\widehat S_{h_n,\delta_n}(q,f)-\widehat S^\pi_{h_n,\delta_n}(q,f)|\le\frac1n\sum_{i=1}^n|\widetilde\Pi_{n,i}^{-1}-\Pi_i^{-1}|=\Xi_{\pi,n}.
\]
Similarly, since $|\psi_{m,i}|\le C_\psi$ almost surely,
\[
\max_{m\in\mathcal I}\sup_{f\in\mathcal F}|\widehat{\mathcal B}_{m,h_n}(f)-\widehat{\mathcal B}^\pi_{m,h_n}(f)|\le C_\psi\Xi_{\pi,n}.
\]
These inequalities hold pathwise and do not require the estimated propensity scores to be independent of the observations used in the empirical criterion.

Combining the bounds above with Step 4 and the uniform population smoothing bound in Eq.~\eqref{eq:uniform_primitive_approximation}, we obtain
\[
\begin{aligned}
\sup_{(q,f)\in\Theta}|\widehat S_{h_n,\delta_n}(q,f)-S(q,d_f)|
&\le E_n^0+\Xi_{\pi,n}+Ca_{h_n,\delta_n},\\
\max_{m\in\mathcal I}\sup_{f\in\mathcal F}|\widehat{\mathcal B}_{m,h_n}(f)-\mathcal B_m(d_f)|
&\le E_n^0+C_\psi\Xi_{\pi,n}+Ca_{h_n,\delta_n}.
\end{aligned}
\]
Finally, the positive-part map is $1$-Lipschitz, so the difference between each empirical and hard population constraint violation is bounded by the corresponding primitive approximation error. For risk constraints, replacing $b_m$ with $\widehat b_{m,n}$ contributes at most $\Delta_{b,n}$. Therefore, for a constant $C<\infty$ independent of $n$,
\[
e_n=\max_{j\in\mathcal J}\sup_{\theta\in\Theta}|\widehat v_{j,n}(\theta)-v_j(\theta)|
\le E_n^0+C\Xi_{\pi,n}+Ca_{h_n,\delta_n}+\Delta_{b,n}.
\]
The proof of Lemma~\ref{lem:empirical_uniform} is complete.
\end{proof}

\subsection{Proof of Lemma~\ref{lem:oracle_continuity}}
\label{pf:Lemma4}
\begin{proof}
We prove Lemma~\ref{lem:oracle_continuity} in five steps.

\medskip\noindent\textit{Step 1: Construction of an approximating sequence in $\mathcal A_\dagger$.} By Theorem~\ref{thm:Thm2}(i), the restricted oracle set $\mathcal A_\dagger$ is nonempty and compact. Since $\operatorname{dist}(\theta_n,\mathcal A_\dagger)\to0$, for each $n$ we can choose $\theta_n^*=(q_\dagger,f_n^*)\in\mathcal A_\dagger$ attaining the minimum distance, so that $\rho_\Theta(\theta_n,\theta_n^*)=\operatorname{dist}(\theta_n,\mathcal A_\dagger)\to0$. By the definition of $\rho_\Theta\{(q,f),(q',f')\}=|q-q'|+\rho_{\mathcal F}(f,f')$, we have $|q_n-q_\dagger|\to0$ and $\rho_{\mathcal F}(f_n,f_n^*)\to0$. Moreover, Theorem~\ref{thm:Thm2}(i) establishes that every $(q_\dagger,f^*)\in\mathcal A_\dagger$ corresponds to a restricted quantile-optimal feasible regime. Therefore,
\[
Q_\tau\{U(d_{f_n^*})\}=q_\dagger,\ \mathcal B_m(d_{f_n^*})\le b_m,\ m\in\mathcal I.
\]
These properties will be used in the remaining steps.

\medskip\noindent\textit{Step 2: Decision-boundary control.} Fix $t\in\{1,2\}$. On the event $\{d_{f_{n,t}}(\H_t)\ne d_{f_{n,t}^*}(\H_t)\}$, let $a=d_{f_{n,t}}(\H_t)=\argmax_{k\in\mathcal A}\langle V_k,f_{n,t}(\H_t)\rangle$ and $a^*=d_{f_{n,t}^*}(\H_t)=\argmax_{k\in\mathcal A}\langle V_k,f_{n,t}^*(\H_t)\rangle$, so that $a\ne a^*$. Define the treatment-score margin under $f_{n,t}^*$ at $\H_t$ by
\[
\Delta_t(\H_t;f_{n,t}^*)=\langle V_{a^*},f_{n,t}^*(\H_t)\rangle-\max_{k\ne a^*}\langle V_k,f_{n,t}^*(\H_t)\rangle.
\]
By optimality of $a^*$ under $f_{n,t}^*$ and of $a$ under $f_{n,t}$,
\[
\begin{aligned}
\Delta_t(\H_t;f_{n,t}^*)
&\le\langle V_{a^*},f_{n,t}^*(\H_t)\rangle-\langle V_a,f_{n,t}^*(\H_t)\rangle\\
&=\underbrace{\langle V_{a^*},f_{n,t}^*(\H_t)-f_{n,t}(\H_t)\rangle}_{(\mathrm I)} +\underbrace{\langle V_{a^*}-V_a,f_{n,t}(\H_t)\rangle}_{(\mathrm{II})\le0} +\underbrace{\langle V_a,f_{n,t}(\H_t)-f_{n,t}^*(\H_t)\rangle}_{(\mathrm{III})}\\
&\le|(\mathrm I)|+|(\mathrm{III})|.
\end{aligned}
\]
By the Cauchy--Schwarz inequality,
\[
|(\mathrm I)| \le\|V_{a^*}\|\|f_{n,t}^*(\H_t)-f_{n,t}(\H_t)\| \le C_V\sup_{h_t\in\mathcal H_t}\|f_{n,t}^*(h_t)-f_{n,t}(h_t)\| \le C_V\rho_{\mathcal F}(f_n,f_n^*),
\]
and similarly,
\[
|(\mathrm{III})| \le\|V_a\|\|f_{n,t}(\H_t)-f_{n,t}^*(\H_t)\| \le C_V\rho_{\mathcal F}(f_n,f_n^*),
\]
where $C_V=\max_{a\in\mathcal A}\|V_a\|<\infty$ depends only on the fixed simplex vertices. Consequently,
\[
\Delta_t(\H_t;f_{n,t}^*)\le2C_V\rho_{\mathcal F}(f_n,f_n^*).
\]
Hence,
\[
\{d_{f_{n,t}}(\H_t)\ne d_{f_{n,t}^*}(\H_t)\}\subseteq\{\Delta_t(\H_t;f_{n,t}^*)\le2C_V\rho_{\mathcal F}(f_n,f_n^*)\}.
\]
Let $r_n=2C_V\rho_{\mathcal F}(f_n,f_n^*)$. By Step 1, $r_n\to0$. If $r_n=0$, the two score functions coincide and induce the same treatment rule. For all sufficiently large $n$ with $0<r_n\le r_t$, Assumption~\ref{ass:boundary}(i) gives
\begin{equation}
\label{eq:score_mismatch}
\begin{aligned}
\PP\{d_{f_{n,t}}(\H_t)\ne d_{f_{n,t}^*}(\H_t)\}
&\le \PP\{\Delta_t(\H_t;f_{n,t}^*)\le r_n\}\\
&\le \sup_{f_t\in\mathcal F_t^\dagger}\PP\{\Delta_t(\H_t;f_t)\le r_n\}
\le C_t r_n^{\alpha_t}\to0.
\end{aligned}
\end{equation}
The same bound holds when $r_n=0$, since both sides are then zero. Thus, the probability that the two score functions induce different treatment decisions converges to zero at each stage.

\medskip\noindent\textit{Step 3: Uniform continuity of population survival functions and risk burdens.} By Theorem~\ref{thm:Thm1}(i), the counterfactual survival function admits the IPW representation. For any observed trajectory, the difference between the treatment indicators under the two regimes satisfies
\[
\left|\prod_{t=1}^2I\{A_t=d_{f_{n,t}}(\H_t)\}-\prod_{t=1}^2I\{A_t=d_{f_{n,t}^*}(\H_t)\}\right| \le\sum_{t=1}^2I\{d_{f_{n,t}}(\H_t)\ne d_{f_{n,t}^*}(\H_t)\}.
\]
Since $I(U>q)\le1$ and $\Pi^{-1}\le\pi_{\min}^{-2}$ by uniform positivity, for every $q\in\mathbb R$,
\[
\begin{aligned}
|S(q,d_{f_n})-S(q,d_{f_n^*})|
&\le\mathbb E\left[\frac{I(U>q)}{\Pi}\left|\prod_{t=1}^2I\{A_t=d_{f_{n,t}}(\H_t)\}-\prod_{t=1}^2I\{A_t=d_{f_{n,t}^*}(\H_t)\}\right|\right]\\
&\le\pi_{\min}^{-2}\sum_{t=1}^2\PP\{d_{f_{n,t}}(\H_t)\ne d_{f_{n,t}^*}(\H_t)\}.
\end{aligned}
\]
The right-hand side does not depend on $q$. Thus, by Eq.~\eqref{eq:score_mismatch},
\begin{equation}
\label{eq:policy_continuity_S}
\sup_{q\in\mathbb R}|S(q,d_{f_n})-S(q,d_{f_n^*})|\le\pi_{\min}^{-2}\sum_{t=1}^2\PP\{d_{f_{n,t}}(\H_t)\ne d_{f_{n,t}^*}(\H_t)\}\to0.
\end{equation}
Similarly, using the IPW representation of the population risk burdens and the boundedness $|\psi_m|\le C_\psi$ under Assumption~\ref{ass:restricted_regular}, for each $m\in\mathcal I$,
\[
|\mathcal B_m(d_{f_n})-\mathcal B_m(d_{f_n^*})| \le C_\psi\pi_{\min}^{-2}\sum_{t=1}^2\PP\{d_{f_{n,t}}(\H_t)\ne d_{f_{n,t}^*}(\H_t)\}.
\]
By Eq.~\eqref{eq:score_mismatch}, we have
\begin{equation}
\label{eq:policy_continuity}
\max_{m\in\mathcal I}|\mathcal B_m(d_{f_n})-\mathcal B_m(d_{f_n^*})|\le C_\psi\pi_{\min}^{-2}\sum_{t=1}^2\PP\{d_{f_{n,t}}(\H_t)\ne d_{f_{n,t}^*}(\H_t)\}\to0.
\end{equation}

\medskip\noindent\textit{Step 4: Asymptotic risk feasibility.} Fix $m\in\mathcal I$. By $[x+y]_+\le[x]_++[y]_+$ and $[x]_+\le|x|$,
\[
\begin{aligned}
[\mathcal B_m(d_{f_n})-b_m]_+
&=[\mathcal B_m(d_{f_n})-\mathcal B_m(d_{f_n^*})+\mathcal B_m(d_{f_n^*})-b_m]_+\\
&\le[\mathcal B_m(d_{f_n})-\mathcal B_m(d_{f_n^*})]_++[\mathcal B_m(d_{f_n^*})-b_m]_+\\
&\le|\mathcal B_m(d_{f_n})-\mathcal B_m(d_{f_n^*})|+[\mathcal B_m(d_{f_n^*})-b_m]_+.
\end{aligned}
\]
Since $\mathcal B_m(d_{f_n^*})\le b_m$ by Step 1, the second term equals zero. Moreover, Eq.~\eqref{eq:policy_continuity} implies $|\mathcal B_m(d_{f_n})-\mathcal B_m(d_{f_n^*})|\to0$. Thus, $[\mathcal B_m(d_{f_n})-b_m]_+\to0$. Taking the maximum over the finite index set $\mathcal I$ yields
\[
\max_{m\in\mathcal I}[\mathcal B_m(d_{f_n})-b_m]_+\to0.
\]
When $\mathcal I=\varnothing$, the conclusion holds by the stated empty-set convention. This establishes asymptotic risk feasibility without requiring Assumption~\ref{ass:oracle_crossing}.

\medskip\noindent\textit{Step 5: Quantile convergence.} Now suppose Assumption~\ref{ass:oracle_crossing} holds. Fix any $0<\epsilon\le r_0$. By this assumption, there exists $\eta(\epsilon)>0$ such that, for every $(q_\dagger,f^*)\in\mathcal A_\dagger$,
\[
S(q_\dagger-\epsilon,d_{f^*})\ge c_\tau+\eta(\epsilon),\ S(q_\dagger+\epsilon,d_{f^*})\le c_\tau-\eta(\epsilon).
\]
Specializing to $\theta_n^*=(q_\dagger,f_n^*)\in\mathcal A_\dagger$, we obtain
\[
S(q_\dagger-\epsilon,d_{f_n^*})\ge c_\tau+\eta(\epsilon),\ S(q_\dagger+\epsilon,d_{f_n^*})\le c_\tau-\eta(\epsilon).
\]
By Step 3, $\sup_{q\in\mathbb R}|S(q,d_{f_n})-S(q,d_{f_n^*})|\to0$. Thus, there exists $N_\epsilon\in\mathbb N$ such that, for all $n\ge N_\epsilon$, $\sup_{q\in\mathbb R}|S(q,d_{f_n})-S(q,d_{f_n^*})|\le\eta(\epsilon)/2$. Consequently, for all $n\ge N_\epsilon$,
\[
S(q_\dagger-\epsilon,d_{f_n}) \ge S(q_\dagger-\epsilon,d_{f_n^*})-\eta(\epsilon)/2 \ge c_\tau+\eta(\epsilon)/2>c_\tau,
\]
\[
S(q_\dagger+\epsilon,d_{f_n}) \le S(q_\dagger+\epsilon,d_{f_n^*})+\eta(\epsilon)/2 \le c_\tau-\eta(\epsilon)/2<c_\tau.
\]
Recall that the survival-based quantile is defined by $Q_\tau\{U(d_{f_n})\}=\sup\{q\in\mathbb R:S(q,d_{f_n})\ge c_\tau\}$. The first inequality implies that $q_\dagger-\epsilon$ satisfies the survival constraint, so $Q_\tau\{U(d_{f_n})\}\ge q_\dagger-\epsilon$. Since $S(\cdot,d_{f_n})$ is nonincreasing, the second inequality implies $S(q,d_{f_n})<c_\tau$ for every $q\ge q_\dagger+\epsilon$. Hence, $Q_\tau\{U(d_{f_n})\}\le q_\dagger+\epsilon$. Then, we have
\[
q_\dagger-\epsilon\le Q_\tau\{U(d_{f_n})\}\le q_\dagger+\epsilon \ \text{for all } n\ge N_\epsilon.
\]
Since $\epsilon>0$ can be arbitrarily small, we conclude that $Q_\tau\{U(d_{f_n})\}\to q_\dagger$. 

The proof of Lemma~\ref{lem:oracle_continuity} is complete.
\end{proof}

\subsection{Proof of Theorem~\ref{thm:Thm3}}
\label{pf:Thm3}
\begin{proof}
We prove the theorem in three steps.

\medskip\noindent\textit{Step 1: Oracle comparison and asymptotic feasibility.} By Lemma~\ref{lem:empirical_uniform} and Eq.~\eqref{eq:primitive_error},
\[
e_n\le E_n^0+C\Xi_{\pi,n}+Ca_{h_n,\delta_n}+\Delta_{b,n},\ E_n^0=O_p(r_{n,\mathrm{emp}}).
\]
Since all four error terms converge to zero in probability, $e_n=o_p(1)$. Moreover, Eq.~\eqref{eq:qualitative_compatibility} implies $\overline\mu_n(E_n^0)^2=O_p(\overline\mu_nr_{n,\mathrm{emp}}^2)=o_p(1)$. Using $(x_1+x_2+x_3+x_4)^2\le4\sum_{k=1}^4x_k^2$ and applying Eq.~\eqref{eq:qualitative_compatibility} to the remaining terms, we obtain
\[
\overline\mu_ne_n^2\le C\overline\mu_n\{(E_n^0)^2+\Xi_{\pi,n}^2+a_{h_n,\delta_n}^2+\Delta_{b,n}^2\}=o_p(1).
\]

Define the nonnegative comparison error as
\[
\widehat D_n=|\mathcal J|\overline\mu_ne_n^2+r_{n,\mathrm{reg}}+\varepsilon_n,
\]
which combines the penalty-weighted constraint approximation error, regularization error, and approximate optimization error. By Assumption~\ref{ass:primitive_entropy}(iii) and Assumption~\ref{ass:approx_opt}, $r_{n,\mathrm{reg}}\to0$ and $\varepsilon_n=o_p(1)$, so $\widehat D_n=o_p(1)$.

By Theorem~\ref{thm:Thm2}(i), $\mathcal A_\dagger$ is nonempty. Choose any fixed $\theta^\dagger=(q_\dagger,f^\dagger)\in\mathcal A_\dagger$. Since $v_j(\theta^\dagger)=0$ for every $j\in\mathcal J$, the definition of $e_n$ gives $0\le\widehat v_{j,n}(\theta^\dagger)=|\widehat v_{j,n}(\theta^\dagger)-v_j(\theta^\dagger)|\le e_n$. Thus,
\[
\widehat{\mathcal L}_n(\theta^\dagger) =q_\dagger-\sum_{j\in\mathcal J}\mu_{j,n}\widehat v_{j,n}(\theta^\dagger)^2-\lambda_nJ(f^\dagger) \ge q_\dagger-|\mathcal J|\overline\mu_ne_n^2-r_{n,\mathrm{reg}}.
\]
The approximate-maximizer condition in Assumption~\ref{ass:approx_opt} therefore yields
\[
\widehat q_n-\sum_{j\in\mathcal J}\mu_{j,n}\widehat v_{j,n}(\widehat\theta_n)^2-\lambda_nJ(\widehat f_n) =\widehat{\mathcal L}_n(\widehat\theta_n) \ge\sup_{\theta\in\Theta}\widehat{\mathcal L}_n(\theta)-\varepsilon_n \ge\widehat{\mathcal L}_n(\theta^\dagger)-\varepsilon_n \ge q_\dagger-\widehat D_n.
\]
Since the quadratic penalties and the regularization term are nonnegative, $\widehat q_n\ge q_\dagger-\widehat D_n$. Rearranging the same inequality and using $\widehat q_n\le q_U$ also gives
\[
\sum_{j\in\mathcal J}\mu_{j,n}\widehat v_{j,n}(\widehat\theta_n)^2\le\widehat q_n-q_\dagger+\widehat D_n\le q_U-q_\dagger+\widehat D_n.
\]
For every $j\in\mathcal J$, $\mu_{j,n}\ge\underline\mu_n>0$ and $v_j(\widehat\theta_n)\le\widehat v_{j,n}(\widehat\theta_n)+e_n$. Hence,
\begin{equation}
\label{eq:empirical_oracle_bounds}
(q_\dagger-\widehat q_n)_+ \le\widehat D_n\xrightarrow{p}0,\ 
\max_{j\in\mathcal J}v_j(\widehat\theta_n) \le e_n+\left(\frac{q_U-q_\dagger+\widehat D_n}{\underline\mu_n}\right)^{1/2}\xrightarrow{p}0.
\end{equation}
The second convergence follows from $e_n=o_p(1)$, $\widehat D_n=o_p(1)$, and $\underline\mu_n\to\infty$. In particular, since $\mathcal I\subseteq\mathcal J$,
\[
\max_{m\in\mathcal I}[\mathcal B_m(d_{\widehat f_n})-b_m]_+=\max_{m\in\mathcal I}v_m(\widehat\theta_n)\xrightarrow{p}0.
\]
Thus, asymptotic risk feasibility holds without Assumption~\ref{ass:oracle_crossing}.

\medskip\noindent\textit{Step 2: Distance to the restricted oracle set.} Fix any $\eta>0$ and define the open neighborhood
\[
G_\eta=\{\theta\in\Theta:\operatorname{dist}(\theta,\mathcal A_\dagger)<\eta\}.
\]
If $\Theta\setminus G_\eta$ is empty, then $\widehat\theta_n\in G_\eta$ for every $n$. Otherwise, $\Theta\setminus G_\eta$ is nonempty and compact. By Theorem~\ref{thm:Thm2}(i), the hard constraint violations are continuous. Therefore, the following minimum is attained:
\[
\kappa_\eta=\min_{\theta=(q,f)\in\Theta\setminus G_\eta}\left\{(q_\dagger-q)_++\max_{j\in\mathcal J}v_j(\theta)\right\}.
\]
We next claim that $\kappa_\eta>0$. Suppose instead that the minimum were zero at some $\theta=(q,f)\in\Theta\setminus G_\eta$. Since every term is nonnegative, $v_j(\theta)=0$ for all $j\in\mathcal J$ and $q\ge q_\dagger$. Thus, $\theta\in\Theta_b$. The definition $q_\dagger=\max_{(q,f)\in\Theta_b}q$ then gives $q\le q_\dagger$. Consequently, $q=q_\dagger$ and $\theta\in\mathcal A_\dagger$, contradicting $\theta\notin G_\eta$. Hence, $\kappa_\eta>0$.

Define the event
\[
\Omega_n=\left\{(q_\dagger-\widehat q_n)_++\max_{j\in\mathcal J}v_j(\widehat\theta_n)<\kappa_\eta\right\}.
\]
For this fixed $\eta$, Eq.~\eqref{eq:empirical_oracle_bounds} implies $\PP(\Omega_n)\to1$. On $\Omega_n$, the estimator must belong to $G_\eta$. Otherwise, the definition of $\kappa_\eta$ would imply $(q_\dagger-\widehat q_n)_++\max_{j\in\mathcal J}v_j(\widehat\theta_n)\ge\kappa_\eta$, contradicting the definition of $\Omega_n$. Thus,
\[
\PP\{\operatorname{dist}(\widehat\theta_n,\mathcal A_\dagger)\ge\eta\}\le\PP(\Omega_n^c)\to0.
\]
Since $\eta>0$ was arbitrary, $\operatorname{dist}(\widehat\theta_n,\mathcal A_\dagger)\xrightarrow{p}0$. Moreover, every element of $\mathcal A_\dagger$ has threshold $q_\dagger$, so the definition of $\rho_\Theta$ gives
\[
|\widehat q_n-q_\dagger| \le\inf_{(q_\dagger,f^*)\in\mathcal A_\dagger}\left\{|\widehat q_n-q_\dagger|+\rho_{\mathcal F}(\widehat f_n,f^*)\right\} =\operatorname{dist}(\widehat\theta_n,\mathcal A_\dagger)\xrightarrow{p}0.
\]

\medskip\noindent\textit{Step 3: Quantile consistency via a subsequence argument.} Suppose Assumption~\ref{ass:oracle_crossing} holds. Define
\[
W_n=Q_\tau\{U(d_{\widehat f_n})\}-q_\dagger.
\]
We next show $W_n\xrightarrow{p}0$ using the subsequence characterization of convergence in probability: a sequence converges to zero in probability if and only if every subsequence admits a further subsequence converging to zero almost surely.

Fix any subsequence $\{n_k\}$. By Step 2 and this characterization, there exists a further subsequence $\{n_{k_\ell}\}$ such that $\operatorname{dist}(\widehat\theta_{n_{k_\ell}},\mathcal A_\dagger)\to0$ almost surely. On the probability-one event where this convergence holds, fix a sample outcome. The resulting sequence of fitted parameters is a deterministic sequence in $\Theta$ whose distance to $\mathcal A_\dagger$ tends to zero. Lemma~\ref{lem:oracle_continuity} therefore gives $Q_\tau\{U(d_{\widehat f_{n_{k_\ell}}})\}\to q_\dagger$ on this event. Hence, $W_{n_{k_\ell}}\to0$ almost surely. Since the original subsequence was arbitrary, the same subsequence characterization yields $W_n\xrightarrow{p}0$, or equivalently,
\[
Q_\tau\{U(d_{\widehat f_n})\}\xrightarrow{p}q_\dagger=Q_{\tau,\dagger}^*.
\]

In addition, if the representability condition in Theorem~\ref{thm:Thm2}(iii) holds, then $q_\dagger=Q_\tau^*$. Thus, under this additional condition, the actual quantile value also converges to the unrestricted optimal value. The proof of Theorem~\ref{thm:Thm3} is complete.
\end{proof}

\subsection{Proof of Theorem~\ref{thm:Thm4}}
\label{pf:Thm4}
\begin{proof}
Throughout the proof, constants denoted by $C$ are finite and independent of $n$, and may change from line to line.

\medskip\noindent\textit{Step 1: Direct threshold and constraint-violation bounds.} By Lemma~\ref{lem:empirical_uniform}, Assumption~\ref{ass:emp_rate}, and the definition of $s_n=r_{n,\mathrm{emp}}+r_{\pi,n}+r_{b,n}+a_{h_n,\delta_n}$,
\[
e_n\le E_n^0+C\Xi_{\pi,n}+Ca_{h_n,\delta_n}+\Delta_{b,n}
=O_p(r_{n,\mathrm{emp}})+O_p(r_{\pi,n})+Ca_{h_n,\delta_n}+O_p(r_{b,n})
=O_p(s_n).
\]
Consequently,
\[
\widehat D_n=|\mathcal J|\overline\mu_ne_n^2+r_{n,\mathrm{reg}}+\varepsilon_n=O_p(\overline\mu_ns_n^2+r_{n,\mathrm{reg}}+r_{n,\mathrm{opt}})=O_p(d_n).
\]
Since $\overline\mu_ns_n^2\to0$, $r_{n,\mathrm{reg}}\to0$, and $r_{n,\mathrm{opt}}\to0$, we have $d_n\to0$, and hence $\widehat D_n=o_p(1)$. Also, we have $s_n\to0$, $t_n\to0$, and $\Delta_n\to0$.

Applying the oracle comparison inequalities established in Step 1 of the proof of Theorem~\ref{thm:Thm3}, we obtain
\[
(q_\dagger-\widehat q_n)_+\le\widehat D_n,\ v_j(\widehat\theta_n)\le e_n+\left(\frac{q_U-q_\dagger+\widehat D_n}{\mu_{j,n}}\right)^{1/2},\ j\in\mathcal J.
\]
These inequalities hold for every $n$ and do not require quantitative oracle regularity. Since $e_n=O_p(s_n)$, $\widehat D_n=O_p(d_n)=o_p(1)$, and $\underline\mu_n\to\infty$, it follows that
\[
(q_\dagger-\widehat q_n)_+ =O_p(d_n),\ \max_{j\in\mathcal J}v_j(\widehat\theta_n) =O_p(s_n+\underline\mu_n^{-1/2}) =O_p(t_n).
\]
Consequently, $G_\dagger(\widehat\theta_n)=O_p(d_n+t_n)=O_p(\Delta_n)$.

\medskip\noindent\textit{Step 2: Distance to the restricted oracle set.} 
Suppose Assumption~\ref{ass:quant_oracle}(i) holds. By the nuisance convergence rates in Assumption~\ref{ass:emp_rate} and the definition of $s_n$,
\[
a_{h_n,\delta_n}^2+r_{n,\mathrm{emp}}^2+\Xi_{\pi,n}^2+\Delta_{b,n}^2=O_p\left(a_{h_n,\delta_n}^2+r_{n,\mathrm{emp}}^2+r_{\pi,n}^2+r_{b,n}^2\right)=O_p(s_n^2).
\]
Together with $\overline\mu_ns_n^2\to0$, this yields
\[
\overline\mu_n\{a_{h_n,\delta_n}^2+r_{n,\mathrm{emp}}^2+\Xi_{\pi,n}^2+\Delta_{b,n}^2\}=O_p(\overline\mu_ns_n^2)=o_p(1).
\]
Thus, Eq.~\eqref{eq:qualitative_compatibility} holds, and Theorem~\ref{thm:Thm3} implies $\operatorname{dist}(\widehat\theta_n,\mathcal A_\dagger)\xrightarrow{p}0$. Consequently, $\PP\{\operatorname{dist}(\widehat\theta_n,\mathcal A_\dagger)\le r_{\mathrm{sep}}\}\to1$. On this event, Assumption~\ref{ass:quant_oracle}(i) gives
\[
c_{\mathrm{sep}}\operatorname{dist}(\widehat\theta_n,\mathcal A_\dagger)^\gamma\le G_\dagger(\widehat\theta_n).
\]

The local error bound applies only when $\operatorname{dist}(\widehat\theta_n,\mathcal A_\dagger)\le r_{\mathrm{sep}}$. Separating this event from its complement, we obtain, for any $L>0$,
\[
\PP\{\operatorname{dist}(\widehat\theta_n,\mathcal A_\dagger)>L\Delta_n^{1/\gamma}\} \le\PP\{\operatorname{dist}(\widehat\theta_n,\mathcal A_\dagger)>r_{\mathrm{sep}}\} +\PP\{G_\dagger(\widehat\theta_n)>c_{\mathrm{sep}}L^\gamma\Delta_n\}.
\]
The first term tends to zero by Theorem~\ref{thm:Thm3}. Since $G_\dagger(\widehat\theta_n)=O_p(\Delta_n)$, the second term can be made arbitrarily small in limit superior by choosing $L$ sufficiently large. Therefore,
\[
\operatorname{dist}(\widehat\theta_n,\mathcal A_\dagger)=O_p(\Delta_n^{1/\gamma}).
\]
Since every element of $\mathcal A_\dagger$ has threshold $q_\dagger$, $|\widehat q_n-q_\dagger|\le\operatorname{dist}(\widehat\theta_n,\mathcal A_\dagger)=O_p(\Delta_n^{1/\gamma})$.

\medskip\noindent\textit{Step 3: Quantitative control of treatment-regime differences.} We first establish deterministic inequalities for an arbitrary $\theta=(q,f)\in\Theta$ close to $\mathcal A_\dagger$. By compactness of $\mathcal A_\dagger$, choose $\theta^*=(q_\dagger,f^*)\in\mathcal A_\dagger$ attaining the minimum distance. Then
\[
\rho_{\mathcal F}(f,f^*)\le\rho_\Theta(\theta,\theta^*)=\operatorname{dist}(\theta,\mathcal A_\dagger).
\]
If $\operatorname{dist}(\theta,\mathcal A_\dagger)=0$, then $\theta\in\mathcal A_\dagger$, so its induced regime is already restricted-optimal and feasible. Otherwise, let $C_V=\max_{a\in\mathcal A}\|V_a\|$. By the decision-boundary argument in Step 2 of the proof of Lemma~\ref{lem:oracle_continuity}, for each $t=1,2$,
\[
\{d_{f_t}(\H_t)\ne d_{f_t^*}(\H_t)\} \subseteq\{\Delta_t(\H_t;f_t^*)\le2C_V\rho_{\mathcal F}(f,f^*)\} \subseteq\{\Delta_t(\H_t;f_t^*)\le2C_V\operatorname{dist}(\theta,\mathcal A_\dagger)\}.
\]
For $\operatorname{dist}(\theta,\mathcal A_\dagger)$ sufficiently small that $2C_V\operatorname{dist}(\theta,\mathcal A_\dagger)\le\min(r_1,r_2)$, Assumption~\ref{ass:boundary}(i) implies
\[
\PP\{d_{f_t}(\H_t)\ne d_{f_t^*}(\H_t)\}\le C_t\{2C_V\operatorname{dist}(\theta,\mathcal A_\dagger)\}^{\alpha_t}.
\]
Define $\zeta=\sum_{t=1}^2\PP\{d_{f_t}(\H_t)\ne d_{f_t^*}(\H_t)\}$ and let $\alpha=\min(\alpha_1,\alpha_2)$. For $0<\operatorname{dist}(\theta,\mathcal A_\dagger)\le\min\{1,r_1/(2C_V),r_2/(2C_V)\}$, Assumption~\ref{ass:boundary}(i) gives
\[
\zeta \le\sum_{t=1}^2C_t(2C_V)^{\alpha_t}\operatorname{dist}(\theta,\mathcal A_\dagger)^{\alpha_t} \le C\operatorname{dist}(\theta,\mathcal A_\dagger)^\alpha.
\]
Since $\Delta_n\to0$ and $\operatorname{dist}(\widehat\theta_n,\mathcal A_\dagger)=O_p(\Delta_n^{1/\gamma})$, the estimated parameter belongs to this neighborhood with probability tending to one. The bound holds uniformly over a sufficiently small neighborhood of $\mathcal A_\dagger$ and across all choices of closest oracles. It can therefore be applied to $\widehat\theta_n$ without requiring a measurable oracle selection.

\medskip\noindent\textit{Step 4: Risk-violation rate.} Continue with the fixed pair $(\theta,\theta^*)$ from Step 3. 

Since $\theta^*=(q_\dagger,f^*)\in\mathcal A_\dagger\subseteq\Theta_b$, we have $\mathcal B_m(d_{f^*})\le b_m$ for every $m\in\mathcal I$. The IPW comparison in Eq.~\eqref{eq:policy_continuity} yields
\[
\max_{m\in\mathcal I}[\mathcal B_m(d_f)-b_m]_+
\le\max_{m\in\mathcal I}|\mathcal B_m(d_f)-\mathcal B_m(d_{f^*})| \le C_\psi\pi_{\min}^{-2}\zeta \le C\operatorname{dist}(\theta,\mathcal A_\dagger)^\alpha.
\]
Since this bound holds uniformly for all $\theta$ sufficiently close to $\mathcal A_\dagger$, it applies to $\widehat\theta_n$ on an event whose probability tends to one. By Step 2,
\[
\max_{m\in\mathcal I}[\mathcal B_m(d_{\widehat f_n})-b_m]_+=O_p(\Delta_n^{\alpha/\gamma}).
\]
Since $\mathcal I\subseteq\mathcal J$, the direct bound in part (i) also gives $\max_{m\in\mathcal I}[\mathcal B_m(d_{\widehat f_n})-b_m]_+=O_p(t_n)$. Unlike the rate derived above, this direct bound does not require Assumption~\ref{ass:quant_oracle}(i).

\medskip\noindent\textit{Step 5: Actual quantile error and comparison with the unrestricted oracle.} Suppose Assumption~\ref{ass:quant_oracle}(ii) also holds. For the fixed pair $(\theta,\theta^*)$ in Step 3, Eq.~\eqref{eq:policy_continuity_S} and the bound on $\zeta$ give
\[
\sup_{u\in\mathbb R}|S(u,d_f)-S(u,d_{f^*})|\le\pi_{\min}^{-2}\zeta\le C_S\operatorname{dist}(\theta,\mathcal A_\dagger)^\alpha
\]
for a fixed constant $C_S>0$. If $\operatorname{dist}(\theta,\mathcal A_\dagger)=0$, then $Q_\tau\{U(d_f)\}=q_\dagger$. Otherwise, set $w=2C_S\operatorname{dist}(\theta,\mathcal A_\dagger)^\alpha/c_{\mathrm{cross}}$, and restrict $\theta$ to a sufficiently small neighborhood of $\mathcal A_\dagger$ so that $w\le r_0$. Then, we have
\[
\begin{aligned}
S(q_\dagger-w,d_f)
&\ge S(q_\dagger-w,d_{f^*})-C_S\operatorname{dist}(\theta,\mathcal A_\dagger)^\alpha\\
&\ge c_\tau+c_{\mathrm{cross}}w-C_S\operatorname{dist}(\theta,\mathcal A_\dagger)^\alpha\\
&=c_\tau+C_S\operatorname{dist}(\theta,\mathcal A_\dagger)^\alpha>c_\tau,
\end{aligned}
\]
\[
\begin{aligned}
S(q_\dagger+w,d_f)
&\le S(q_\dagger+w,d_{f^*})+C_S\operatorname{dist}(\theta,\mathcal A_\dagger)^\alpha\\
&\le c_\tau-c_{\mathrm{cross}}w+C_S\operatorname{dist}(\theta,\mathcal A_\dagger)^\alpha\\
&=c_\tau-C_S\operatorname{dist}(\theta,\mathcal A_\dagger)^\alpha<c_\tau.
\end{aligned}
\]
The first inequality makes $q_\dagger-w$ an admissible threshold in the survival-based quantile definition. By monotonicity, the second excludes every threshold at or above $q_\dagger+w$. Hence,
\[
q_\dagger-w\le Q_\tau\{U(d_f)\}\le q_\dagger+w.
\]
Therefore, there exist constants $r_*=\min\{1,r_1/(2C_V),r_2/(2C_V),(\frac{c_{\mathrm{cross}}r_0}{2C_S})^{1/\alpha} \}>0$ and $C=2C_S/c_{\mathrm{cross}}<\infty$ such that
\begin{equation}
\label{eq:local_quantile_error}
|Q_\tau\{U(d_f)\}-q_\dagger|\le C\operatorname{dist}(\theta,\mathcal A_\dagger)^\alpha
\end{equation}
for every $\theta\in\Theta$ satisfying $\operatorname{dist}(\theta,\mathcal A_\dagger)\le r_*$.

Applying Eq.~\eqref{eq:local_quantile_error} to $\widehat\theta_n$, for every $L>0$,
\[
\PP\{|Q_\tau\{U(d_{\widehat f_n})\}-q_\dagger|>L\Delta_n^{\alpha/\gamma}\} \le\PP\{\operatorname{dist}(\widehat\theta_n,\mathcal A_\dagger)>r_*\} +\PP\{C\operatorname{dist}(\widehat\theta_n,\mathcal A_\dagger)^\alpha>L\Delta_n^{\alpha/\gamma}\}.
\]
The first term tends to zero because $\operatorname{dist}(\widehat\theta_n,\mathcal A_\dagger)=o_p(1)$. Since $\operatorname{dist}(\widehat\theta_n,\mathcal A_\dagger)=O_p(\Delta_n^{1/\gamma})$, the second term can be made arbitrarily small in limit superior by taking $L$ sufficiently large. Thus,
\[
|Q_\tau\{U(d_{\widehat f_n})\}-q_\dagger|=O_p(\Delta_n^{\alpha/\gamma}).
\]
Under the representability condition in Theorem~\ref{thm:Thm2}(iii), $q_\dagger=Q_\tau^*$, giving the unrestricted-oracle rate. More generally, whenever $Q_\tau^*$ is finite, the class approximation gap $Q_\tau^*-q_\dagger$ is nonnegative because the represented feasible regimes form a subclass of $\mathcal D_{b,\mathcal I}$. Using $[a+x]_+\le a+[x]_+$ for $a\ge0$,
\[
\begin{aligned}
[Q_\tau^*-Q_\tau\{U(d_{\widehat f_n})\}]_+
&=[(Q_\tau^*-q_\dagger)
  +q_\dagger-Q_\tau\{U(d_{\widehat f_n})\}]_+\\
&\le (Q_\tau^*-q_\dagger)
+[q_\dagger-Q_\tau\{U(d_{\widehat f_n})\}]_+.
\end{aligned}
\]
The second term is bounded by the absolute quantile error just established. The proof of Theorem~\ref{thm:Thm4} is complete.
\end{proof}

\section{Complete Data Generating Processes for Risk-Free Scenarios} 
\subsection{Scenario A (Two stages, binary treatments)}
The stage-1/baseline covariates $\bm{X}_1 = (X_{1,1}, X_{1,2}, X_{1,3})^\top$ follow $N(\bm 0, \bm I_3)$. The stage-1 treatment $A_1 \in \{0, 1\}$ is assigned according to $$\pi_1^*(A_1 = 1 | \bm{X}_1) = \operatorname{expit}\left(-0.5 + 0.5X_{1,1} - 0.3X_{1,2}\right).$$ 
Based on $(\bm{X}_1, A_1)$, the stage-2 covariates $\bm{X}_2 = (X_{2,1}, X_{2,2})^\top$ are generated by
$$X_{2,1}=0.5X_{1,1}+0.3A_1+\varepsilon_{2,1},\
X_{2,2}=-0.2X_{1,2}-0.2A_1+\varepsilon_{2,2}.$$
Given the history $\bm H_2=(\bm X_1,A_1,\bm X_2)$, the stage-2 treatment $A_2\in\{0,1\}$ is assigned according to 
$$\pi_2^*(A_2 = 1 | \bm{H}_2) = \operatorname{expit} \left(-0.3 + 0.4X_{2,1} - 0.3X_{1,1} + 0.2A_1\right).$$ 
The final outcome is
$$Y=1-2X_{1,2} +2X_{1,3} +A_1(X_{1,1} +X_{1,2}) +0.5X_{2,2} +A_2(X_{2,1} -0.5X_{1,1}) +\varepsilon_Y,$$
where $\varepsilon_{2,1},\varepsilon_{2,2},\varepsilon_Y$ are mutually independent $N(0,1)$ random variables, independent of $\bm X_1$.

\textit{\textbf{Population quantile oracle:}}
The population oracle is defined over all measurable deterministic treatment regimes based on the available stage-specific histories, without restricting them to the score class used for estimation.

Since the stage-2 treatment affects the outcome through $A_2(X_{2,1}-0.5X_{1,1})$, the pointwise optimal stage-2 rule is $d_2^*(\bm H_2)=I\{X_{2,1}-0.5X_{1,1}>0\}$. For $a\in\{0,1\}$, define $\ell_a(\bm X_1)=1-2.1X_{1,2}+2X_{1,3}+a(X_{1,1}+X_{1,2}-0.1)$. Here \(\ell_a(\bm X_1)\) collects the deterministic terms in the potential outcome under the stage-1 treatment \(a\), including its effect on the stage-2 covariates. Under $A_1=a$ and the optimal stage-2 rule $d_2^*$, the potential outcome can be expressed as
\begin{equation*}\label{eq:oracle-A-outcome}
Y(a,d_2^*)=\ell_a(\bm X_1)+(\varepsilon_{2,1}+0.3a)_++0.5\varepsilon_{2,2}+\varepsilon_Y,
\end{equation*}
where $(u)_+=\max(u,0)$. This representation accounts for the effects of the stage-1 treatment on both stage-2 covariates and the subsequent optimal treatment benefit.

Let $W=0.5\varepsilon_{2,2}+\varepsilon_Y\sim N(0,1.25)$ and $Z=\varepsilon_{2,1}\sim N(0,1)$, with $W$ and $Z$ independent. Since $W+(Z+0.3a)_+=\max\{W,W+Z+0.3a\}$, the conditional residual CDF is
$$G_a(u) =P\{W\le u,\ W+Z\le u-0.3a\} =\Phi_2\left(\frac{u}{\sqrt{1.25}},\frac{u-0.3a}{1.5};\frac{\sqrt5}{3}\right),$$
where $\Phi_2(\cdot,\cdot;\rho)$ denotes the standard bivariate normal CDF with correlation $\rho$. Consequently, $P\{Y(a,d_2^*)\le q|\bm X_1\}=G_a\{q-\ell_a(\bm X_1)\}$.

For a common population threshold $q$, define $C_A(q)=E\left[\min_{a\in\{0,1\}}G_a\{q-\ell_a(\bm X_1)\}\right]$. Let $q_{\tau,A}^*$ denote the unique solution to $C_A(q)=\tau$. The marginal quantile-optimal regime $d_\tau^{A,*}=(d_{1,\tau}^{A,*},d_{2,\tau}^{A,*})$ is given by
\begin{equation*}\label{eq:oracle-A-rule}
d_{1,\tau}^{A,*}(\bm X_1) =I\left\{G_1\{q_{\tau,A}^*-\ell_1(\bm X_1)\}<G_0\{q_{\tau,A}^*-\ell_0(\bm X_1)\}\right\},\ d_{2,\tau}^{A,*}(\bm H_2) =d_2^*(\bm H_2).
\end{equation*}
Note that this regime minimizes the marginal CDF at $q_{\tau,A}^*$ and attains $Q_\tau\{Y(d_\tau^{A,*})\}=q_{\tau,A}^*=\sup_d Q_\tau\{Y(d)\}$. Ties are resolved in favor of treatment $0$. The population oracle threshold $q_{\tau,A}^*$ is computed under the known data-generating process and held fixed across simulation replications. Oracle performance is assessed using empirical quantiles from independently generated test samples.

\textit{\textbf{Population mean oracle and comparison:}}
Under the optimal stage-2 rule $d_2^*$, the conditional expected outcome for the stage-1 treatment $a\in\{0,1\}$ is
\[
E\{Y(a,d_2^*)|\bm X_1\}=\ell_a(\bm X_1)+g(0.3a),
\]
where $g(c)=E\{(Z+c)_+\}=\phi(c)+c\Phi(c)$, and $\Phi$ and $\phi$ denote the standard normal CDF and density, respectively. The conditional mean contrast between the two stage-1 treatments is $E\{Y(1,d_2^*)-Y(0,d_2^*)|\bm X_1\}
=X_{1,1}+X_{1,2}+\kappa$, 
where $\kappa=-0.1+g(0.3)-g(0)\approx0.067819$. Hence, the mean-optimal regime $d_{\mathrm{mean}}^{A,*}=(d_{1,\mathrm{mean}}^{A,*},d_2^*)$ has the stage-1 rule
\[
d_{1,\mathrm{mean}}^{A,*}(\bm X_1)
=I\{X_{1,1}+X_{1,2}+\kappa>0\}.
\]

The mean- and marginal quantile-optimal regimes differ in this scenario. At the mean-optimal decision boundary $X_{1,1}+X_{1,2}=-\kappa$, the two conditional outcome distributions have equal means but different variances, since $\operatorname{Var}\{(Z+0.3)_+\}>\operatorname{Var}(Z_+)$. Their conditional CDFs therefore cross. Fix any finite threshold $q$ and a pair $(X_{1,1},X_{1,2})$ on this boundary. Varying $X_{1,3}$ shifts both conditional distributions by the same arbitrary real amount, without changing the conditional mean contrast. Hence, there is a baseline history on the boundary at which the two conditional CDFs differ strictly at $q$. By continuity, this ordering persists on an open set on the side of the boundary where the mean-optimal rule selects the treatment with the larger conditional CDF. The Gaussian baseline distribution assigns positive probability to this set. Consequently, $P\{Y(d_{\mathrm{mean}}^{A,*})\le q\}>C_A(q)$ for every finite $q$. In particular, at $q=q_{\tau,A}^*$, the left-hand side exceeds $\tau$. Since the outcome CDF under the mean-optimal regime is continuous and strictly increasing, this implies
$$Q_\tau\{Y(d_{\mathrm{mean}}^{A,*})\} < Q_\tau\{Y(d_\tau^{A,*})\} =q_{\tau,A}^*, \ \tau\in(0,1).$$

\subsection{Scenario B (Two stages, $K=4$ treatments)}
We extend Scenario A to accommodate four treatment options at each stage. The stage-1 covariates $\bm X_1=(X_{1,1},\ldots,X_{1,5})^\top$ follow $N(\bm 0, \bm I_5)$. The stage-1 treatment $A_1\in\{1,2,3,4\}$ is assigned according to a multinomial logistic model,
\[
\pi_1^*(A_1=k|\bm X_1)=\frac{\exp(\eta_{1,k})}{\sum_{j=1}^4\exp(\eta_{1,j})},\ k=1,\ldots,4,
\]
where $(\eta_{1,1},\eta_{1,2},\eta_{1,3},\eta_{1,4})=(0.4X_{1,1}-0.2X_{1,2},\,0.2X_{1,3},\,-0.3X_{1,1}+0.3X_{1,4},\,0)$.

The stage-2 covariates $\bm X_2=(X_{2,1},X_{2,2})^\top$ are generated by
\[
X_{2,1}=\gamma_{A_1}+0.3X_{1,1}+1.5\varepsilon_{2,1},\
X_{2,2}=-0.2X_{1,2}+0.2I\{A_1=3\}+\varepsilon_{2,2},
\]
where $(\gamma_1,\gamma_2,\gamma_3,\gamma_4)=(-1.5,0.5,2.0,0.0)$. Given the history $\bm H_2=(\bm X_1,A_1,\bm X_2)$, the stage-2 treatment $A_2\in\{1,2,3,4\}$ is assigned according to
\[
\pi_2^*(A_2=k|\bm H_2)=\frac{\exp(\eta_{2,k})}{\sum_{j=1}^4\exp(\eta_{2,j})},\ k=1,\ldots,4,
\]
where $(\eta_{2,1},\eta_{2,2},\eta_{2,3},\eta_{2,4})=(0.5X_{2,1},\,0.2X_{2,2},\,-0.4X_{2,1}+0.3X_{1,1},\,0)$.

To induce heterogeneous treatment effects, let $J\in\{1,2,3,4\}$ denote the population quartile group of $X_{1,1}$, defined by $J=1+\sum_{r=1}^3 I\{X_{1,1}>\Phi^{-1}(r/4)\}$, where $\Phi$ is the standard normal CDF. The stage-specific treatment effects $\beta_t(a,J)$, for $t=1,2$ and $a\in\{1,2,3,4\}$, are specified by the following matrices, with rows corresponding to quartile groups and columns to treatments:
\[
\bm B_1=
\begin{pmatrix}
2.0&0.8&-0.5&0.0\\
1.5&1.8&0.5&0.9\\
0.5&1.0&1.5&1.8\\
-0.5&0.0&2.0&1.4
\end{pmatrix},
\
\bm B_2=
\begin{pmatrix}
1.8&0.9&-0.6&0.2\\
1.1&1.7&0.1&1.0\\
0.3&1.2&1.5&1.7\\
-0.6&0.2&2.1&1.6
\end{pmatrix},
\]
where $\beta_t(a,J)=(\bm B_t)_{J,a}$. The final outcome is
\[
Y=1-X_{1,2}+0.5X_{1,3}+\beta_1(A_1,J)+0.3X_{2,2}+\beta_2(A_2,J)+\varepsilon_Y,
\]
where $\varepsilon_{2,1},\varepsilon_{2,2},\varepsilon_Y$ are mutually independent $N(0,1)$ random variables, independent of $\bm X_1$.

\textit{\textbf{Population quantile oracle:}}
As in Scenario A, the population oracle is defined over all measurable deterministic treatment regimes based on the available stage-specific histories, without restricting them to the score class used for estimation.

For treatments $a,b\in\{1,2,3,4\}$ at the two stages, substituting the stage-2 covariate equations into the outcome gives
\[
Y(a,b)=1-1.06X_{1,2}+0.5X_{1,3}+\widetilde\beta_1(a,J)+\beta_2(b,J)+0.3\varepsilon_{2,2}+\varepsilon_Y,
\]
where $\widetilde\beta_1(a,J)=\beta_1(a,J)+0.06I\{a=3\}$. The additional term accounts for the effect of the stage-1 treatment on $X_{2,2}$ and hence the final outcome. Although the stage-1 treatment also affects $X_{2,1}$, this covariate does not enter the outcome directly, and the optimal stage-2 treatment effect depends only on $J$. Consequently, the population quantile-optimal regime $d^{B,*}=(d_1^{B,*},d_2^{B,*})$ is
\begin{equation*}\label{eq:oracle-B-rule}
d_1^{B,*}(\bm X_1) =\arg\max_{a\in\{1,2,3,4\}}\widetilde\beta_1(a,J),\ d_2^{B,*}(\bm H_2) =\arg\max_{b\in\{1,2,3,4\}}\beta_2(b,J).
\end{equation*}
For the specified coefficients, both stages select treatments $(1,2,4,3)$ across quartile groups $J=1,\ldots,4$, respectively. For any admissible regime $d$, under a common coupling of the exogenous variables,
\[
\widetilde\beta_1\{d_1(\bm X_1),J\}+\beta_2\{d_2(\bm H_2^d),J\}
\le \max_a\widetilde\beta_1(a,J)+\max_b\beta_2(b,J).
\]
The proposed regime $d^{B,*}$ attains both maxima, so $Y(d^{B,*})\ge Y(d)$ almost surely under this coupling. Thus, it attains both the optimal mean and every marginal quantile:
\[
E\{Y(d^{B,*})\}=\sup_d E\{Y(d)\},\
Q_\tau\{Y(d^{B,*})\}=\sup_d Q_\tau\{Y(d)\},\ \tau\in(0,1).
\]
Unlike Scenario A, the quantile-optimal regime coincides with the mean-optimal regime. The oracle regime is held fixed across simulation replications, and its performance is assessed using empirical quantiles from independently generated test samples.

\section{Complete Data Generating Processes for Risk-Aware Scenarios} 
\subsection{Scenario C (Two stages, binary treatments)}
We extend Scenario A to incorporate treatment-related risks. The stage-1 covariates $\bm X_1=(X_{1,1},X_{1,2},X_{1,3})^\top$ follow $N(\bm 0,\bm I_3)$. The stage-1 treatment $A_1\in\{0,1\}$ is assigned according to
$$\pi_1^*(A_1 = 1 | \bm{X}_1) = \operatorname{expit}\left(-0.5 + 0.5X_{1,1} - 0.3X_{1,2}\right).$$ 
To incorporate treatment-related risks, the stage-1 side-effect indicators $\bm Z_1=(Z_1^{(1)},Z_1^{(2)})^\top$ are generated according to $Z_1^{(m)}|\bm X_1,A_1\sim\operatorname{Bernoulli}(p_{1,m})$, for $m=1,2$, where
$$\operatorname{logit}(p_{1,1}) = -1.2+0.5X_{1,1}-0.4X_{1,2}+0.9A_1,$$
$$\operatorname{logit}(p_{1,2}) = -1.0+0.3X_{1,1}-0.2X_{1,2}+A_1.$$
The stage-1 risk score is defined as $\mathcal R_1=(Z_1^{(1)}+Z_1^{(2)})/2$.

The stage-2 covariates $\bm X_2=(X_{2,1},X_{2,2})^\top$ are generated by
\[
X_{2,1}=0.5X_{1,1}+0.3A_1+\varepsilon_{2,1},\
X_{2,2}=-0.2X_{1,2}-0.2A_1+\varepsilon_{2,2}.
\]
Given the history $\bm H_2=(\bm X_1,A_1,\bm X_2,\bm Z_1)$, the stage-2 treatment $A_2\in\{0,1\}$ is assigned according to
\[
\pi_2^*(A_2=1|\bm H_2)=\operatorname{expit}(-0.3+0.4X_{2,1}-0.3X_{1,1}+0.2A_1-0.5\mathcal R_1).
\]
The stage-2 side-effect indicators $\bm Z_2=(Z_2^{(1)},Z_2^{(2)})^\top$ are generated according to $Z_2^{(m)}|\bm H_2,A_2\sim\operatorname{Bernoulli}(p_{2,m})$, for $m=1,2$, where
$$\operatorname{logit}(p_{2,1}) = -1.2+0.4X_{2,1}-0.3X_{2,2}+0.7A_2+0.6\mathcal{R}_1,$$
$$\operatorname{logit}(p_{2,2}) = -0.8+0.2X_{2,1}-0.2X_{2,2}+0.9A_2+0.5\mathcal{R}_1.$$
The stage-2 risk score is defined as $\mathcal R_2=(Z_2^{(1)}+Z_2^{(2)})/2$.

To accommodate stagewise utility construction, the total efficacy outcome is decomposed as $Y=Y_1+Y_2$, where
$$Y_1 =1-2X_{1,2}+2X_{1,3}+A_1(X_{1,1}+X_{1,2})+\varepsilon_1,$$
$$Y_2=0.5X_{2,2}+A_2(X_{2,1}-0.5X_{1,1}-0.6\mathcal R_1)+\varepsilon_2.$$
Here $\varepsilon_{2,1},\varepsilon_{2,2}\sim N(0,1)$ and $\varepsilon_1,\varepsilon_2\sim N(0,0.5)$ are mutually independent and independent of the baseline covariates.

For evaluation, we define a common reference utility $U_t^{\mathrm{ref}}=Y_t-\phi_{\mathrm{SE}}(\mathcal R_t;\rho=1,\xi=2)$, for $t=1,2$, with total utility $U^{\mathrm{ref}}=U_1^{\mathrm{ref}}+U_2^{\mathrm{ref}}$, where $\phi_{\mathrm{SE}}$ is defined in Section~2.3 of the main manuscript. This specification fixes $\lambda_{\mathcal R}=1$ and $(\rho,\xi)=(1,2)$. RQDTR-U uses this prespecified utility for policy learning and cross-validation, with only the regularization parameter $\lambda$ selected by cross-validation. All learned regimes, including efficacy-only and constraint-based regimes, are evaluated using the same efficacy, reference utility, and risk measures.

\subsection{Scenario D (Two stages, $K=4$ treatments)}
We extend Scenario B to incorporate treatment-related risks. The stage-1 covariates $\bm X_1=(X_{1,1},\ldots,X_{1,5})^\top$ follow $N(\bm 0,\bm I_5)$. The stage-1 treatment $A_1\in\{1,2,3,4\}$ is assigned according to a multinomial logistic model,
\[
\pi_1^*(A_1=k|\bm X_1)=\frac{\exp(\eta_{1,k})}{\sum_{j=1}^4\exp(\eta_{1,j})},\ k=1,\ldots,4,
\]
where $(\eta_{1,1},\eta_{1,2},\eta_{1,3},\eta_{1,4})=(0.4X_{1,1}-0.2X_{1,2},\,0.2X_{1,3},\,-0.3X_{1,1}+0.3X_{1,4},\,0)$. 
To incorporate treatment-related risks, the stage-1 side-effect indicators $\bm Z_1=(Z_1^{(1)},Z_1^{(2)})^\top$ are generated according to $Z_1^{(m)}|\bm X_1,A_1\sim\operatorname{Bernoulli}(p_{1,m})$, for $m=1,2$, where
$$\operatorname{logit}(p_{1,1}) =-1+0.3X_{1,1}-0.2X_{1,2}+0.7I\{A_1=1\}+0.1I\{A_1=2\}+0.9I\{A_1=3\},$$
$$\operatorname{logit}(p_{1,2})=-0.8+0.2X_{1,1}+0.2X_{1,4}+0.5I\{A_1=1\}+0.2I\{A_1=2\}+I\{A_1=3\}.$$
The stage-1 risk score is defined as $\mathcal R_1=(Z_1^{(1)}+Z_1^{(2)})/2$.

The stage-2 covariates $\bm X_2=(X_{2,1},X_{2,2})^\top$ are generated by
\[
X_{2,1}=\gamma_{A_1}+0.3X_{1,1}+1.5\varepsilon_{2,1},\ X_{2,2}=-0.2X_{1,2}+0.2I\{A_1=3\}+\varepsilon_{2,2},
\]
where $(\gamma_1,\gamma_2,\gamma_3,\gamma_4)=(-1.5,0.5,2.0,0.0)$. Given the history $\bm H_2=(\bm X_1,A_1,\bm X_2,\bm Z_1)$, the stage-2 treatment $A_2\in\{1,2,3,4\}$ is assigned according to
\[
\pi_2^*(A_2=k|\bm H_2)=\frac{\exp(\eta_{2,k})}{\sum_{j=1}^4\exp(\eta_{2,j})},\ k=1,\ldots,4,
\]
where $(\eta_{2,1},\eta_{2,2},\eta_{2,3},\eta_{2,4})=(0.5X_{2,1}-0.3\mathcal R_1,\,0.2X_{2,2},\,-0.4X_{2,1}+0.3X_{1,1}+0.4\mathcal R_1,\,0)$.
The stage-2 side-effect indicators $\bm Z_2=(Z_2^{(1)},Z_2^{(2)})^\top$ are generated according to $Z_2^{(m)}|\bm H_2,A_2\sim\operatorname{Bernoulli}(p_{2,m})$, for $m=1,2$, where
$$\operatorname{logit}(p_{2,1}) = -1+0.3X_{2,1}-0.2X_{2,2}+0.6 I \{A_2=1\}+0.1 I \{A_2=2\}+ 0.8 I \{A_2=3\} +0.6 \mathcal{R}_1,$$
$$\operatorname{logit}(p_{2,2}) = -0.8+0.2X_{2,1}+0.1X_{1,1}+0.5 I \{A_2=1\}+0.2 I \{A_2=2\}+  I \{A_2=3\} +0.7 \mathcal{R}_1.$$
The stage-2 risk score is defined as $\mathcal R_2=(Z_2^{(1)}+Z_2^{(2)})/2$.

To induce heterogeneous treatment effects, let $J\in\{1,2,3,4\}$ denote the population quartile group of $X_{1,1}$, defined by $J=1+\sum_{r=1}^3 I\{X_{1,1}>\Phi^{-1}(r/4)\}$, where $\Phi$ is the standard normal CDF. The stage-specific treatment effects $\beta_t(a,J)$, for $t=1,2$ and $a\in\{1,2,3,4\}$, are specified by the following matrices, with rows corresponding to quartile groups and columns to treatments:
\[
\bm B_1=
\begin{pmatrix}
2.0&0.8&-0.5&0.0\\
1.5&1.8&0.5&0.9\\
0.5&1.0&1.5&1.8\\
-0.5&0.0&2.0&1.4
\end{pmatrix}, \
\bm B_2=
\begin{pmatrix}
1.8&0.9&-0.6&0.2\\
1.1&1.7&0.1&1.0\\
0.3&1.2&1.5&1.7\\
-0.6&0.2&2.1&1.6
\end{pmatrix},
\]
where $\beta_t(a,J)=(\bm B_t)_{J,a}$.

To accommodate stagewise utility construction, the total efficacy outcome is decomposed as $Y=Y_1+Y_2$, where
$$Y_1=1-X_{1,2} +0.5X_{1,3} +\beta_1(A_1,J) +\varepsilon_1,$$
$$Y_2=0.3X_{2,2}+\beta_2(A_2,J)-0.5\mathcal R_1 I\{A_2=3\}+\varepsilon_2.$$
Here $\varepsilon_{2,1},\varepsilon_{2,2}\sim N(0,1)$ and $\varepsilon_1,\varepsilon_2\sim N(0,0.5)$ are mutually independent and independent of the baseline covariates and side-effect innovations.

For evaluation, we define a common reference utility $U_t^{\mathrm{ref}}=Y_t-\phi_{\mathrm{SE}}(\mathcal R_t;\rho=1,\xi=2)$, for $t=1,2$, with total utility $U^{\mathrm{ref}}=U_1^{\mathrm{ref}}+U_2^{\mathrm{ref}}$, where $\phi_{\mathrm{SE}}$ is defined in Section~2.3 of the main manuscript. This specification fixes $\lambda_{\mathcal R}=1$ and $(\rho,\xi)=(1,2)$. RQDTR-U uses this prespecified utility for policy learning and cross-validation, with only the regularization parameter $\lambda$ selected by cross-validation. All learned regimes, including efficacy-only and constraint-based regimes, are evaluated using the same efficacy, reference utility, and risk measures.

\section{Three-Stage Simulation}
\label{app:three-stage}
The methodology and theoretical results developed in Sections~2--3 of the main manuscript are stated for a two-stage decision problem, which is the canonical setting in the existing quantile DTR literature \citep{linn2017interactive, wang2018quantile, xia2026successive}. To examine whether the proposed framework remains operational beyond the two-stage setting, we conduct a proof-of-feasibility study under a three-stage data-generating process. The empirical objective admits a natural finite-horizon analogue, in which the two-stage softmax product $p_{1,A_1}(H_1;f_1,h)\,p_{2,A_2}(H_2;f_2,h)$ is replaced by the cumulative product $\prod_{t=1}^{T} p_{t,A_t}(H_t;f_t,h)$, with the corresponding inverse-propensity weights extended to all $T$ stages in the empirical survival estimator $\widehat{S}(q,f)$, and the two-stage block-coordinate updates are replaced by alternating updates over $(f_1,\ldots,f_T)$ and $q$. The data-generating process below adopts a short-memory/Markov-type covariate transition structure, in which the conditional distribution of $X_t$ depends on the past only through the immediately preceding state-action pair $(X_{t-1},A_{t-1})$. However, this does not restrict the quantile-optimal treatment rules to depend only on the current state, as illustrated by the dependence of the stage-2 oracle on the residual threshold $q_{\tau,E}^*-Y_1$. This choice isolates the question of multistage execution from the orthogonal question of long-range history dependence. A complete $T$-stage methodological and theoretical development is left to future work.

We restrict the three-stage comparison to RQDTR-E and the oracle benchmark to illustrate the feasibility of implementing the proposed framework with $T=3$. A broader comparative study at $T\ge3$ would require multistage quantile DTR baselines, which to our knowledge are not yet available in the literature. Such a study is left to future work. In Scenario E, each setting is replicated $500$ times with $n_{\text{train}}=500$ and $n_{\text{test}}=5000$ at quantile levels $\tau\in\{0.10, 0.25,0.50\}$. Implementation choices (propensity model, kernel, tuning, and optimizer) are kept identical to those used in Scenarios~A--D. We report (i) the estimated quantile efficacy value $\widehat{Q}_{\tau,\text{test}}^{Y}(\widehat{d})$; (ii) the mean efficacy value $\widehat{V}_{Y,\text{test}}(\widehat{d})$; and (iii) the stagewise correct-decision rate $n_{\text{test}}^{-1}\sum_i I\{\widehat{d}_t(H_{ti})=d_t^{*}(H_{ti})\}$ for $t=1,2,3$. The first two quantities, together with their oracle counterparts, assess whether the proposed three-stage procedure attains a value close to the oracle benchmark; the third quantity diagnoses whether the estimation quality is stable across stages or concentrated at particular decision points.

\subsection{Scenario E (Three stages)} 
The stage-1 covariates $\bm X_1=(X_{1,1},X_{1,2},X_{1,3})^\top$ follow $N(\bm 0,\bm I_3)$. The stage-1 treatment $A_1\in\{0,1\}$ is assigned according to
\[
\pi_1^*(A_1=1|\bm X_1)=\operatorname{expit}(-0.4+0.4X_{1,1}-0.3X_{1,2}).
\]
For $t=2,3$, the stage-$t$ covariates $\bm X_t=(X_{t,1},X_{t,2})^\top$ are generated by
\[
X_{t,1}=0.5X_{t-1,1}+0.4A_{t-1}+\varepsilon_{t,1},\
X_{t,2}=-0.3X_{t-1,2}+0.3A_{t-1}+\varepsilon_{t,2}.
\]
Given the history $\bm H_t=(\bm X_1,A_1,Y_1,\ldots,\bm X_{t-1},A_{t-1},Y_{t-1},\bm X_t)$, the stage-$t$ treatment $A_t\in\{0,1\}$ is assigned according to
\[
\pi_t^*(A_t=1|\bm H_t)=\operatorname{expit}(-0.3+0.4X_{t,1}-0.3X_{t,2}+0.2A_{t-1}),\ t=2,3.
\]

The total efficacy outcome is $Y=Y_1+Y_2+Y_3$, where
\[
\begin{aligned}
Y_1&=0.5-0.5X_{1,2}+X_{1,3}+A_1(X_{1,1}+0.5X_{1,2})+\varepsilon_1,\\
Y_2&=0.3X_{2,2}+A_2(X_{2,1}-0.3X_{2,2})+\varepsilon_2,\\
Y_3&=0.3X_{3,2}+A_3(X_{3,1}-0.3X_{3,2}+0.2)+\varepsilon_3.
\end{aligned}
\]
Here $\varepsilon_{t,1},\varepsilon_{t,2}\sim N(0,1)$, for $t=2,3$, and $\varepsilon_t\sim N(0,0.5)$, for $t=1,2,3$, are mutually independent and independent of the baseline covariates.

\textit{\textbf{Population quantile oracle:}}
As in Scenarios A and B, the population oracle is defined over all measurable deterministic treatment regimes based on the available stage-specific histories, without restricting them to the score class used for estimation.

Since the stage-3 treatment affects the outcome through $A_3(X_{3,1}-0.3X_{3,2}+0.2)$, the pointwise optimal stage-3 rule is $d_3^{E,*}(\bm H_3)=I\{X_{3,1}-0.3X_{3,2}+0.2>0\}$.
Under this rule, the stage-3 outcome is $Y_3=\max\{0.3X_{3,2},X_{3,1}+0.2\}+\varepsilon_3$.

For the stage-2 treatment $A_2=a$ and covariates $\bm X_2=(u,v)^\top$, define $r_a(u,v)=0.3v+a(u-0.3v)$ and 
\[
m_{a,0}(u,v)=r_a(u,v)-0.09v+0.09a,\
m_{a,1}(u,v)=r_a(u,v)+0.5u+0.4a+0.2.
\]
Under the optimal stage-3 rule, $Y_2+Y_3$ is the maximum of two correlated Gaussian variables with means $m_{a,0}(u,v)$ and $m_{a,1}(u,v)$, variances $1.09$ and $2$, and covariance $1$. Thus, its conditional CDF is
\[
F_{2,a}^E(z| u,v)
=\Phi_2\left(
\frac{z-m_{a,0}(u,v)}{\sqrt{1.09}},
\frac{z-m_{a,1}(u,v)}{\sqrt{2}};
\frac{1}{\sqrt{2.18}}
\right),
\]
where $\Phi_2(\cdot,\cdot;\rho)$ denotes the standard bivariate normal CDF with correlation $\rho$.

For a common population threshold $q$, define
\[
C_{1,a}^E(\bm x_1,q)
=E\left[
\min_{b\in\{0,1\}}F_{2,b}^E(q-Y_1^a|\bm X_2^a)
|\bm X_1=\bm x_1
\right],
\]
where $(Y_1^a,\bm X_2^a)$ are generated under the stage-1 treatment $A_1=a$. The minimum determines the optimal stage-2 treatment conditional on the observed history, while the expectation integrates over the stage-1 outcome and subsequent covariates before the stage-1 decision. The minimum marginal CDF is
\[
C_E(q)=E\left[\min_{a\in\{0,1\}}C_{1,a}^E(\bm X_1,q)\right]
=\inf_d P\{Y(d)\le q\}.
\]
Let $q_{\tau,E}^*$ denote the unique solution to $C_E(q)=\tau$. The marginal quantile-optimal regime $d_\tau^{E,*}=(d_{1,\tau}^{E,*},d_{2,\tau}^{E,*},d_{3,\tau}^{E,*})$ is
\begin{equation*}\label{eq:oracle-E-rule}
\begin{aligned}
d_{1,\tau}^{E,*}(\bm X_1)
&=I\left\{C_{1,1}^E(\bm X_1,q_{\tau,E}^*)<C_{1,0}^E(\bm X_1,q_{\tau,E}^*)\right\},\\
d_{2,\tau}^{E,*}(\bm H_2)
&=I\left\{F_{2,1}^E(q_{\tau,E}^*-Y_1|\bm X_2)
<F_{2,0}^E(q_{\tau,E}^*-Y_1|\bm X_2)\right\},\\
d_{3,\tau}^{E,*}(\bm H_3)
&=d_3^{E,*}(\bm H_3).
\end{aligned}
\end{equation*}
Note that this regime minimizes the marginal CDF at $q_{\tau,E}^*$ and attains $Q_\tau\{Y(d_\tau^{E,*})\}=q_{\tau,E}^*=\sup_d Q_\tau\{Y(d)\}$. Ties are resolved in favor of treatment $0$. The population oracle threshold $q_{\tau,E}^*$ is computed numerically under the known data-generating process and held fixed across simulation replications. Oracle performance is assessed using empirical quantiles from independently generated test samples.


\subsection{Results} 
Table~\ref{tab:scenario_E} shows that RQDTR-E remains operational under $T = 3$, with both the mean efficacy value $V_Y$ and the quantile efficacy value $Q_\tau^Y$ within $0.26$--$0.66$ of the oracle benchmark across all three quantile levels. The stagewise correct-decision rates exhibit a clear monotone decay from stage~1 to stage~3 ($82.85\%$, $75.51\%$ and $70.16\%$ at $\tau = 0.25$, with analogous patterns at $\tau \in \{0.10, 0.50\}$). The lower-tail value $Q_\tau^Y$ at $\tau = 0.10$ shows the largest absolute quantile gap to the oracle, consistent with the additional difficulty of estimating extreme quantiles in finite samples. Overall, this proof-of-feasibility study supports the conceptual extension of RQDTR to three stages while empirically illustrating the technical challenges.

\begin{table}[tb]
\centering
\footnotesize
\caption{Results for Scenario E. Means of $V_{Y}$ and $Q_\tau^Y$ with standard deviation in parentheses and mean of $\mathrm{CDR}_1$, $\mathrm{CDR}_2$, and $\mathrm{CDR}_3$ are summarized at quantile levels $\tau \in \{0.10, 0.25, 0.50\}$. ``CDR'' denotes the stagewise correct-decision rate, defined as $\mathrm{CDR}_t = \tfrac{1}{n}\sum_i I\{\widehat{d}_t(H_{ti})=d_t^{*}(H_{ti})\} \times 100\%$.}
\label{tab:scenario_E}
\begin{tabular}{@{}llccccc@{}}
\toprule
$\tau$ & Method & $V_{Y}$ (SD) $\uparrow$ & $Q_\tau^Y$ (SD) $\uparrow$ & $\mathrm{CDR}_1$ $\uparrow$ & $\mathrm{CDR}_2$ $\uparrow$ & $\mathrm{CDR}_3$ $\uparrow$ \\
\midrule
\multirow{2}{*}{0.10} & Oracle   & 2.423 (0.033) & -0.493 (0.045) & 100.00\% & 100.00\% & 100.00\% \\
& RQDTR-E & 1.759 (0.208) & -1.081 (0.222) & 78.56\%  & 71.52\%  & 66.40\%  \\
\midrule
\multirow{2}{*}{0.25} & Oracle   & 2.426 (0.033) & 0.758 (0.039) & 100.00\% & 100.00\% & 100.00\% \\
& RQDTR-E & 1.927 (0.147) & 0.405 (0.138) & 82.85\%  & 75.51\%  & 70.16\%  \\
\midrule
\multirow{2}{*}{0.50} & Oracle   & 2.425 (0.033) & 2.266 (0.041) & 100.00\% & 100.00\% & 100.00\% \\
& RQDTR-E & 1.971 (0.127) & 2.006 (0.092) & 81.35\%  & 73.76\%  & 70.84\%  \\
\bottomrule
\end{tabular}
\end{table}

This simulation study demonstrates the feasibility of implementing RQDTR beyond the two-stage setting. Extending the theoretical guarantees to general $T\ge3$ requires controlling the accumulation of smoothing approximation and estimation errors across stages, together with the increased complexity of multistage optimization. Establishing these guarantees remains an important direction for future research.



\section{Comparison with Mean-Based Dynamic Treatment Regimes}
Mean DTR methods aim to improve average outcomes across the entire population. In contrast, quantile DTR methods target a prespecified part of the outcome distribution. For example, when $\tau<0.5$, a quantile criterion can prioritize patients in the lower tail of the outcome distribution, who are often those at greatest clinical risk. Thus, treatment recommendations may differ substantially for vulnerable patients even when average efficacy remains similar. This distinction is particularly important in precision medicine, where preventing poor outcomes is often more clinically relevant than maximizing average improvement. In Scenario F, we compare the proposed method with four existing mean-based DTR methods: Q-learning \citep{zhao2009reinforcement}, BOWL \citep{zhao2015new}, augmented-IPW estimator (AIPWE, \citet{zhang2013robust}), and RWL \citep{zhou2017residual}. We evaluate the estimated regimes using the mean efficacy-based value $V_Y$ and quantile efficacy-based value $Q_\tau^Y$ at quantile levels $\tau \in \{0.10, 0.25, 0.50\}$ over 500 independent replications. In each replication, we generate an independent training sample of size $n_{\mathrm{train}}=500$ and evaluate the estimated regime on an independent test set of size $n_{\mathrm{test}}=5000$.

\subsection{Scenario F (Mean DTR Methods vs.\ Quantile DTR Methods)} 
We consider a two-stage binary treatment setting with $A_t\in\{0,1\}$ for $t=1,2$. Let $\bm X_1=(X_{11},X_{12})^\top$ with $X_{11},X_{12}\overset{\mathrm{i.i.d.}}{\sim}N(0,1)$. The stage-1 treatment is generated from
$$A_1 | \bm X_1 \sim \mathrm{Bernoulli}\{\pi_1(\bm X_1)\}, \
\pi_1(\bm X_1)=\operatorname{expit}(0.3X_{11}-0.2X_{12}).$$
The stage-2 covariates are generated as
$$X_{21}=0.6X_{11}+0.4A_1+\eta_1,\
X_{22}=0.6X_{12}-0.3A_1+\eta_2,$$
where $\eta_1,\eta_2\overset{\mathrm{i.i.d.}}{\sim}N(0,1)$. The stage-2 treatment is generated from
$$A_2 | \bm H_2 \sim \mathrm{Bernoulli}\{\pi_2(\bm H_2)\},\ \pi_2(\bm H_2)=\operatorname{expit}(0.2X_{21}-0.3X_{22}+0.2A_1),$$
where $\bm H_2=(\bm X_1,A_1,\bm X_2)$ denotes the observed history available before the stage-2 treatment decision.

To induce heterogeneous treatment response, we define a severity-based subgroup variable $G\in\{1,2,3\}$ as a deterministic function of the observed baseline severity score $S=X_{11}+0.5X_{12}$:
$$G=
\begin{cases}
1, & S\le q_{0.33},\\
2, & q_{0.33}<S\le q_{0.67},\\
3, & S>q_{0.67},
\end{cases}$$
where $q_{0.33}$ and $q_{0.67}$ are the population 33rd and 67th percentiles of $S$, respectively. These severity-defined subgroups induce different stage-2 treatment-effect and outcome-distribution patterns.

The final outcome is generated by $Y=\mu(\bm H_2,A_1,A_2,G)+\varepsilon(A_2,G)$ with
$$\mu(\bm H_2,A_1,A_2,G) = 1+0.5X_{11}+0.5X_{21} +\beta_1(X_{11})A_1 +\beta_2(X_{21},G)A_2,$$
where $\beta_1(X_{11})=0.2+0.5I(X_{11}>0)-0.2I(X_{11}\le 0)$ and
$$\beta_2(X_{21},G)=
\begin{cases}
0.8+0.6I(X_{21}>0), & G=1,\\
0.3+0.3I(X_{21}>0), & G=2,\\
0.9+0.4I(X_{21}>0), & G=3.
\end{cases}$$
The error distribution varies across subgroups and stage-2 treatments:
$$\varepsilon(A_2,1)\sim N(0,1),\
\varepsilon(A_2,2)\sim N(0,1.2^2),$$
$$\varepsilon(0,3)\sim N(0,1), \ \varepsilon(1,3)\sim
0.8\,N(0.8,1^2)+0.2\,N(-5,1.2^2).$$

\textit{\textbf{Population mean and quantile oracles:}}
The population oracles are defined over all measurable deterministic treatment regimes based on the available stage-specific histories. Let $m(\H_2)=1+0.5X_{11}+0.5X_{21}+\beta_1(X_{11})A_1$ and $b(\H_2)=\beta_2(X_{21},G)$. For $a\in\{0,1\}$, define the conditional outcome CDF $F_a(q|\H_2)=P(Y\le q|\H_2,A_2=a)$.

For $G\in\{1,2\}$, with $\sigma_1=1$ and $\sigma_2=1.2$, the conditional CDFs are
\[
F_0(q|\H_2)=\Phi\left(\frac{q-m(\H_2)}{\sigma_G}\right),\
F_1(q|\H_2)=\Phi\left(\frac{q-m(\H_2)-b(\H_2)}{\sigma_G}\right).
\]
For $G=3$, they are $F_0(q|\H_2)=\Phi\{q-m(\H_2)\}$ and
\[
F_1(q|\H_2)=0.8\Phi\{q-m(\H_2)-b(\H_2)-0.8\}
+0.2\Phi\left(\frac{q-m(\H_2)-b(\H_2)+5}{1.2}\right),
\]
where $\Phi$ denotes the standard normal CDF.

Since the conditional mean contrast is $E(Y| \H_2,A_2=1)-E(Y|\H_2,A_2=0)=b(\H_2)-0.36I(G=3)>0$, the mean-optimal stage-2 rule always selects treatment 1. Moreover, under a common coupling of the exogenous disturbances, changing $A_1$ from 0 to 1 increases $X_{21}$ by $0.4$ and $m(\H_2)$ by $0.2+\beta_1(X_{11})>0$, without decreasing $b(\H_2)$. Thus, the optimal stage-1 treatment is 1 for both the mean and marginal quantile objectives, and the population mean-optimal regime is $d_{\mathrm{mean}}^{F,*}=(1,1)$.

For marginal quantile optimization, define
\[
C_F(q)=E\left[\min_{a\in\{0,1\}}F_a(q|\H_2^{(1)})\right],
\]
where $\H_2^{(1)}$ denotes the stage-2 history under $A_1=1$. Let $q_{\tau,F}^*$ be the unique solution to $C_F(q)=\tau$. The marginal quantile-optimal regime $d_\tau^{F,*}=(d_{1,\tau}^{F,*},d_{2,\tau}^{F,*})$ is
\[
d_{1,\tau}^{F,*}(\bm X_1)=1,\
d_{2,\tau}^{F,*}(\H_2)
=I\{F_1(q_{\tau,F}^*|\H_2)<F_0(q_{\tau,F}^*|\H_2)\}.
\]
This regime minimizes the marginal outcome CDF at $q_{\tau,F}^*$ and attains $Q_\tau\{Y(d_\tau^{F,*})\}=q_{\tau,F}^*=\sup_d Q_\tau\{Y(d)\}$. Ties are resolved in favor of treatment 0.

Unlike the mean-optimal rule, the quantile-optimal stage-2 rule may select treatment 0 in subgroup 3. For example, when $G=3$, $A_1=1$, $X_{11}=1$, and $X_{21}=3$, the conditional mean favors treatment 1, whereas at a threshold near the mean-optimal regime's median, $q=2.296$, the conditional CDFs satisfy $F_0(q|\H_2)\approx0.080<F_1(q|\H_2)\approx0.195$. This difference persists over a set of histories with positive probability, establishing that the mean-optimal and median-optimal regimes do not coincide.


\subsection{Results} 
Table~\ref{tab:scenario_F_optimal} summarizes the performance of population-optimal regimes under different criteria in Scenario F, evaluated using a Monte Carlo sample of size $n=10^5$. The mean-optimal regime achieves the highest mean efficacy ($V_Y=2.356$) but exhibits relatively poor lower-tail performance, with $Q^Y_{0.10}=-0.327$ and $Q^Y_{0.25}=0.886$. In contrast, the lower-quantile-optimal regimes sacrifice some mean efficacy but substantially improve lower-tail outcomes. Specifically, the 0.10-quantile-optimal regime increases $Q^Y_{0.10}$ from $-0.327$ to $0.126$, while the 0.25-quantile-optimal regime increases $Q^Y_{0.25}$ from $0.886$ to $1.099$. The median-optimal regime also differs from the mean-optimal regime, achieving a higher median ($2.332$ versus $2.296$) with a modest reduction in mean efficacy ($2.282$ versus $2.356$). These results demonstrate that optimizing different distributional criteria can yield distinct treatment regimes and corresponding trade-offs in efficacy performance.

Table~\ref{tab:scenario_F} shows that the mean DTR methods perform well for the mean and median criteria. For example, Q-learning, the strongest mean DTR method in this setting, achieves the largest estimated mean value ($V_Y=2.343$), and the largest median value ($Q^Y_{0.50}=2.276$). However, these mean-based methods do not improve the lower tail: their estimated $Q^Y_{0.10}$ values remain around $-0.32$ to $-0.45$, and their estimated $Q^Y_{0.25}$ values are below $0.88$. In contrast, RQDTR-E with $\tau=0.25$ achieves the best lower-tail performance, with $Q^Y_{0.10}=-0.082$ and $Q^Y_{0.25}=0.972$. These values are close to the corresponding population lower-quantile-optimal values in Table~\ref{tab:scenario_F_optimal} and are substantially higher than those obtained by the mean-based competitors.

Overall, Scenario F indicates an objective-specific trade-off rather than a universal ranking of mean-based and quantile-based DTR methods. In this deliberately constructed setting, Q-learning attains the largest estimated mean and median values, whereas RQDTR-E attains substantially higher lower-tail values at $Q_{0.10}^Y$ and $Q_{0.25}^Y$. At the population level, the mean-optimal regime differs from the quantile-optimal regimes at all three quantile levels. These results demonstrate, within Scenario F, that optimizing average performance can lead to a regime with inferior lower-tail efficacy when treatment effects are heterogeneous across severity-defined subgroups. They support choosing the optimization criterion according to the scientific target, but do not imply that one criterion is uniformly preferable across applications.

\begin{table}[tb]
\centering
\caption{Results for optimal dynamic treatment regimes under different criteria in Scenario F, evaluated using a Monte Carlo sample of size $n=10^5$. Columns 2--5 report the mean and the 0.10, 0.25, and 0.50 quantiles of the potential efficacy outcomes under each regime.}
\label{tab:scenario_F_optimal}
\footnotesize
\setlength{\tabcolsep}{8pt}
\begin{tabular}{@{}lcccc@{}}
\toprule 
  & $V_Y$ & $Q^Y_{0.10}$ & $Q^Y_{0.25}$ & $Q^Y_{0.50}$\\ 
\midrule 
Mean criterion & 2.356 & $-0.327$ & 0.886 & 2.296 \\
0.10-quantile criterion & 2.072 & 0.126 & 1.067 & 2.111 \\
0.25-quantile criterion & 2.102 & 0.078 & 1.099 & 2.180 \\
0.50-quantile criterion & 2.282 & $-0.251$ & 0.959 & 2.332 \\
\bottomrule
\end{tabular}
\end{table}

\begin{table}[tb]
\centering
\footnotesize
\setlength{\tabcolsep}{3pt}
\caption{Results for Scenario F. Evaluation metrics include the mean efficacy-based value $V_Y$ and quantile efficacy-based value $Q_\tau^Y$ for $\tau \in \{0.10, 0.25, 0.50\}$. Means of all metrics, with standard deviations in parentheses, are summarized over 500 independent replications. The best result for each metric is highlighted in bold.}
\label{tab:scenario_F}
\begin{tabular}{@{}lcccc@{}}
\toprule
Method & $V_Y$ (SD) $\uparrow$ & $Q_{0.1}^Y$ (SD) $\uparrow$ & $Q_{0.25}^Y$ (SD) $\uparrow$ & $Q_{0.5}^Y$ (SD) $\uparrow$ \\
\midrule
\multicolumn{5}{l}{\textit{Mean DTRs}} \\
\addlinespace[2pt]
Q-learning & \textbf{2.343 (0.053)} & -0.325 (0.066) & 0.876 (0.048) & \textbf{2.276 (0.056)} \\
BOWL       & 2.323 (0.090) & -0.320 (0.077) & 0.866 (0.053) & 2.267 (0.078) \\
AIPWE      & 2.025 (0.068) & -0.452 (0.093) & 0.668 (0.078) & 1.955 (0.079) \\
RWL        & 2.338 (0.047) & -0.329 (0.065) & 0.868 (0.060) & 2.270 (0.061) \\
\midrule
\multicolumn{5}{l}{\textit{Quantile DTRs}} \\
\addlinespace[2pt]
RQDTR-E ($\tau=0.1$)  & 1.899 (0.143) & -0.090 (0.126) & 0.874 (0.132) & 1.934 (0.156) \\
RQDTR-E ($\tau=0.25$) & 2.062 (0.095) & \textbf{-0.082 (0.108)} & \textbf{0.972 (0.087)} & 2.116 (0.106) \\
RQDTR-E ($\tau=0.5$)  & 2.225 (0.063) & -0.223 (0.094) & 0.944 (0.068) & 2.256 (0.063) \\
\bottomrule
\end{tabular}
\end{table}

\section{Sensitivity Analysis for Smoothing Parameters}
\label{app:sensitivity_h_delta}
\begin{figure}[ht]
\centering
\includegraphics[width=\linewidth]{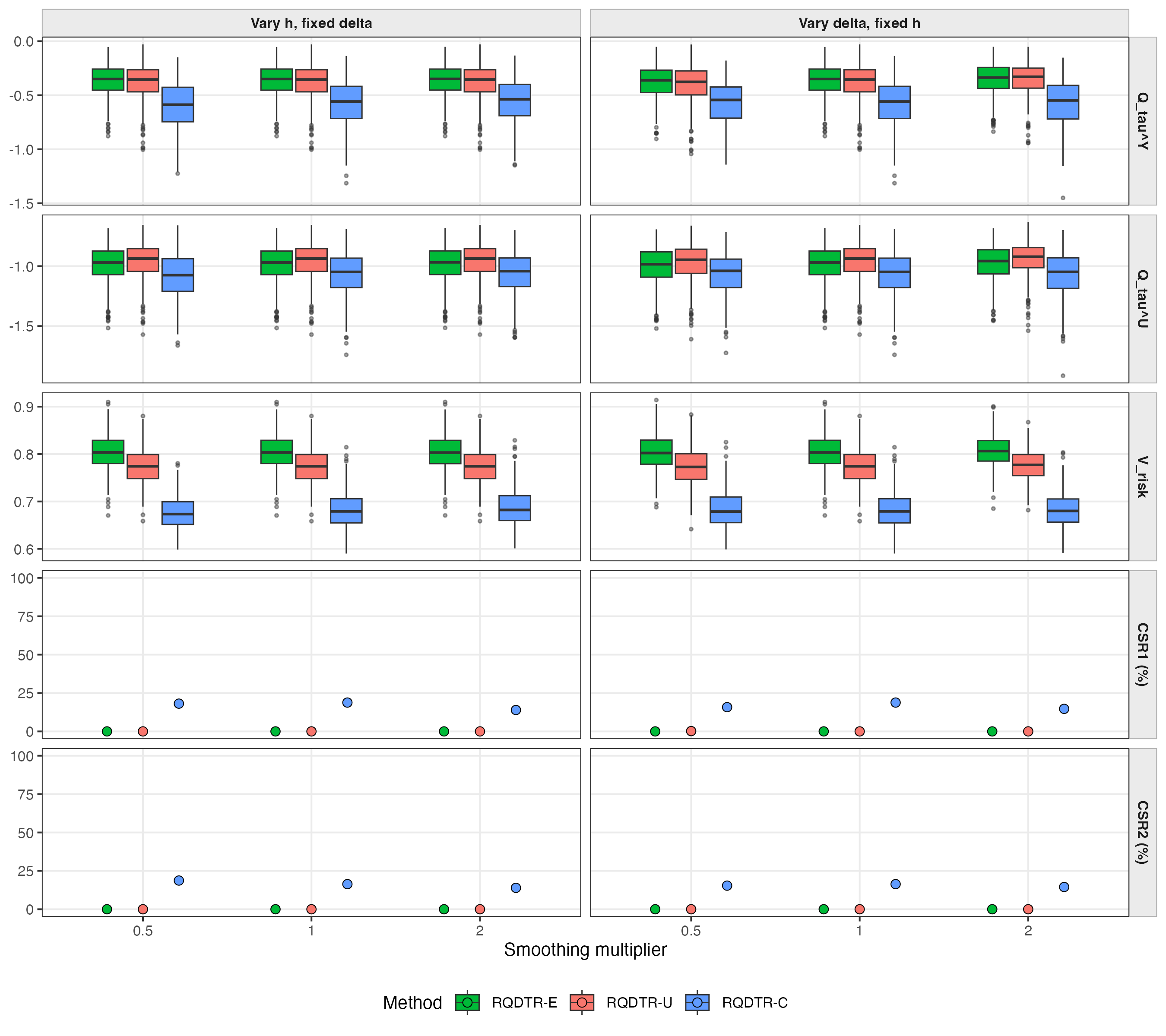}
\caption{Sensitivity of RQDTR variants to smoothing parameters $(h, \delta)$ in Scenario~C. Left column: vary $h$ at fixed $\delta = 0.1 \, \mathrm{SD}(U)$. Right column: vary $\delta$ at fixed $h = 0.2/\log n$. Rows from top to bottom: $Q_\tau^Y$, $Q_\tau^U$, $V_{\mathrm{risk}}$, $\widehat{\mathrm{CSR}}_1$, $\widehat{\mathrm{CSR}}_2$.}
\label{fig:sensitivity_boxplots_h_delta}
\end{figure}

The proposed RQDTR framework involves two smoothing parameters: the softmax bandwidth $h$ for approximating the treatment-assignment indicator and the logistic bandwidth $\delta$ for approximating the tail indicator. In Sections~4--5 of the main manuscript, we use the default choices $h=0.2/\log n$ for all RQDTR variants and $\delta=0.1\,\mathrm{SD}(Y)$ for RQDTR-E/C or $\delta=0.1\,\mathrm{SD}(U)$ for RQDTR-U. To assess the impact of these choices on efficacy, utility, and risk outcomes, we conduct a sensitivity analysis in Scenario C of the simulation study. We vary $h$ over the grid $\{c_h(0.2/\log n): c_h\in\{0.5,1,2\}\}$ and $\delta$ over the grid $\{c_\delta(0.1\,\mathrm{SD}(Y)): c_\delta\in\{0.5,1,2\}\}$ for RQDTR-E/C or $\{c_\delta(0.1\,\mathrm{SD}(U)): c_\delta\in\{0.5,1,2\}\}$ for RQDTR-U. For each setting and each of the three RQDTR variants, we report the efficacy-based value $Q_\tau^Y$, utility-based value $Q_\tau^U$, average observed risk burden $V_{\mathrm{risk}}$, and constraint satisfaction rate $\mathrm{CSR}_m=\frac{1}{|\mathcal I_{\mathrm{valid}}|}\sum_{r\in\mathcal I_{\mathrm{valid}}} I\{\widehat{\mathcal B}_m(\hat d^{(r)})\le b_m^{(r)}\}$, $m=1,\ldots,M$, over 500 replications with training sample size $n_{\mathrm{train}}=500$ and testing sample size $n_{\mathrm{test}}=5000$ at quantile level $\tau=0.25$. Here $\mathcal I_{\mathrm{valid}}$ denotes the set of valid replications, $\widehat{\mathcal B}_{m,\mathrm{test}}^{(r)}(\hat d^{(r)})$ is the attribute-specific risk burden estimated by direct Monte Carlo averaging on an independent test sample generated under the learned regime, and $b_m^{(r)}=0.85\,\mathcal B_m^{\mathrm{obs},(r)}$ is the corresponding risk tolerance calculated from the training sample. Table~\ref{tab:sensitivity_vary_delta} summarizes the results when varying $\delta$ with $c_h=1$, and Table~\ref{tab:sensitivity_vary_h} summarizes the results when varying $h$ with $c_\delta=1$. Figure~\ref{fig:sensitivity_boxplots_h_delta} provides a visual summary of the sensitivity analysis across all metrics and all three RQDTR variants.

\begin{table}[tb]
\centering
\caption{Sensitivity to $\delta$: results for $c_h = 1$ and varying $c_\delta \in \{0.5, 1.0, 2.0\}$.}
\label{tab:sensitivity_vary_delta}
\footnotesize
\begin{tabular}{@{}ccccccc@{}}
\toprule
$c_\delta$ & Method & $Q_\tau^Y$ (SD) $\uparrow$ & $Q_\tau^U$ (SD) $\uparrow$ & $V_{\text{risk}}$ (SD) $\downarrow$ & CSR1 $\uparrow$ & CSR2 $\uparrow$ \\
\midrule
\multirow{3}{*}{0.5} & RQDTR-E & -0.383 (0.153) & -1.004 (0.165) & 0.803 (0.038) & 0.00\% & 0.00\% \\
 & RQDTR-U & -0.393 (0.169) & -0.968 (0.154) & 0.774 (0.038) & 0.20\% & 0.00\% \\
 & RQDTR-C & -0.571 (0.202) & -1.068 (0.176) & 0.683 (0.038) & 15.79\% & 15.38\% \\
\midrule
\multirow{3}{*}{1.0} & RQDTR-E & -0.368 (0.147) & -0.990 (0.159) & 0.804 (0.036) & 0.00\% & 0.00\% \\
 & RQDTR-U & -0.378 (0.159) & -0.955 (0.145) & 0.774 (0.036) & 0.00\% & 0.00\% \\
 & RQDTR-C & -0.576 (0.215) & -1.071 (0.186) & 0.681 (0.037) & 18.79\% & 16.36\% \\
\midrule
\multirow{3}{*}{2.0} & RQDTR-E & -0.352 (0.140) & -0.977 (0.151) & 0.807 (0.032) & 0.00\% & 0.00\% \\
 & RQDTR-U & -0.353 (0.148) & -0.937 (0.138) & 0.777 (0.033) & 0.00\% & 0.00\% \\
 & RQDTR-C & -0.575 (0.210) & -1.069 (0.183) & 0.682 (0.036) & 14.69\% & 14.49\% \\
\bottomrule
\end{tabular}
\end{table}

\begin{table}[tb]
\centering
\caption{Sensitivity to $h$: results for $c_\delta = 1$ and varying $c_h \in \{0.5, 1.0, 2.0\}$.}
\label{tab:sensitivity_vary_h}
\footnotesize
\begin{tabular}{@{}ccccccc@{}}
\toprule
$c_h$ & Method & $Q_\tau^Y$ (SD) $\uparrow$ & $Q_\tau^U$ (SD) $\uparrow$ & $V_{\text{risk}}$ (SD) $\downarrow$ & CSR1 $\uparrow$ & CSR2 $\uparrow$ \\
\midrule
\multirow{3}{*}{0.5} & RQDTR-E & -0.369 (0.147) & -0.990 (0.160) & 0.804 (0.036) & 0.00\% & 0.00\% \\
 & RQDTR-U & -0.378 (0.159) & -0.955 (0.145) & 0.774 (0.036) & 0.00\% & 0.00\% \\
 & RQDTR-C & -0.594 (0.210) & -1.084 (0.183) & 0.677 (0.036) & 18.09\% & 18.70\% \\
\midrule
\multirow{3}{*}{1.0} & RQDTR-E & -0.368 (0.147) & -0.990 (0.159) & 0.804 (0.036) & 0.00\% & 0.00\% \\
 & RQDTR-U & -0.378 (0.159) & -0.955 (0.145) & 0.774 (0.036) & 0.00\% & 0.00\% \\
 & RQDTR-C & -0.576 (0.215) & -1.071 (0.186) & 0.681 (0.037) & 18.79\% & 16.36\% \\
\midrule
\multirow{3}{*}{2.0} & RQDTR-E & -0.368 (0.147) & -0.989 (0.159) & 0.804 (0.036) & 0.00\% & 0.00\% \\
 & RQDTR-U & -0.378 (0.159) & -0.955 (0.145) & 0.774 (0.036) & 0.00\% & 0.00\% \\
 & RQDTR-C & -0.566 (0.218) & -1.067 (0.188) & 0.686 (0.039) & 13.93\% & 13.93\% \\
\bottomrule
\end{tabular}
\end{table}

Results show that efficacy and utility outcomes are relatively stable across the smoothing parameter settings examined. Across all settings and all three RQDTR variants, the efficacy quantile $Q_\tau^Y$ varies within $6.61\%$ of its default-parameter value, and the utility quantile $Q_\tau^U$ varies within $1.88\%$. The observed risk burden $V_{\mathrm{risk}}$ is even more stable, varying by less than $0.73\%$ from its default-parameter value. The constraint-satisfaction rates of RQDTR-C exhibit greater sensitivity: $\widehat{\mathrm{CSR}}_1$ ranges from $13.93\%$ to $18.79\%$, and $\widehat{\mathrm{CSR}}_2$ ranges from $13.93\%$ to $18.70\%$. Nevertheless, both rates remain above the largest CSR reported for the competing methods in Scenario~C at $\tau=0.25$ in the main manuscript ($5.64\%$ for SCL). This analysis confirms that the proposed framework is stable to the specific choice of $(h, \delta)$ within practical ranges, supporting the use of the default values throughout the main manuscript.

\section{Sensitivity Analysis for Tolerance Level $b_m$}
\begin{figure}[tb]
    \centering
    \includegraphics[width= \linewidth]{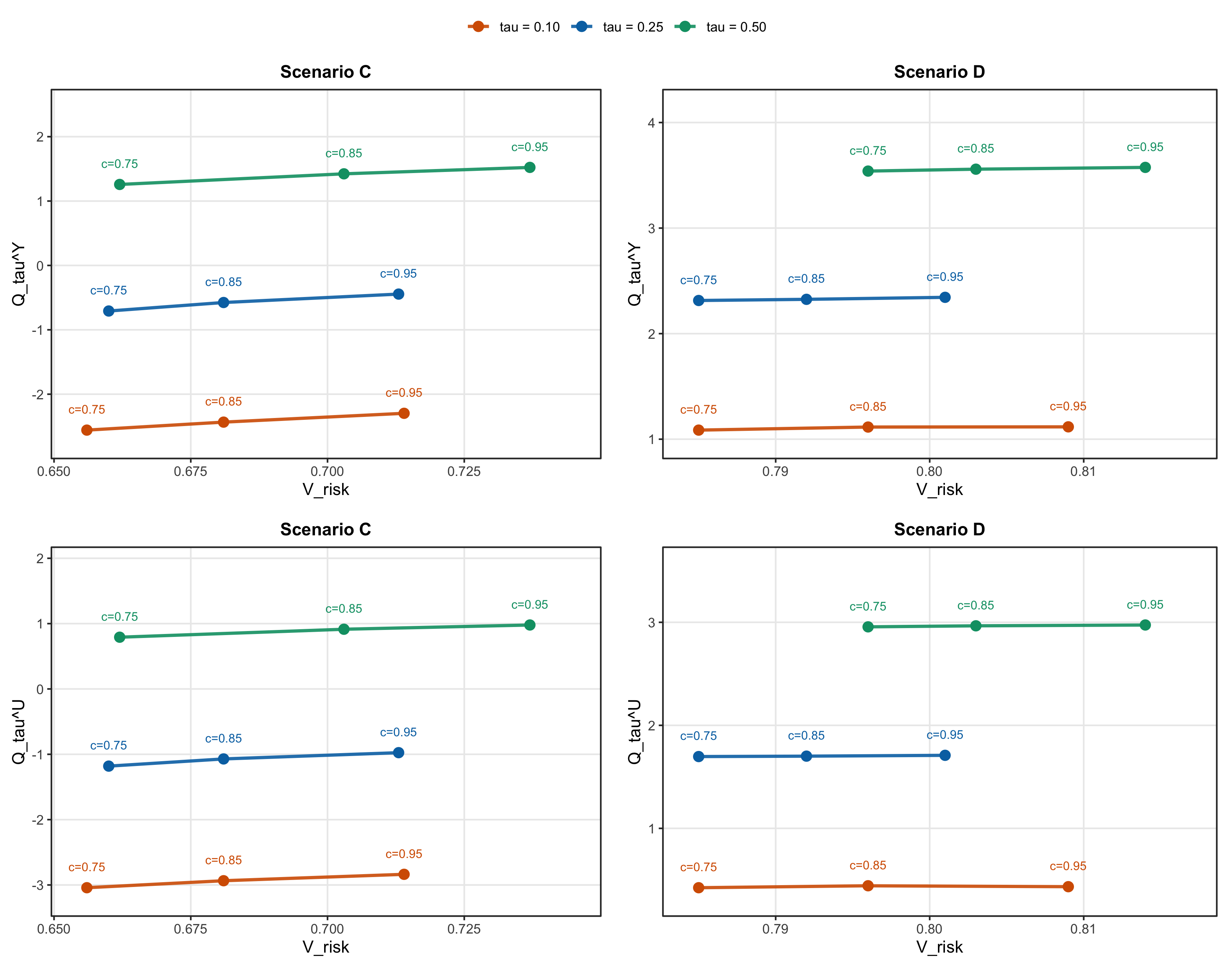}
    \caption{Tolerance-level sensitivity of RQDTR-C at $\tau\in\{0.1,0.25,0.5\}$. Each point reports the mean of the efficacy-based quantile value $Q_\tau^Y$ or the utility-based quantile value $Q_\tau^U$ over 500 independent replications and corresponds to a value of $c\in\{0.75,0.85,0.95\}$ in $b_m=c\,\mathcal{B}_m^{\mathrm{obs}}$.}
    \label{fg:sensitivity_RQDTR-C_all_tau}
\end{figure}

To investigate the role of the risk tolerance level, we consider $b_m=c\,\mathcal{B}_m^{\mathrm{obs}}$ with
$c\in\{0.75,0.85,0.95\}$, where a larger value of $c$ corresponds to a less stringent population-level risk constraint. Figure~\ref{fg:sensitivity_RQDTR-C_all_tau} reports the performance of RQDTR-C at $\tau\in\{0.1,0.25,0.5\}$ in both Scenarios C and D. Results are summarized over 500 independent replications, each with a training sample size $n_{\mathrm{train}}=500$ and testing sample size $n_{\mathrm{test}}=5000$.

In Scenario C, the qualitative pattern is consistent across all three quantile levels. As $c$ increases, the observed risk burden $V_{\mathrm{risk}}$ increases, while both the efficacy quantile value $Q_\tau^Y$ and the utility quantile value $Q_\tau^U$ improve. Thus, relaxing the risk constraint permits additional efficacy and utility gains at the expense of greater population-level risk. A similar but much weaker pattern is observed in Scenario D. Increasing $c$ leads to a higher observed risk burden, whereas the corresponding changes in $Q_\tau^Y$ and $Q_\tau^U$ are comparatively small across the three quantile levels. In particular, the performance gains beyond $c=0.85$ are negligible or nearly flat, despite the further increase in risk. Hence, $c=0.85$ provides a reasonable compromise in this scenario, as relaxing the constraint further yields little additional efficacy or utility benefit. 

\section{Sensitivity Analysis for Utility Specifications}
\label{app:utility_preferences}
We examine the sensitivity of RQDTR-U to the utility parameters $(\rho,\xi)\in \{0.5,1,1.5\} \times \{0.5,1,2\}$ in Scenario C at $\tau=0.25$, holding $\lambda_{\mathcal R}=1$ fixed. Since $\lambda_{\mathcal R}$ and $\rho$ enter the utility through their product $\lambda_{\mathcal R}\rho$, varying $\rho$ with $\lambda_{\mathcal R}=1$ also captures changes in the overall benefit-risk trade-off weight. We therefore fix $\lambda_{\mathcal R}=1$ and vary $(\rho,\xi)$ to examine sensitivity to the magnitude and curvature of the risk penalty. For each specification, RQDTR-U is refitted with the utility held fixed, and only $\lambda$ is selected by 5-fold cross-validation using the validation IPW quantile under that specification. To ensure comparability across utility specifications, all nine resulting regimes are evaluated under a common reference utility, $U^{\mathrm{ref}}=\sum_{t=1}^2\{Y_t-\phi_{\mathrm{SE}}(\mathcal R_t;1,2)\}$. Results are summarized over 500 independent replications, each with a training sample size $n_{\mathrm{train}}=500$ and testing sample size $n_{\mathrm{test}}=5000$.

Table~\ref{tab:utility_preference_sensitivity} shows that increasing $\rho$ leads to lower risk burdens at the expense of efficacy. For example, at $\xi=2$, increasing $\rho$ from $0.5$ to $1.5$ reduces the mean $V_{\mathrm{risk}}$ from $0.790$ to $0.760$, but $Q_\tau^Y$ decreases from $-0.368$ to $-0.402$. In contrast, varying $\xi$ has relatively little effect on the evaluated outcomes. The mean reference-utility quantile remains within a narrow range of $-0.966$ to $-0.951$ across all nine specifications, with $-0.955$ under the reference specification $(1,2)$. These results indicate that the risk-penalty magnitude primarily governs the efficacy-risk trade-off, while the reference-utility performance remains relatively stable across the examined preferences.

\begin{table}[tb]
\centering
\caption{Sensitivity of RQDTR-U to utility specifications used for policy learning in Scenario C at $\tau=0.25$. Means of the efficacy-based quantile $Q_\tau^Y$, reference-utility quantile $Q_\tau^U$, and average risk burden $V_{\mathrm{risk}}$, with standard deviations in parentheses, are summarized over 500 independent replications. All regimes are evaluated using the reference utility with $\lambda_{\mathcal R}=1$ and $(\rho,\xi)=(1,2)$.}
\label{tab:utility_preference_sensitivity}
\footnotesize
\begin{tabular}{@{}ccccc@{}}
\toprule
$\rho$ & $\xi$ & $Q_\tau^Y$ (SD) $\uparrow$ & $Q_\tau^U$ (SD) $\uparrow$ & $V_{\mathrm{risk}}$ (SD) $\downarrow$ \\
\midrule
0.5 & 0.5 & -0.368 (0.153) & -0.964 (0.150) & 0.788 (0.035) \\
0.5 & 1.0 & -0.368 (0.153) & -0.965 (0.150) & 0.789 (0.035) \\
0.5 & 2.0 & -0.368 (0.153) & -0.966 (0.151) & 0.790 (0.035) \\
\midrule
1.0 & 0.5 & -0.378 (0.158) & -0.951 (0.143) & 0.773 (0.035) \\
1.0 & 1.0 & -0.377 (0.160) & -0.952 (0.145) & 0.773 (0.036) \\
1.0 & 2.0 & -0.378 (0.159) & -0.955 (0.145) & 0.774 (0.036) \\
\midrule
1.5 & 0.5 & -0.407 (0.177) & -0.957 (0.149) & 0.756 (0.036) \\
1.5 & 1.0 & -0.405 (0.176) & -0.957 (0.149) & 0.757 (0.036) \\
1.5 & 2.0 & -0.402 (0.174) & -0.958 (0.149) & 0.760 (0.037) \\
\bottomrule
\end{tabular}
\end{table}

\section{Sensitivity Analysis for Training Sample Size $n_{\mathrm{train}}$}
\begin{figure}[tb]
    \centering
    \includegraphics[width= \linewidth]{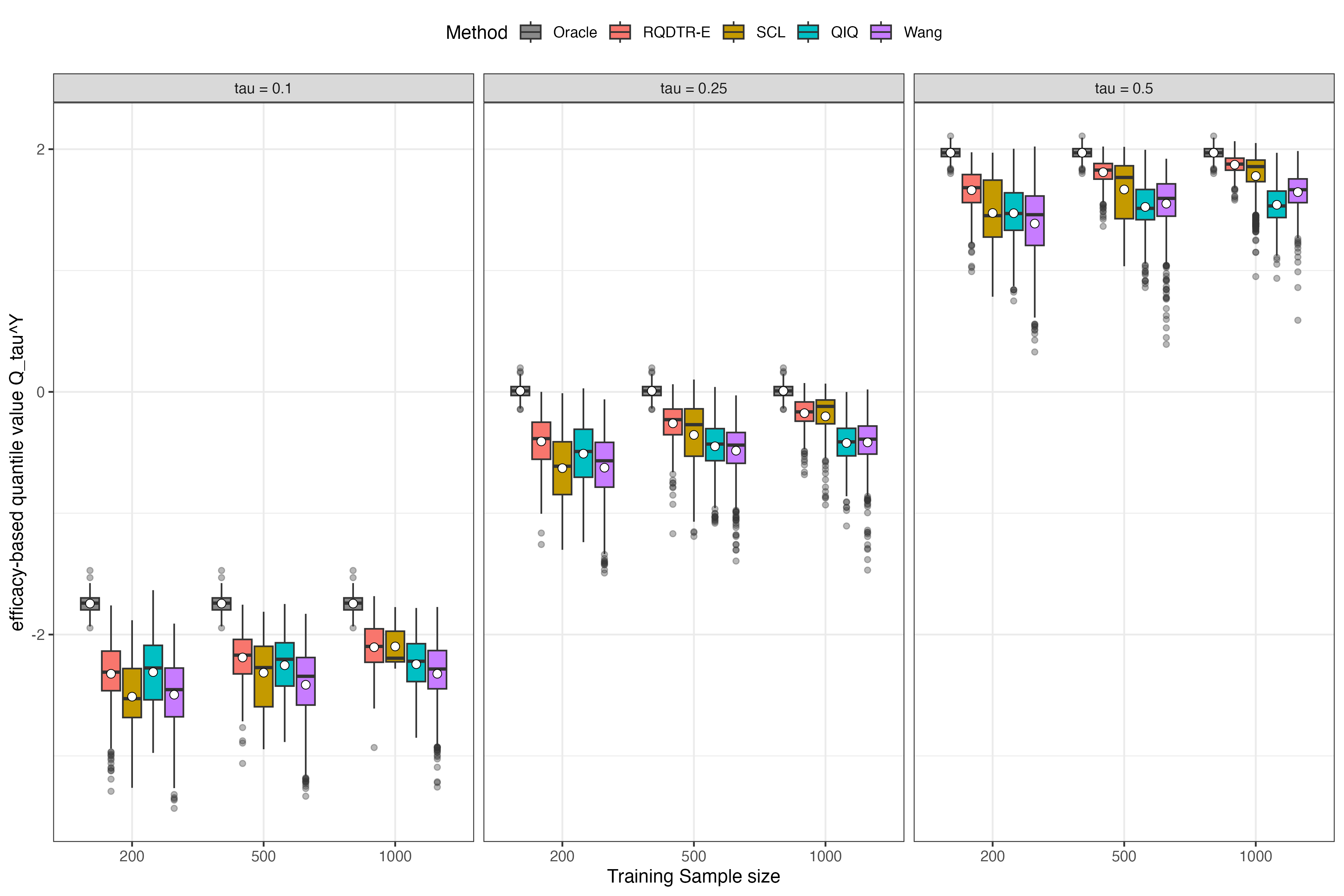}
    \caption{Sensitivity to the training sample size in Scenario A. Boxplots show the efficacy quantile value $Q_\tau^Y$ for $\tau\in\{0.1,0.25,0.5\}$ and $n_{\mathrm{train}}\in\{200,500,1000\}$, based on 500 independent replications. Results are reported for the oracle regime, RQDTR-E, SCL, QIQ, and Wang's method. The white point in each box denotes the replication mean.}
    \label{fg:sensitivity_scenarioA_sample size}
\end{figure}

Figure~\ref{fg:sensitivity_scenarioA_sample size} examines the finite-sample performance of the efficacy-oriented methods as the training sample size increases from $200$ to $1000$ in Scenario A. Across all three quantile levels, increasing $n_{\mathrm{train}}$ generally shifts the estimated quantile values toward the oracle benchmark and reduces their between-replication variability. This pattern is expected because larger training samples provide more information for estimating lower-tail treatment heterogeneity. When $\tau=0.25$ and $0.5$, RQDTR-E attains the highest quantile value among all competing methods at each sample size. At the more challenging lower-tail level $\tau=0.1$, all methods exhibit a larger gap from the oracle and greater dispersion, reflecting the increased difficulty of learning a treatment regime from relatively limited information in the lower tail. However, RQDTR-E remains competitive across all sample sizes: it outperforms or is at least comparable to SCL, QIQ-learning, and Wang's method.

\begin{figure}[tb]
    \centering
    \includegraphics[width= \linewidth]{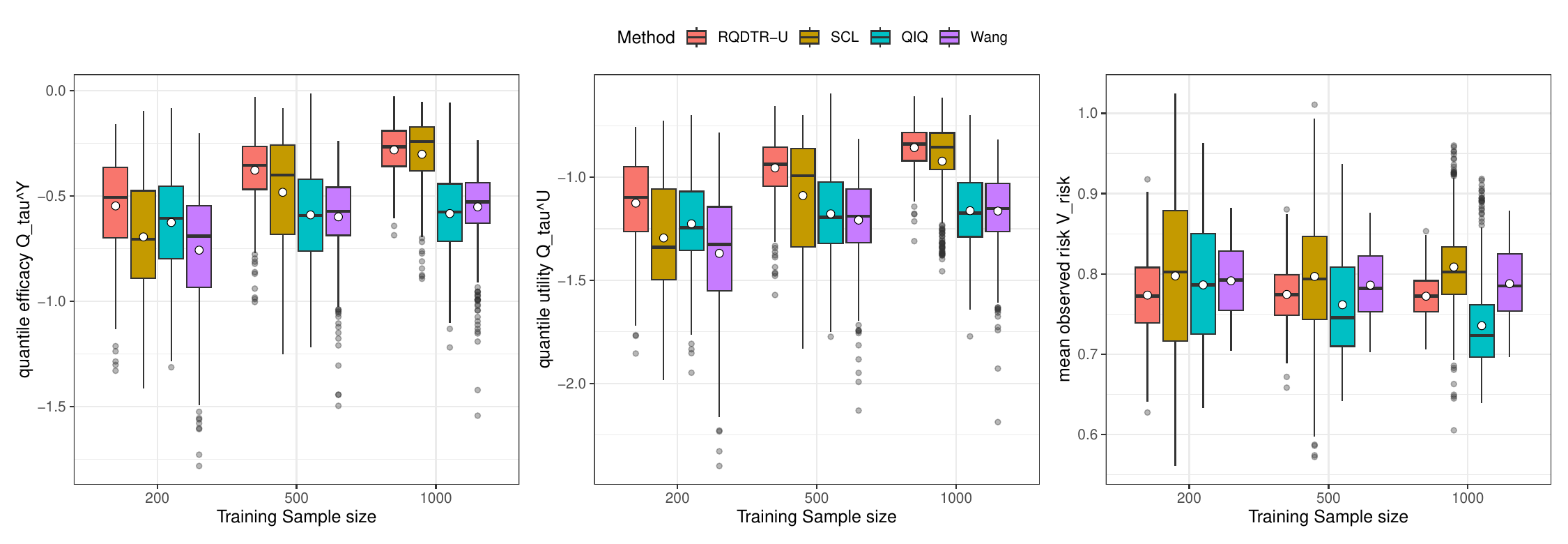}
    \caption{Sample-size sensitivity in Scenario C at $\tau=0.25$ for $n_{\mathrm{train}} \in \{200,500,1000\}$. Boxplots of the efficacy-based value $Q_\tau^Y$, utility-based value $Q_\tau^U$, and average observed risk burden $V_{\mathrm{risk}}$ for RQDTR-U, SCL, QIQ-learning, and Wang's method. The white dot in each box indicates the mean over 500 independent replications.}
    \label{fg:scenarioC_samplesize_boxplot}
\end{figure}

In Scenario C, we compare the performance of three RQDTR variants with three existing quantile DTR methods. To avoid overcrowding the figure, we focus on RQDTR-U and the three existing quantile DTR competitors. 
Figure~\ref{fg:scenarioC_samplesize_boxplot} presents the sample-size sensitivity results at $\tau=0.25$ for training sample size $n_{\mathrm{train}}\in\{200,500,1000\}$. As $n_{\mathrm{train}}$ increases, the between-replication variability generally decreases and the efficacy- and utility-based quantile values of all methods tend to improve. Across all three sample sizes, RQDTR-U consistently attains the largest $Q_\tau^Y$ and $Q_\tau^U$. Its advantages are most pronounced at the smaller training sample sizes, where the competing methods exhibit both greater dispersion and lower central values. At $n_{\mathrm{train}}=1000$, the efficacy performance of SCL becomes closer to that of RQDTR-U, but RQDTR-U continues to attain a higher utility-based quantile value. Although QIQ-learning attains a lower observed risk burden than RQDTR-U when the training sample size is large, its $Q_\tau^Y$ and $Q_\tau^U$ remain substantially lower. Overall, RQDTR-U consistently achieves higher efficacy- and utility-based quantile values across all three sample sizes while maintaining a competitive risk burden, demonstrating a favorable benefit-risk trade-off.

\section{Application to Major Depressive Disorder Data}

\subsection{Variable Contributions to the Learned RQDTR-U Regime}
\label{sc:Variable Contribution_MDD}
To further interpret the treatment regimes learned in the MDD application, we conduct a descriptive analysis of the fitted RQDTR-U regimes. Since all real-data analyses used linear angle-based score functions, the fitted stage-$t$ score function can be written as $f_t(\H_t)=\{f_{t,1}(\H_t),\ldots,f_{t,K-1}(\H_t)\}^{\top}$, where $f_{t,r}(\H_t)=\beta_{t,r}^{\top}\H_t$ for $r=1,\ldots,K-1$. For the $j$th covariate at stage $t$, let $\hat\beta_{t,\cdot j} = (\hat\beta_{t,1j},\ldots,\hat\beta_{t,K-1,j})^{\top}$ denote its estimated coefficient vector across the $K-1$ angle-based score dimensions. We define the covariate contribution score as $\mathrm{CS}_{t,j} = \lVert\hat\beta_{t,\cdot j}\rVert_2 (\frac{1}{n}\sum_{i=1}^n |\H_{t,ij}|),$ where $\H_{t,ij}$ is the value of the $j$th covariate in the stage-$t$ history for patient $i$. This score measures the average magnitude of the contribution of covariate $j$ to the fitted angle-based treatment-score vector. Thus, larger values indicate that the covariate plays a stronger role in determining the treatment recommendation under the learned regime. For each fitted regime, we rank covariates according to $\mathrm{CS}_{t,j}$ within each treatment stage. Note that this measure is descriptive and should not be interpreted as a causal effect of the covariate.

\begin{figure}
\centering
\includegraphics[width=\linewidth]{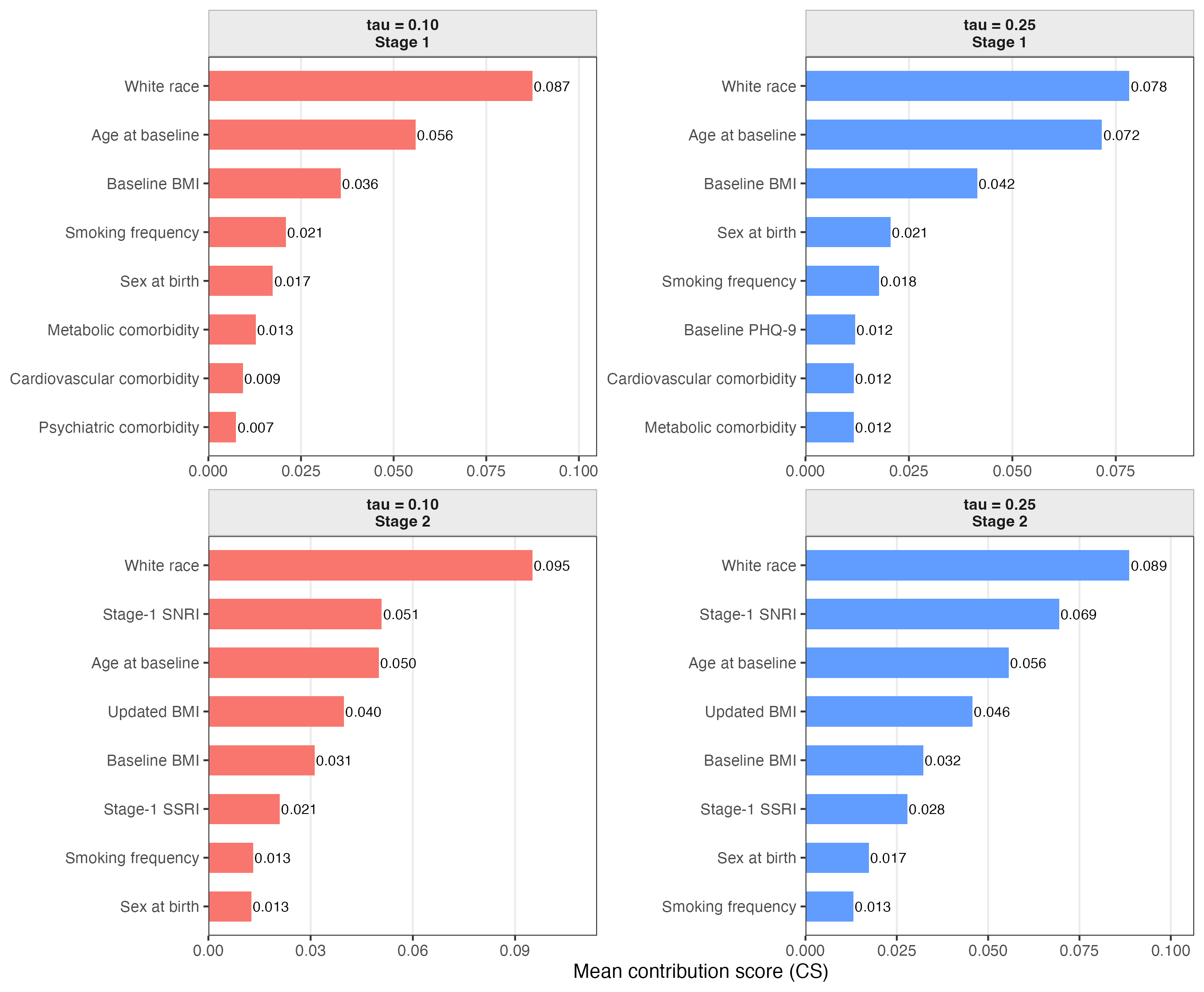}
\caption{Top covariates contributing to the learned RQDTR-U regime in the MDD application. ``CS'' is defined as the product of the Euclidean norm of the estimated angle-based score coefficients and the average absolute covariate value in the training sample.}
\label{fig:mdd_variable_contribution}
\end{figure}

Figure~\ref{fig:mdd_variable_contribution} shows the mean contribution score obtained from 100 independent random splits. Across the two quantile levels and the two stages, white-race indicator, age at baseline, and baseline BMI appeared among the top-ranked covariates for the learned treatment rules. Smoking frequency, sex at birth, and metabolic or cardiovascular comorbidity also contributed to the stage-1 rule, suggesting that baseline health status and comorbidity burden influenced the initial antidepressant recommendation. In the stage-2 rule, baseline characteristics remained important. At the same time, the prior stage-1 treatment indicators, especially SNRI and SSRI, also ranked highly, indicating that the learned dynamic regime incorporated treatment history when updating stage-2 recommendations. Overall, these patterns indicate that the learned RQDTR-U regime used demographic information, baseline clinical burden, BMI, and prior antidepressant exposure when assigning individualized antidepressant treatment.

\subsection{Counterfactual Policy Evaluation in the Test Set}
\label{app:mdd_counterfactual_evaluation}

For each random split of the MDD data, a candidate regime $\widehat d=(\widehat d_1,\widehat d_2)$ is learned from the training set and evaluated on the held-out test set using inverse-probability weighting (IPW). Define the treatment-path matching indicator $I_i(\widehat d)=I\{A_{1i}=\widehat d_1(\bm H_{1i})\}I\{A_{2i}=\widehat d_2(\bm H_{2i})\}$ and the corresponding IPW weight
\[
W_i(\widehat d)=\frac{I_i(\widehat d)}{\widetilde\pi_1(A_{1i}|\bm H_{1i})\widetilde\pi_2(A_{2i}|\bm H_{2i})},
\]
where $\widetilde\pi_t$ denotes the estimated and clipped propensity score at stage $t$.

Let $Y_i=Y_{1i}+Y_{2i}$ denote cumulative efficacy. For all methods, we use a common reference utility $U_i^{\mathrm{ref}}=\sum_{t=1}^2\{Y_{ti}-\phi_{\mathrm{SE}}(\mathcal R_{ti};1,2)\}$, corresponding to $\lambda_{\mathcal R}=1$ and $(\rho,\xi)=(1,2)$, with equal weights $\omega_m=1$ for all risk attributes. For $G\in\{Y,U^{\mathrm{ref}}\}$, the test-set quantile is estimated by intercept-only IPW-weighted quantile regression:
\[
\widehat Q_{\tau,\mathrm{test}}^G(\widehat d)
\in\argmin_{q\in\mathbb R}
\sum_{i=1}^{n_{\mathrm{test}}}
W_i(\widehat d)\ell_\tau(G_i-q),
\]
where $\ell_\tau(u)=u\{\tau-I(u<0)\}$ is the quantile check-loss function. Normalizing the IPW weights does not change the minimizer.

The average risk burden is estimated using the self-normalized IPW mean:
\[
\widehat V_{\mathrm{risk,test}}(\widehat d)
=\frac{\sum_{i=1}^{n_{\mathrm{test}}}W_i(\widehat d)(\mathcal R_{1i}+\mathcal R_{2i})}{\sum_{i=1}^{n_{\mathrm{test}}}W_i(\widehat d)}.
\]
These estimators follow from the population-level IPW identification result in Theorem~\ref{thm:Thm1}(i) under Assumptions~1--3 of the main manuscript.

\subsection{Ties and Target Quantile Definition in the Discrete Utility}
\label{app:mdd_ties}

In the major depressive disorder application, the PHQ-9-based efficacy outcome is discrete and the adverse-burden indicators are binary. Consequently, the empirical distributions of the cumulative efficacy outcome $Y$ and the constructed utility $U$ may contain ties. The population target follows the quantile definition in Section~2.3 of the main manuscript:
\[
Q_\tau\{G(d)\}=\sup\{q:\mathbb P(G(d)>q)\ge1-\tau\},
\]
where $G(d)$ denotes a scalar counterfactual outcome. 
In finite-sample evaluation, quantiles are computed by the IPW-weighted check-loss minimization described in Section~S.13.2. When the cumulative normalized weight equals $\tau$ at an observed outcome value, the minimizer may be nonunique. In such cases, the numerical solution returned by \texttt{quantreg::rq()} is used without explicitly enforcing the upper-endpoint convention.

The asymptotic results in Section~S.4 impose additional distributional regularity conditions, including Assumption~\ref{ass:boundary}(ii), which need not hold for discrete outcomes.



\section{Application to MIMIC-III Sepsis Data}
\subsection{Variable Contributions to the Learned RQDTR-C Regime}

In the MIMIC-III sepsis application, we conduct the same descriptive analysis on the fitted RQDTR-C regimes as in Section~\ref{sc:Variable Contribution_MDD}. 

\begin{figure}
\centering
\includegraphics[width=\linewidth]{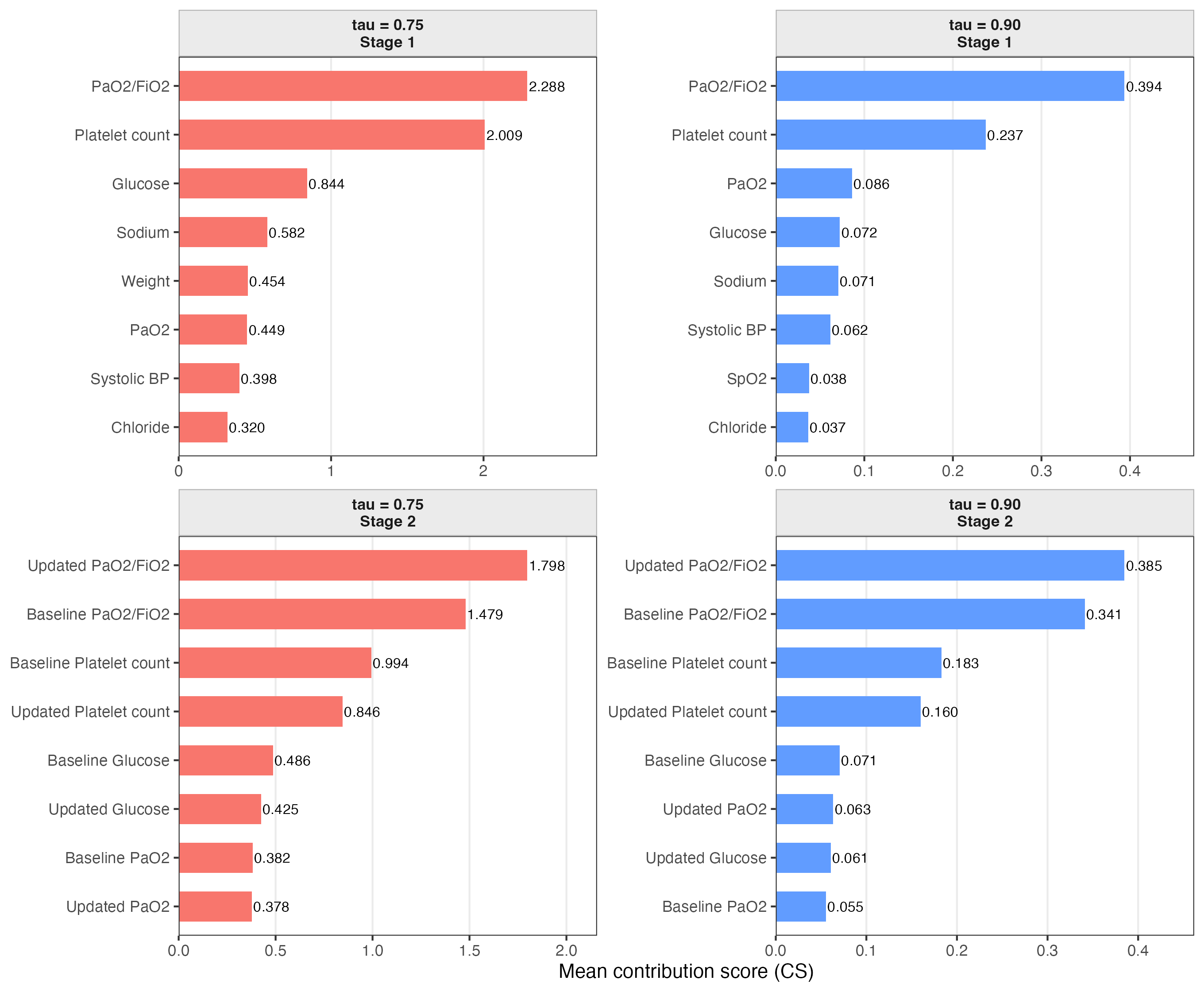}
\caption{Top covariates contributing to the learned RQDTR-C regime in the MIMIC-III sepsis application. ``CS'' is defined as the product of the Euclidean norm of the estimated angle-based score coefficients and the average absolute covariate value in the training sample.}
\label{fig:mimic_variable_contribution}
\end{figure}

Figure~\ref{fig:mimic_variable_contribution} shows the mean contribution score obtained from 100 independent random splits.
Across the two upper-tail quantile levels, variables related to oxygenation and hematologic function were consistently important in the learned RQDTR-C treatment rule. In particular, the PaO$_2$/FiO$_2$ ratio and platelet count appeared among the top-ranked covariates at both stages and both quantile levels. Glucose and PaO$_2$ also contributed substantially across stages, especially through both baseline and updated measurements in the stage-2 rule. In contrast, sodium, chloride, and systolic blood pressure were more important in the stage-1 rule, suggesting that electrolyte balance and hemodynamic status contributed more strongly to the initial resuscitation recommendation than to the stage-2 recommendation. Overall, these patterns indicate that the learned RQDTR-C regime used clinically relevant information on oxygenation status, hematologic function, metabolic burden, electrolyte balance, and hemodynamic status when assigning individualized resuscitation intensity.

\section{Implementation Details}
Across all simulation studies, real-data applications, and sensitivity analyses, RQDTR uses linear-kernel RKHS decision functions. This choice keeps the implementation simple and ensures that the fitted regimes are directly interpretable through their estimated coefficient vectors. Nonlinear Gaussian-kernel extensions can be implemented within the same framework. For the competing methods, SCL uses linear-kernel RKHS decision functions, whereas QIQ-learning and Wang's method use their default linear specifications.

At each treatment stage, propensity scores are estimated using multinomial logistic regression on the full training sample, clipped below at $\pi_{\min}=0.01$, and reused across cross-validation folds. The smoothing parameters are set to $h=0.2/\log n$ for all RQDTR variants, with $\delta=0.1\,\mathrm{SD}(Y)$ for RQDTR-E/C and $\delta=0.1\,\mathrm{SD}(U)$ for RQDTR-U. For RQDTR-C, the risk tolerance is specified as $b_m=c\,\mathcal B_m^{\mathrm{obs}}$, where $\mathcal B_m^{\mathrm{obs}}=\frac{1}{n}\sum_{i=1}^n\{Z_{1i}^{(m)}+Z_{2i}^{(m)}\}$ denotes the empirical average risk burden for the $m$th risk attribute under the observed treatment assignments. Unless otherwise stated, we set $c=0.85$, targeting a 15\% reduction relative to the observed risk burden. In simulations, the learned deterministic regime $\hat d^{(r)}$ is evaluated on an independent test sample, with covariates and risk outcomes generated sequentially under the known data-generating mechanism. The attribute-specific risk burden is estimated by direct Monte Carlo averaging:
\[
\widehat{\mathcal B}_{m,\mathrm{test}}^{(r)}(\hat d^{(r)})
=\frac{1}{n_{\mathrm{test}}} \sum_{i=1}^{n_{\mathrm{test}}}
\left\{Z_{1i}^{(m,r)}(\hat d^{(r)})+Z_{2i}^{(m,r)}(\hat d^{(r)})\right\}.
\]
Constraint satisfaction is determined by comparing this estimate with the training-sample tolerance $b_m^{(r)}$, without an additional feasibility tolerance. Unlike the smoothed IPW estimators used during training, the simulation evaluation uses direct Monte Carlo averaging.

The alternating optimization uses L-BFGS-B through \texttt{optim()} for updating $f$ and \texttt{optimize()} for updating $q$. The positive penalty weights are initialized at $10$ and doubled when the empirical stopping conditions are not met, subject to an upper cap of $10^6$. For the final fit, the algorithm allows up to 20 penalty stages with 100 alternating updates per stage, and terminates successfully when the smoothed survival constraint is satisfied within $\epsilon_S=1/(2\sqrt n)$ and, for RQDTR-C, all smoothed risk constraints are satisfied within $\epsilon_R=1/(2\sqrt n)$. Fits that fail to meet these conditions within the iteration budget or encounter numerical errors are excluded from performance summaries. For cross-validation, the algorithm uses at most 5 penalty stages with 30 alternating updates per stage and tolerances of $\max\{\epsilon_S,0.02\}$ and $\max\{\epsilon_R,0.02\}$, respectively.

The regularization parameter $\lambda$ is selected from a 5-point logarithmically spaced grid over $[10^{-4},1]$ using 5-fold cross-validation with method-specific validation criteria. For RQDTR-E, the criterion is the smoothed IPW tail-probability surrogate evaluated at the quantile threshold estimated from the corresponding training fold. For the main RQDTR-U analyses, we prespecify $\lambda_{\mathcal R}=1$ and $(\rho,\xi)=(1,2)$, yielding the reference cumulative utility $U^{\mathrm{ref}}=\sum_{t=1}^2\{Y_t-\phi_{\mathrm{SE}}(\mathcal R_t;1,2)\}$. The validation utility quantile is estimated using IPW-weighted quantile regression based on hard treatment-path matching, and the criterion is its average across five folds. Thus, all candidate values of $\lambda$ are evaluated under the same utility specification. For RQDTR-C, the criterion is a penalized validation objective consisting of the training-fold quantile threshold minus quadratic penalties for violations of the tail-probability and risk-burden constraints evaluated on the validation fold. Competing methods are tuned according to their default procedures.


For Scenarios A--D, each simulation setting consists of 500 replications, while each real-data application involves 100 training--test splits. Replications in which a method fails to converge are excluded from the performance summaries for that method.
Tables~\ref{tab:simulation_contributing_counts} and \ref{tab:application_contributing_counts} present the number of successful model fits, broken down by setting, method, and quantile level.

\begin{table}[tb]
\centering
\caption{Numbers of valid fits contributing to the simulation results for Scenarios A--D, by method and quantile level. Each setting involves 500 attempted replications.}
\label{tab:simulation_contributing_counts}
\footnotesize
\setlength{\tabcolsep}{12pt}
\begin{tabular}{@{}lrrr@{}}
\toprule
Method & $\tau=0.10$ & $\tau=0.25$ & $\tau=0.50$ \\
\midrule
\multicolumn{4}{@{}l}{\textit{Scenario A}} \\
\addlinespace[2pt]
SCL & 71 & 321 & 487 \\
QIQ & 500 & 500 & 500 \\
Wang & 500 & 500 & 500 \\
RQDTR-E & 498 & 499 & 500 \\
\midrule
\multicolumn{4}{@{}l}{\textit{Scenario B}} \\
\addlinespace[2pt]
OVO-SCL & 500 & 500 & 500 \\
OVO-QIQ & 500 & 500 & 500 \\
OVO-Wang & 500 & 500 & 500 \\
RQDTR-E & 500 & 500 & 500 \\
\midrule
\multicolumn{4}{@{}l}{\textit{Scenario C}} \\
\addlinespace[2pt]
SCL & 54 & 319 & 482 \\
QIQ & 500 & 500 & 500 \\
Wang & 500 & 500 & 500 \\
RQDTR-E & 499 & 492 & 500 \\
RQDTR-U & 494 & 493 & 500 \\
RQDTR-C & 492 & 495 & 500 \\
\midrule
\multicolumn{4}{@{}l}{\textit{Scenario D}} \\
\addlinespace[2pt]
OVO-SCL & 500 & 500 & 500 \\
OVO-QIQ & 500 & 500 & 500 \\
OVO-Wang & 500 & 500 & 500 \\
RQDTR-E & 500 & 500 & 500 \\
RQDTR-U & 460 & 479 & 498 \\
RQDTR-C & 500 & 500 & 500 \\
\bottomrule
\end{tabular}
\end{table}

\begin{table}[tb]
\centering
\caption{Numbers of valid fits contributing to the All of Us MDD and MIMIC-III sepsis results, by method and quantile level. Each application involves 100 attempted training--test splits.}
\label{tab:application_contributing_counts}
\footnotesize
\setlength{\tabcolsep}{10pt}
\begin{tabular}{@{}lrrrr@{}}
\toprule
& \multicolumn{2}{c}{MDD} & \multicolumn{2}{c}{MIMIC-III sepsis} \\
\cmidrule(lr){2-3}\cmidrule(lr){4-5}
Method & $\tau=0.10$ & $\tau=0.25$ & $\tau=0.75$ & $\tau=0.90$ \\
\midrule
OVO-SCL & 100 & 100 & 100 & 100 \\
OVO-QIQ & 100 & 100 & 85 & 70 \\
OVO-Wang & 100 & 100 & 100 & 100 \\
RQDTR-E & 100 & 100 & 100 & 100 \\
RQDTR-U & 100 & 100 & 100 & 100 \\
RQDTR-C & 66 & 88 & 93 & 98 \\
\bottomrule
\end{tabular}
\end{table}

\section{Extensions to Subgroup-Specific and Conditional Risk Constraints}
\label{sec:risk_constraint_extensions}
The risk constraints in the main manuscript bound the expected cumulative risk burden in the target population. Although the treatment decisions $d_t(\bm H_t)$ depend on the observed patient history, the constraints average the resulting risk over patients. Consequently, a regime satisfying $\mathcal B_m(d)\le b_m$ may still yield substantial expected risk in particular patient subgroups when this is offset by lower burdens elsewhere.

A natural extension would impose separate bounds for a fixed, finite collection of prespecified baseline subgroups $\mathcal G_\ell\subseteq\mathcal H_1$, $\ell=1,\ldots,L$, with $p_\ell=\mathbb P(\bm H_1\in\mathcal G_\ell)>0$. For risk attribute $m$, define
\[
\mathcal B_{m,\ell}(d)=\mathbb E\left[\psi_{\mathrm{agg}}\{Z_1^{(m)}(d),Z_2^{(m)}(d)\}|\bm H_1\in\mathcal G_\ell\right],
\]
and impose $\mathcal B_{m,\ell}(d)\le b_{m,\ell}$, where $b_{m,\ell}$ specifies the subgroup-specific risk tolerance. Related subgroup-specific constraints have been considered for single-stage treatment rules \citep{wang2018learning}. Under the identification assumptions in Section~2.1 of the main
manuscript, the subgroup-specific burden admits the representation
$$\mathcal B_{m,\ell}(d) =
\frac{1}{p_\ell} \mathbb E\left[
\frac{I\{\bm H_1\in\mathcal G_\ell\}
\psi_{\mathrm{agg}}\{Z_1^{(m)},Z_2^{(m)}\}
\prod_{t=1}^2 I\{A_t=d_t(\bm H_t)\}}{\pi_1(A_1|\bm H_1)\pi_2(A_2|\bm H_2)}\right].$$
The resulting constraints preserve the finite expectation-constraint structure of the framework without changing the population quantile objective. Establishing corresponding theoretical guarantees requires feasibility and appropriate regularity conditions, while small subgroups and limited treatment overlap may reduce estimation precision.

A further extension would impose covariate-conditional expected-risk constraints of the form
\[
\mathbb E\left[\psi_{\mathrm{agg}}\{Z_1^{(m)}(d),Z_2^{(m)}(d)\}|\bm H_1=h_1\right]\le b_m(h_1)
\]
for $P_{\bm H_1}$-almost every $h_1$, where $b_m(h_1)$ specifies a covariate-dependent risk tolerance. Unlike finite subgroup constraints, this formulation generally involves infinitely many constraints for continuous baseline histories and requires additional methods for conditional risk estimation and uniform error control.


\bibliographystyle{imsart-nameyear} 
\bibliography{references}